# Mid-Infrared Observations of Planetary Nebulae detected in the GLIMPSE 3D Survey

J.A. Quino-Mendoza[1], J.P. Phillips[1], G. Ramos-Larios[1]

[1]Instituto de Astronomía y Meteorología, Av. Vallarta No. 2602, Col. Arcos Vallarta, C.P. 44130 Guadalajara, Jalisco, México   e-mail : jpp@astro.iam.udg.mx

**Resumen**

Presentamos mapas, perfiles y fotometría de 24 nebulosas planetarias (NPs) detectadas en el estudio del plano galáctico en el infrarrojo medio (MIR) de GLIMPSE 3D. Las NPs muestran muchas de las propiedades observadas en estudios previos de estas fuentes, incluyendo la evidencia de emisión a mayores longitudes de onda afuera de las zonas ionizadas, una consecuencia probable por la emisión de hidrocarbonos aromáticos policíclicos (PAHs) dentro de las regiones de fotodisociación (PDRs). Notamos también variaciones en los cocientes de flujo $5.8\mu m/4.5\mu m$ y $8.0\mu m/4.5\mu m$ con respecto a la distancia del núcleo y presentamos evidencia de un aumento en la emisión MIR en los halos de las fuentes – el brillo superficial del halo a $5.8\mu m$ y $8.0\mu m$, cuando se compara con los valores máximos dentro del núcleo parecen ser > 10 veces más grandes que los observados en el visible. Finalmente, presentamos la fotometría de esta muestra inicial de NPs extendidas, encontrando evidencia de posibles variaciones en color con la evolución nebular.

**Abstract**

We present mapping, profiles and photometry for 24 planetary nebulae (PNe) detected in the GLIMPSE 3D mid-infrared (MIR) survey of the Galactic plane. The PNe show many of the properties observed in previous studies of these sources, including evidence for longer wave emission from outside of the ionised zones, a likely consequence of emission from polycyclic aromatic hydrocarbons (PAHs) within the nebular photo-dissociation regimes (PDRs). We also note variations in $5.8\mu m/4.5\mu m$ and $8.0\mu m/4.5\mu m$ flux ratios with distance from the nuclei, and present evidence for enhanced MIR emission in the halos of the sources – 5.8 and 8.0 $\mu m$ halo surface brightnesses, when compared to maximal values within the nuclei, appear to be > 10 times greater than are observed in the visible. We finally present photometry for this initial sample of extended PNe, noting evidence for possible variations in colour with nebular evolution.

**Key Words:** (ISM:) planetary nebulae: general --- infrared: ISM --- ISM: jets and outflows

## 1. Introduction

The late evolution of moderate mass stars is known to lead to an appreciable loss of mass, and the formation of planetary nebular shells. Whilst these envelopes are mostly neutral during the earliest phases of expansion, subsequent evolution of the central star leads to increased HII/HI mass ratios.

Much of the prior investigation of these sources in the radio and visible has concentrated on defining the physical properties and kinematics of the ionised shells. More recent observations in the infrared, however, have also extended our understanding of the neutral and dust emission regimes. Early bolometric observations, for instance, resulted in the detection of emission excesses attributable to hot and warm dust (see e.g. Barlow (1983) for a listing of earlier ground-based photometry). Subsequent advances in technology, and in the sensitivity of detector systems, have also enabled spectroscopy and mapping using ground- and space-based observatories. These include low resolution photometry and mapping undertaken with the Infrared Astronomical Satellite (IRAS, e.g. Leene, Zhang & Pottasch 1987), and spectroscopy of a broad range of sources using the Infrared Space Observatory (ISO; e.g. Liu et al. 2001; Szczerba et al. 1998; Beintema 1997).

Perhaps the greatest advance to have occurred in recent years has been through near-to-far infrared observations with the Spitzer Space Telescope (Spitzer; Werner et al. 2004), however, which has enabled spectroscopy and mapping at high spatial and spectral resolutions. Individually pointed observations have been reported by Hora et al. (2004), Guiles et al. (2007), Su et al. (2007), Ueta (2006), Phillips & Ramos-Laros (2010), and Ramos-Larios et al. (2008), among many others. Similarly, the results of the more extended GLIMPSE and MIPSGAL surveys are reported by Phillips & Ramos-Larios (2008), Cohen et al. (2007), Ramos-Larios & Phillips (2008), Zhang & Kwok (2009), Phillips & Marquez-Lugo (2010), Chu et al. (2009) and Mizuno et al. (2010), These show that much of the MIR emission is attributable to polycyclic aromatic hydrocarbons (PAHs) and broad variety of ionic and molecular transitions, whilst much of the longer wave emission appears to arise from cool dust continua, as well as transitions such as [ArII] $\lambda 6.985$ $\mu$m, [NeVI] $\lambda 7.643$ $\mu$m, [ArIII] $\lambda 8.991$ $\mu$m, [Ne V] $\lambda 24.32$ $\mu$m, [OIV] $\lambda 25.87$ $\mu$m and so forth. The extension of PAH and dust emission components outside of the ionised shells suggests the presence of enveloping photo-dissociation regimes (PDRs), likely associated with neutral gas in the AGB mass-loss halos. Individual structures in these halos have been investigated by Phillips et al. (2009), Ramos-Larios & Phillips (2009), Phillips & Ramos-Larios (2010), and Ramos-Larios, Phillips & Cuesta (2010).

Whilst much of the MIPSGAL survey material is available for public use, the 70 $\mu$m data is inaccessible at present, and the associated catalogues are in process of being compiled. Similarly, although the GLIMPSE I and II data is fully available, only ~75 % of the GLIMPSE 3D survey is accessible at present. Nevertheless, although interesting portions of the Galactic plane are still missing from the latter survey, we are in a position to analyse the properties of the PNe which have been detected so far. The coordinates of these sources are taken from Acker et al. (1992), Parker et al. (2006) and Miszalski et al. (2008), the latter two of which are referred to as the MASH I and II catalogues. These catalogues include the majority of Galactic PNe to have been identified so far, although the sampling of bulge PNe is less complete. A fuller discussion of the latter component can be found in Phillips & Ramos-Larios (2009).

We shall here concentrate upon extended nebular shells having reasonably high S/N detections. This will permit us to provide mapping, profiles and photometry for 24 PNe in the 3.6, 4.5, 5.8 and 8.0 $\mu$m photometric channels. It will become apparent that the sources have interesting characteristics in the MIR which are often at variance with what is observed in other wavelength regimes, and attributable to the combined influence of neutral and ionised gas and dust components.

## 2. Observations

One of the major legacy programs of Spitzer was to map the inner region of the Galactic plane, a program which was undertaken using the Infrared Array Camera (IRAC; Fazio et al. 2004) in four photometric channels. The first results of this survey (denominated GLIMPSE I) covered the longitude and latitude ranges $\ell$ = 10°-65°, 295°-350°, $b$ = ± 1°, and data products were delivered between April 2005 and August 2006. The second extension of this survey (GLIMPSE II) covered regions close to the Galactic centre (GC), including the longitude range 10° < ℓ < 350°, and latitudes extending up to b = ± 2° – although it excludes the region about the G.C. itself (i.e. ℓ = ± 1°, b = ± 0.75°). Products for this survey became available in May 2007. Finally, the GLIMPSE 3D survey represents a vertical extension of the GLIMPSE I & II surveys, and covers higher Galactic latitudes in selected longitude ranges. For the most part, latitudes |b| are < 3 deg, but this extends to |b| < 4.2 deg in the centre of the Galaxy. The data has been delivered in several tranches between 2007 and the present date, and will (when fully available) cover an area of 199.6 deg$^2$; although data for b > 1°, 8° < ℓ < 355°, and 5° < ℓ < 8°, b < -1° is still not available for general use – a region which corresponds to ~25 % of the final survey area. We shall be investigating the properties of planetary nebulae detected in the available 3D regime.

The IRAC employed filters having isophotal wavelengths (and bandwidths Δλ) of 3.550 μm (Δλ = 0.75 μm), 4.493 μm (Δλ = 1.9015 μm), 5.731 μm (Δλ = 1.425 μm) and 7.872 μm (Δλ = 2.905 μm). The normal spatial resolution for this instrument varied between ~1.7 and ~2 arcsec (Fazio et al. 2004), and is reasonably similar in all of the bands, although there is a stronger diffraction halo at 8 μm than in the other IRAC bands. This leads to differences between the point source functions (PSFs) at ~0.1 peak flux. The observations have a spatial resolution of 0.6 arcsec/pixel.

We have used these data to produce colour-coded combined images in three of the IRAC bands, where 3.6 μm results are represented as blue, 4.5 μm emission is green, and the 8.0 μm results are indicated by red. We have also undertaken contour mapping in all four of the IRAC bands, where the lowest contours correspond to $3\sigma_{rms}$ background noise levels or greater, and the intrinsic flux $E_n$ for contour $n$ has a value

$$E_n = A\,10^{(n-1)C} - B \quad \text{MJy sr}^{-1} \quad .......... (1)$$

where $A$ is a constant, $n = 1$ corresponds to the lowest (i.e. the outermost) contour level, $B$ is the background, and $C$ is the logarithmic separation between contours. Contour parameters $A$, $B$ and $C$ are indicated in the relevant figure captions.

We have additionally produced profiles through the sources after the elimination of background emission (particularly strong in the 5.8 and 8.0 $\mu$m IRAC channels), and the removal of linear spatial gradients. These results were then processed to indicate the variation of 3.6$\mu$m/4.5$\mu$m, 5.8$\mu$m/4.5$\mu$m and 8.0$\mu$m/4.5$\mu$m ratios with distance from the nebular nuclei. The rationale behind this is based on the fact that many PNe possess strong polycyclic aromatic hydrocarbon (PAH) emission bands at 3.3, 6.2, 7.7 and 8.6 $\mu$m, located in the 3.6, 5.8 and 8.0 $\mu$m IRAC bands. Given that the 4.5 $\mu$m band is usually dominated by the bremsstrahlung continuum, as well as by a variety of molecular and ionic transitions, such ratios enable us to gain an insight into the distribution of PAHs within the nebular shells (see e.g. Phillips & Ramos-Larios 2008a, 2008b).

Contour mapping of the 3.6$\mu$m/4.5$\mu$m, 5.8$\mu$m/4.5$\mu$m and 8.0$\mu$m/4.5$\mu$m flux ratios was undertaken by estimating levels of background emission, removing these from the images, and subsequently setting values at < 3$\sigma_{rms}$ noise levels to zero. The maps were then ratiod on a pixel-by-pixel basis, and the results contoured using standard IRAF programs. Contour levels are given through $R_n = D10^{(n-1)E}$, where the parameters (*D, E*) are again detailed in the figure captions.

The IRAC fluxes of extended sources are open to errors deriving from instrumental scattering, and require to be corrected by factors of 0.91 at 3.6 $\mu$m, 0.94 at 4.5 $\mu$m, 0.66-0.73 at 5.8 $\mu$m (we take a mean value of 0.70), and 0.74 at 8.0 $\mu$m - although this correction also depends upon the surface brightness distribution of the source under consideration. A discussion of this question is to be found in Sect. 4.11 of the IRAC Instrument Handbook (version 1.0, 2010), located at http://ssc.spitzer.caltech.edu/irac/iracinstrumenthandbook/IRAC_Instrument_Handbook.pdf.

We have, in the face of these uncertainties, chosen to leave the flux ratio mapping and profiles unchanged. The maximum correction factors for the ratios are likely to be > 0.74, but less than unity, and ignoring this correction has little effect upon our interpretation of the results.

We have, finally, provided photometry for all of these sources, where the values were estimated by integrating within circular or polygonal apertures centred upon the nebulae; estimating the levels of background at various regions about the nebulae; applying the instrumental scattering corrections described above; and determined magnitudes using the α Lyrae calibration of Reach et al. (2006). The results are presented in Table 1.

## 3. Discussion of Individual PNe Detected in the GLIMPSE 3D survey

We have obtained mapping and profiles for 14 PNe detected in the GLIMPSE 3D survey, and we shall present details of these sources in our analysis below. Images of six of the regions are also plotted in Fig. 1, where we have combined 3.6 μm (blue), 4.5 μm (green) and 8.0 μm (red) IRAC imaging of the sources. It will be seen that most of the nebulae are relatively compact and ill-defined, although we see evidence for ring-like structures in IC 4674, H 1-6 and M 1-29, and for bipolarity in G009.8-01.1 (otherwise known as PHR1811-2100). Although the range of bright high S/N sources is therefore less extensive than for GLIMPSE I & II (see e.g. Phillips & Ramos-Larios 2008; Ramos-Larios & Phillips 2008), we are able to draw interesting conclusions concerning the properties of these nebulae.

### 3.1 Pe 2-12

This is a relatively little studied PN which appears to have an elliptical morphology (Fig. 1), and an optical diameter of ~ 5 arcsec (Cahn & Kaler 1971). Shell densities appear to be of order $1.1\ 10^3$ cm$^{-3}$ (although this value has a large range of uncertainty; see Chiappini et al. 2009), whilst the He/H abundance ratio is ≅ 0.072, and N/O ≅ 0.714 (Chiappini etal. 2009).

It is apparent that whilst the optical dimensions are comparable to those at 3.6 and 4.5 μm (see Fig. 2), they are significantly less than those observed at larger MIR wavelengths. Such trends are partly affected by the differing instrumental sensitivities in the various IRAC channels, the values of contour level cut-off, and variation of source flux with wavelength, all of which may influence the apparent size of the envelope. There is nevertheless clear evidence that this increase in size reflects a real trend in source dimensions. Logarithmic profiles through the centre of the source permit integration over several pixels, for instance, and lead to higher levels of S/N. These also show an increase in width with increasing IRAC wavelength. Where one ignores the trends observed at 3.6 μm (background field star contamination is extremely high in this channel), then we find 1/10th peak emission widths of 10.0 arcsec at 4.5 μm, 13.0 arcsec at 5.8 μm, and 14.7 arcsec at 8.0 μm. Such a trend is closely comparable to what has been seen in other PNe (e.g. Phillips & Ramos-Larios 2008a; Ramos-Larios & Phillips 2008), and may be attributable to PAH band emission within the nebular PDR; the v=0-0 S(4)-S(7) transitions of $H_2$; and warm dust continua arising from stochastic heating of small grains (Phillips & Ramos-Larios 2008a). It is also conceivable, although perhaps less likely, that lower excitation ionic lines contribute to these trends, including the λ8.99 μm transition of [ArIII]. The absence of spectra for the exterior regions of this source precludes us from constraining these mechanisms, or defining which is the most dominant – and a similar situation applies for other nebulae as well. However, the presence of PAH band emission in PNe such as NGC 40 (Ramos-Larios et al. 2010), and in the PDRs about compact HII regions (Phillips & Ramos-Larios 2008c), suggests that the increase in MIR fluxes to longer wavelengths, and extended emission in the 5.8 and 8.0 μm channels may arise from small PAH molecules within the nebular PDRs. This is perhaps the primary mechanism where C/O > 1.

The evolutionary status of this source is quite uncertain, although one might envisage that where the PN is relatively young (and the dimensions of the ionised regime are small), then we may be observing an envelope which is largely neutral and emerging from the proto-planetary phase of evolution. Having said this, we note that Zhang (1995) quotes a statistical distance of 15.4 kpc, which would place Pe 2-12 well on the other side of the Galactic centre. The corresponding physical size (for an optical diameter of θ = 5 arcsec) would then be of the order of ~0.37 pc, making the source reasonably evolved.

There appears to be an increase in ratios 5.8μm/4.5μm and 8.0μm/4.5μm over the central portions of the source (Fig. 3); a trend which is also confirmed through the IRAC band ratio mapping illustrated in Fig. 4, whence it would seem that nuclear 5.8μm/4.5μm and 8.0μm/4.5μm ratios are low, and surrounded by a collar of higher ratio material. This may arise, yet again, as a result of more extended PAH band emission.

We finally note evidence for a dip in the profiles at the centre of the source, suggesting the possible presence of a central cavity in the longer wave (5.8 and 8.0 μm) emission (see Fig. 3), or a narrowing of the structure associated with a nebular waist. On the other hand, the high ellipticity of the central 8.0 μm contours (see Fig. 2) suggest aspect ratios which are relatively large (and of order ~ 2). Both of these results may imply that the nebula is a barely resolved bipolar source.

### 3.2 IC 4673

Early optical observations of this source quoted by Cahn & Kaler (1971) and Chu et al. (1987) showed it to have a diameter of 14.8 arcsec, compared to radio dimensions of 17 arcsec (Zijlstra et al. 1989) Patriarchi & Perinotto (1994) used an ESO B survey plate to determine a size of 18x15 arcsec$^2$, whilst imaging by Schwarz et al. (1992) appears to show a halo extending to ~25x16 arcsec$^2$. The results of the latter authors have been combined in the Planetary Nebula Image Catalogue (Bruce Balick & collaborators, available at http://www.astro.washington.edu/users/balick/PNIC/) to make an excellent false-colour image of the source which is similar to that in Fig. 1. Patriarchi & Perinotto (1994) suggest that IC 4673 is a Type I source, and their abundances He/H ≅ 0.15 and N/O ≅ 5.57 are certainly consistent with this presumption. However, we note that values [He/H, N/O] of [0.0869, 0.565], [0.132, 0.424] and [0.146, 0.281] have been quoted by Chiappini et al. (2009), Exter et al. (2004) and Kingsburgh & Barlow (1994), and it is possible that the Type I designation may have to be revised.

Phillips (2003) points out that the mean HeII Zanstra is 113.3 kK, whilst the corresponding H Zanstra value is 63.5 kK, perhaps implying a significant loss of H ionising photons by the nebular shell. This may suggest that the nebula is density bound, and that there is little neutral material outside of the ionised regime. Such an analysis would be inconsistent with the assumptions of Petriarch & Perinotto (1994), who assumed that the shell was mostly ionisation bound. It *is* however consistent with the results in Fig. 5.

It is apparent, in the latter case, that the 3.6 and 5.8 μm results are noisier than at 4.5 and 8.0 μm, with the latter emission showing dimensions of 19.6x17.0 arcsec$^2$. These values are comparable to the visual sizes noted above, suggesting that very little of the longer wave emission arises outside of the ionised shell – and that any PDR makes a modest overall contribution.

Profiles through the centre of the source (Figs. 6 & 7, upper panels), show evidence for the minor axis surface brightness enhancements seen in the visible (Fig. 6 and Fig. 1), whilst there is also evidence for a sinusoidal variation in the 8.0μm/4.5μm ratio (Fig. 6, lower panel). It will be noted however that there is little evidence for the radial increase in ratios noted in many other PNe (see e.g. Phillips & Ramos-Larios 2008a, 2010, and sources Pe 2-12 and M 1-29 in the present study), whereby 8.0μm/4.5μm and 5.8μm/4.5μm increase with increasing distance from the nucleus. On the contrary, our ratio mapping (Fig. 8) shows that much of the variation is due to a decrease in ratios at the positions of peak surface brightness. A similar decrease is also noted for the 3.6μm/4.5μm mapping, but appears to be absent from the 5.8μm/4.5μm results.

The fact that such variations are present for 3.6μm/4.5μm and 8.0μm/4.5μm, but absent in the 5.8μm/4.5μm map, suggests that the trends are unlikely to be due to PAH emission bands. It is possible however that ionic lines (such as [ArII] λ6.985 μm and [ArIII] λ8.991 μm) are preferentially enhancing 8.0 μm fluxes.

It would therefore appear likely that IC 4673 possesses a highly ionised envelope with little evidence for a PDR or PAH emission bands, leading to a similar shell morphology and dimensions to those observed in the visible.

### 3.3 NGC 6578

NGC 6578 has an elliptical shell with dimensions $\simeq$ 7.5x25 arcsec$^2$ (see e.g. the imaging of Corradi et al. (2003), and the HST [OIII] image illustrated by Hajian et al. (2007)), surrounded by an irregular outer halo with size 14x60 arcsec$^2$ (Corradi et al. 2003). Modelling of the spectral lines in the source has been undertaken by Hajian et al. (2007), whilst Perinotto et al. (1994) determine abundance ratios He/H $\cong$ 0.14, N/O $\cong$ 0.2; values which are also similar to those estimated by Perinotto, Morbidelli & Scatarzi (2004) (He/H $\cong$ 0.198, N/O $\cong$ 0.21). It would therefore appear that whilst the helium abundance is high, the nebula is not a Type I source.

The mapping in Fig. 9 doesn't bear much relation to the symmetric structure noted in the visible, partially because of the presence of a bright point source close to $\alpha$ = 18:16:16.65, $\delta$ = -20:27:04.7. It is possible that this corresponds to a rather red (peaking at $\lambda \sim 6$ $\mu$m) background object which is unrelated to NGC 6578, and is appreciably contaminating the inner source structure. We have therefore attempted to side-step this feature in our profiles in Fig. 10, where it is clear that trends in all of the channels are very much the same. This results in relatively modest variations in the IRAC band ratios (Fig. 10, lower panel), although there is some evidence for peaks in 5.8$\mu$m/4.5$\mu$m and 8.0$\mu$m/4.5$\mu$m close to the limits of the interior shell (i.e. at relative positions ± 5 arcsec). It is fascinating to note that the visual halo about the source has its counterpart at 8.0 $\mu$m as well, where we see evidence for a circular region with size $\sim$11x12 arcsec$^2$. This is, in fact, a lower limit dimension, and profiles show that the halo extends to distances of at least $\sim$17 arcsec from the centre. Such a result is, however, somewhat surprising, given the weakness of the halo in the visible. Corradi et al. (2003) find that the halo trends for H$\alpha$+[NII] and [OIII] are very closely similar, and imply intensities of $\sim$0.15 at RP = 5 arcsec, $\sim$5 10$^{-3}$ at 10 arcsec, and $\sim$10$^{-3}$ at 20 arcsec when compared to peak strengths within the nucleus of the source. This is well below what would ordinarily be detectable with Spitzer. The reason why we are detecting the halo in the present observations is because it is relatively very much brighter in the MIR, and has a relative intensity $\sim$2 10$^{-2}$ at distances of 15 arcsec from the centre.

So, the halo surface brightness in the MIR, when compared to peak values in the nucleus, is >10 times greater than observed in the visual. This unusual behaviour appears typical of many PNe (see e.g. Phillips et al. 2009; Ramos-Larios & Phillips 2009), and may presumably arise due to H$_2$ emission, PAH emission bands, or as a result of exceptionally strong low excitation transitions such as [ArII] $\lambda$6.985 $\mu$m or [ArIII] $\lambda$8.991 $\mu$m. Much depends, in this interpretation, on whether the halo is ionised or neutral, or perhaps even composed of a mixture of the two, and it is still far from clear what the primary emission mechanisms in this present source might be.

### 3.4 Sp 1

Sp 1 is one of the most interesting of the sources to fall within the limits of the GLIMPSE 3D survey. The nebula has a closely circular structure with diameter ~120 arcsec (e.g. Miszalski et al. 2009); a large [OIII] expansion velocity of ~60 km s$^{-1}$ (Meatheringham et al. 1988); and moderate H and HeII Zanstra temperatures of ~32.5 and 82.0 kK (Phillips 2003). The difference between these values may indicate the escape of H ionising photons, and a nebular structure which is density bound. An image of the source (taken by D. Malin) is available at http://www.aao.gov.au/images/captions/aat095.html.

By far the most interesting characteristic of this angularly large PN, however, is that it appears to be centred on a close binary system with period 2.91 days, having an amplitude variation of 0.1 mag in B (e.g. Bond & Livio 1990). Although the binary is a non-eclipsing system, it may be exhibiting radiation effects (e.g. Mendez et al. 1988).

It has popularly been supposed that close binary systems result in bipolar outflows, and Sp 1 would therefore appear to be in conflict with this presumption; a problem which also occurs for SuWt 2 and WeBo 1 (Bond, Ciardullo & Webbink 1996), both of which are non-bipolar sources centred upon close binary systems. One way out of this predicament has been to suppose that Sp 1 may be being viewed head-on, and that the ring corresponds to the toroidal waist of the bipolar PN (Bond & Livio 1990). Similar hypotheses have also been proposed for NGC 6720 (Bryce, Balick & Meaburn 1994) and NGC 6781 (Ramos-Larios, Phillips & Guerrero, in preparation).

The present contour results are shown in Fig. 11, where it is clear that emission from the source is very faint indeed, and barely detected in the MIR. The apparent extension of the 8.0 $\mu$m shell to the left-hand side is a result of the contaminating effects of background emission. Excluding this factor, and taking into account the extreme noisiness of the results, it is clear that the MIR and optical results are likely to be reasonably consistent.

## 3.5 H 1-6

H 1-6 appears to be a reasonably bright PN at infrared wavelengths, although it has been very little studied in other wavelength regimes. The elliptical shell (Figs. 1 & 12) has dimensions of ~16.3x14.4 arcsec$^2$. Profiles through the source are illustrated in Fig. 13, and show a double-humped structure arising from an internal ring. The 3.6-5.8 $\mu$m profiles are strongly dominated by a star at RP ~ -5 arcsec, however, and this causes some distortion of the IRAC band ratios (Fig. 13, lower panel).

## 3.6 H 1-7

This nebula appears to have a rather irregular morphology at optical wavelengths (Schwarz et al. 1992), and has [SII] densities which are moderately high (~ 4.2 10$^3$ cm$^{-3}$; Chappini et al. 2009). The N/O ratio of $\cong$ 0.56, and a value He/H $\cong$ 0.12 may imply that the source is a Type I PN (Chiappini et al. 2009). The optical diameter is quoted as 8.5 arcsec (Perek & Kohoutek 1967), which is significantly smaller than any of the contour maps in Fig. 14, and less than half of the size observed at 8.0 $\mu$m (~ 22 arcsec). This, and the tendency of the source to become larger towards longer IRAC wavelengths, may suggest that much of the longer wave emission is associated with the nebular PDR.

Further evidence for this is provided in Fig. 15, where we show logarithmic profiles through the centre of the source (upper panel), and IRAC band ratios in the central portion of the envelope (lower panel). It is apparent from this that the shorter wave profiles (3.6 and 4.5 $\mu$m) cut-off close to RP $\cong$ -15 arcsec and +19 arcsec (an extension to RP ~ -25 arcsec is likely to be due to an unrelated field star, having as it does stronger emission in the shorter wave IRAC channels), whilst the 8.0 $\mu$m profile extends to RP $\cong$ -36 arcsec and +32 arcsec. There is, apart from this, very little discernible structure, and the appearance is similar to those of many other PNe observed in this survey. Examples of these (mostly more compact) sources are discussed in Sect. 3.11.

Finally, the differing IRAC band ratios are shown in the lower panel of Fig. 15. It is again clear, as for Pe 2-12 (Sect. 3.1), that there is a marked increase in 5.8μm/4.5μm and 8.0μm/4.5μm ratios close to the edges of the central shell, perhaps arising from an increase in PAH emission within an enveloping PDR. There is little comparable variation for 3.6μm/4.5μm, however, which remains (for the most part) close to unity.

### 3.7 M 1-29

M 1-29 is an interesting PN with optical dimensions of ~7.6 arcsec (Cahn & Kaler 1971; Moreno et al. 1988). The kinematics of the shell appear complex, with evidence for velocities of up to ± 50 km s$^{-1}$ (Gesicki et al. 2003), and this, together with the VLA mapping of Zijlstra et al. (1989), has been used to suggest that it may be a poorly resolved bipolar outflow. Perusal of the VLA mapping suggests that it is more likely to have an elliptical structure, however, with higher levels of surface brightness at the minor axis limits of the shell. The size of the envelope at radio wavelengths appears to be ≅ 11.3x6.5 arcsec$^2$. Estimates of the ratios [He/H, N/O] have varied between [0.15, 0.93] (Chiappini et al. 2009), [0.145, 0.68-0.7] (Wang & Liu 2007), and [0.14, 0.32] (Perinotto 1991). It would seem clear, from the more recent estimates, that the nebula should be classified as a Type I source. The only previous analysis of the object in the infrared appears to have been undertaken by Cassasus et al. (2001), who find that the spectrum is flat between 7 and 14 μm, and shows no evidence for dust emission continua.

It appears clear, from our contour mapping (Fig. 16), that there is a marked increase in dimensions between 3.6 and 8.0 μm, with the overall dimensions at the latter wavelength being of order ~ 13 arcsec; a value which is rather larger than is determined for the radio and optical regimes. This is confirmed through logarithmic profiles through the centre of the source, which show a 1/10$^{th}$ peak width of ~11.3 arcsec at 3.6 μm, of 13.7 arcsec at 5.8 μm, and of ~15 arcsec at 8.0 μm. The overall 8.0 μm dimensions at 0.6% peak height – the lowest level at which it appears possible to detect the envelope above background fluctuations and noise – are of the order of ~39.7 arcsec.

The 8.0μm/4.5μm ratio also increases from a value ~2.6 at the centre, to ≈ 8-10 at RP ~ 7 arcsec (Fig. 17); a tendency which is confirmed in the ratio maps illustrated in Fig. 18. It is therefore likely that much of the longer wave emission arises from outside of the ionised shell.

### 3.8 MeWe 1-6

This is a recently discovered example of an old PN with extremely low levels of surface brightness in the visible (Melmer & Weinberger 1990). The structure seems to take the form of a thick-ringed shell with dimensions of order ~16 arcsec. There have however, apart from this, been no other studies of this source.

Our IRAC contour maps are illustrated in Fig. 19, whence it is again evident that surface brightnesses are low, and that the object is located in a region of high field star densities. This makes it difficult discern the shell at 3.6 and 5.8 μm. Profiles through the envelope (Fig. 20) reveal the expected double-humped structure, and suggest 8.0 μm dimensions ~21.6 arcsec which are slightly larger than in the visible.

### 3.9 PN G009.8-01.1 (PHR1811-2100)

This source represents one of the many new PNe detected in the MASH I survey, where Hα images show the presence of a bright rectangular band surmounted by orthogonal fainter wing-like structures. The source is classified as a bipolar PN. The published spectrum appears to be extremely noisy, although Parker et al. (2006) suggest that [NII] > 4Hα, and [OIII] > Hβ.

Our present contour maps (Fig. 21) are dominated the bright interior emission regime, with evidence for 8.0 μm extensions to ~25 arcsec. The minor axis width is ~12 arcsec. This compares to dimensions 32x12 arcsec$^2$ quoted for the Hα image. It would therefore appear that we are detecting most of the region defined through optical imaging.

Profiles through the source are illustrated in Figs. 22 & 23, from which it is clear that there is a small dip in 8.0 μm emission close to the centre of the minor axis traverse (Fig. 23). This may indicate barely resolved evidence for a central, inclined toroidal structure. The fact that this dip is evident 8.0 $\mu$m, but absent at shorter wavelengths may arise where 8.0 $\mu$m fluxes are enhanced within the toroid – perhaps as a result of warm dust emission or fluorescently or shock excited $H_2$. The 3.6 and 4.5 $\mu$m fluxes, by contrast, are more likely to arise from ionic transitions, and to trace components of gas within the limits of this structure.

The IRAC band ratios in the lower panels are highly variable with relative position, and show evidence for lower 8.0$\mu$m/4.5$\mu$m and 5.8$\mu$m/4.5$\mu$m ratios close to the centre of the source. These trends are also evident in our contour mapping in Fig. 24, where we see a minimum in 5.8$\mu$m/4.5$\mu$m and 8.0$\mu$m/4.5$\mu$m ratios close to the centre of the source, and larger values towards the periphery. There is a great deal of similarity between the longer wave ratio maps, suggesting that their fragmentary appearance is real, and corresponds to actual variations in the ratios. A similarly high level of complexity is also noted in the 3.6$\mu$m/4.5$\mu$m map, although the central minimum is less strong.

### 3.10 PN G344.4+01.8 (MPA1654-4041)

This source was first detected in the MASH II survey, and is quoted as having an elliptical morphology with dimensions $\cong$15x9 arcsec$^2$ (Miszalski et al. 2008). It is also defined as being a "likely" PN, through there is no visual spectrum by which this identification can be assessed.

The dimensions of the 8.0 $\mu$m mapping (~15.6x11 arcsec$^2$; see Fig. 25) are closely similar to those determined in the visual, and although there appears to be an increase in dimensions between 3.6 and 8.0 $\mu$m, it is likely that much of this is due to the low S/N of our results. Profiles along the major and minor axes of the source are illustrated in Fig. 26, from which it is clear that there is a bright unresolved nucleus at 3.6-5.8 $\mu$m, and a somewhat broader structure at 8.0 $\mu$m. The minor axis variation is of particular interest, in that it shows a systematic fall-off with distance from the nucleus. This is untypical of what is observed in most other PNe, where we normally see evidence for well defined central shells, rather than continuous stellar-wind type fall-offs. Such variations have only previously been noted in bipolar flows (see e.g. Phillips & Ramos-Larios 2008, 2010), where CS winds are important in defining the kinematics. This trend may therefore place the identification of the source in doubt.

### 3.11 Other PNe Detected in the GLIMPSE 3D survey

It is apparent that the haul of extended, higher surface-brightness PNe is rather slim – particularly when compared to what has been observed in previous surveys (e.g. Phillips & Ramos-Larios 2008; Ramos-Larios & Phillips 2008). This was, in general, broadly to be expected, given that the 3D survey covers regions which are further away from the Galactic plane, and at larger positive and negative latitudes. The spatial density of PNe is very much reduced compared to those observed at lower Galactic latitudes

It is finally worth noting that a good fraction of the PNe detected in this survey were either extremely compact and unresolved, or had a size which was too modest to permit a useful analysis. Four examples of the latter sources are shown in Fig. 27, where we illustrate profiles through the centres of M 1-27, M 1-35, Pe 1-6 and Pe 1-15. The most extended of these sources are M 1-27 and Pe 1-6, both of which have circular/elliptical morphologies. It is clear however that it would be difficult to reliably assess their properties, and/or assess spatial trends in their emission.

### 4. Photometry of the GLIMPSE 3D Sources.

We have finally assessed fluxes for the more extended sources detected in the GLIMPSE 3D survey, and results are provided in Table 1. This represents the first part of a more extended program of photometry which will encompass extended and compact sources alike. Columns 1 and 2 of the table give the Galactic coordinates and normally recognised names, whilst column 3 indicates the catalogue from which the positions were acquired (MI corresponds to MASH I (Parker et al. 2006), MII is MASH II (Miszalski et al. 2008), and Ack corresponds to Acker et al. (1992)). The remaining four columns give the magnitude in each of the IRAC bands, determined using the zero magnitude calibration of Reach et al. (2006) and the techniques described in Sect. 2. Typical errors range from 0.03 mag for the fainter sources through to 0.01 or less for the brightest.

It is of interest to compare these present results with photometry of other PNe, and this is undertaken in the [3.6]-[4-5]/[5.8]-[8.0] colour plane in Fig. 28. In this latter case, MASH I sources are indicated by squares, the Acker et al. PNe are identified through disks, and the single MASH II PN is represented by a triangle. The solid lines indicate the limits for PNe detected through previous photometry, and are taken from the compilation of results by Phillips & Ramos-Larios (2010).

It is finally worth noting that previous photometry of PNe has sometimes used slightly differing correction factors for the diffuse background emission (see Sect. 2). These correction factors have proven extremely difficult to pin down, and officially quoted estimates have varied over time. It follows that care much be taken when comparing differing sets of data.

For the most part, it is clear that the 3.6 and 4.5 $\mu$m corrections factors are similar, however, and the same applies for the 5.8 and 8.0 $\mu$m factors as well. Changes in the values of these factors therefore lead to only minor variations in colour, and a comparison between differing sets of data is expected to be reliable.

It is clear that our present results fall neatly within the range for other PNe, although they occupy a region which is somewhat more restricted. In particular, it would seem that there is a paucity of indices [5.8]-[8.0] < 1. This may, in part, testify to lower levels of stellar contamination, whether this occurs due to unrelated field stars or the PN central stars themselves. It can also however occur where there is strong shock-excited $H_2$ emission, higher excitation gas thermal emission, and/or strong dust continua or PAH emission bands. A further interesting aspect of the results is that there appears to be a tendency for the Acker et al. sources (disks) to have higher values of [5.8]-[8.0]. Specifically, whilst 80% of MASH I sources have indices smaller than [5.8]-[8.0] = 2.05 mag, the corresponding fraction for the Acker et al. nebulae is closer to 25 %. Since the MASH I sources (squares) are normally regarded as being larger, more evolved, and to have lower levels of surface brightness, this may imply that there is an evolutionary trend within the colour plane – a trend which is related to a variation in the strength of the emission mechanisms cited above. It is possible for instance that ionic transitions (such as [ArII] $\lambda$6.985 $\mu$m and [ArIII] $\lambda$8.991 $\mu$m) become weaker as nebulae evolve; although the reduction of nebular excitation with increasing size, and evolution of the central stars to lower temperatures might be expected to have precisely the reverse effect. It is also conceivable that PAH band excitation becomes less strong as nebulae expand, and that radiative excitation of the molecules decreases. This would affect 8.0 $\mu$m band fluxes more than those as 5.8 $\mu$m, and lead to decreases in the [5.8]-[8.0] indices. Finally, the paucity of lower [5.8]-[8.0] sources may reflect a bias in the sample, such that the proportion of more evolved PNe is less than in previous studies of these nebulae.

It is clear however that these evolutionary processes are no more than possible explanations, and further spectroscopy is required before we can understand such trends.

## 5. Conclusions

We have undertaken a provisional survey of PNe detected in the GLIMPSE 3D survey. This has involved mapping the shells in the 3.6, 4.5, 5.8 and 8.0 bands, as well as obtaining profiles through the nuclei, and maps of flux ratio parameters. We have detected many of the signatures discovered in previous mapping of such sources, including increases in the 5.8$\mu$m/4.5$\mu$m and 8.0$\mu$m/4.5$\mu$m ratios with distance from the nuclei, and emission from enveloping PDRs (Pe 2-12, H1-7, M 1-29); and evidence for enhanced levels of emission in the halos of the sources (NGC 6578) when compared to trends observed in the visual wavelength regime. It is therefore clear that the MIR characteristics of the PNe are sometimes radically different from those observed in other wavelength regimes.

We have also presented the first phase of photometry for GLIMPSE 3D sources, limited (in this case) to PNe having evidence for spatial extension. The colour-colour mapping of these results appears similar to what has been observed for other PNe, although we also note evidence for a possible secular evolution in the [5.8]-[8.0] indices. The more evolved GLIMPSE I sources appear to have lower mean colours, whilst the (mostly) less evolved Acker et al. nebulae have larger values of this parameter. This may reflect corresponding variations in nebular emission mechanisms.

**Acknowledgements**

This work is based, in part, on observations made with the Spitzer Space Telescope, which is operated by the Jet Propulsion Laboratory, California Institute of Technology under a contract with NASA. GRL acknowledges support from CONACyT (Mexico) grant 132671 and PROMEP (Mexico).

Table 1

MIR Photometry for Extended GLIMPSE 3D Planetary Nebulae

| PN G | NAME | CAT | [3.6] mag | [4.5] mag | [5.8] mag | [8.0] mag |
|---|---|---|---|---|---|---|
| 002.8-02.2 | Pe 2-12 | ACK | 8.59 | 8.13 | 6.44 | 4.75 |
| 003.5-02.4 | IC 4673 | ACK | 9.47 | 7.66 | 8.45 | 5.64 |
| 003.6-02.3 | M 2-26 | ACK | 10.30 | 9.54 | 9.95 | 7.27 |
| 003.9-02.3 | M 1-35 | ACK | 9.06 | 8.02 | 7.85 | 5.17 |
| 004.2-02.5 | PHR1805-2631 | M1 | 11.68 | 10.43 | 11.09 | 9.64 |
| 009.4-01.2 | PHR1811-2123 | M1 | 9.27 | 9.10 | 9.28 | 7.88 |
| 009.7-01.1 | PHR1811-2102 | M1 | 11.90 | 11.35 | 10.73 | 9.53 |
| 009.8-01.1 | PHR1811-2100 | M1 | 10.11 | 9.28 | 8.79 | 7.75 |
| 010.8-01.8 | NGC 6578 | ACK | 7.81 | 6.99 | 7.43 | 5.06 |
| 025.9-02.1 | Pe 1-15 | ACK | 11.17 | 9.83 | 10.29 | 7.51 |
| 330.6-02.1 | He 2-153 | ACK | 9.63 | 8.76 | 8.57 | 6.86 |
| 335.4-01.9 | PHR1637-4957 | M1 | 9.83 | 9.10 | 9.22 | 7.86 |
| 336.2+01.9 | Pe 1-6 | ACK | 10.39 | 9.19 | 9.78 | 6.91 |
| 344.2-01.2 | H 1-6 | ACK | 9.36 | 8.02 | 7.65 | 6.01 |
| 344.4+01.8 | MPA1654-4041 | M2 | 11.20 | 10.41 | 11.21 | 9.20 |
| 345.2-01.2 | H 1-7 | ACK | 6.67 | 6.18 | 4.79 | 2.81 |
| 350.4+02.0 | PHR1712-3543 | M1 | 10.92 | 9.75 | 10.14 | 8.20 |
| 356.0-01.4 | PPA1741-3302 | M1 | 11.29 | 10.63 | 11.09 | 8.42 |
| 356.5-02.3 | M 1-27 | ACK | 8.09 | 7.36 | 6.67 | 4.39 |
| 356.5-01.8 | PPA1744-3252 | M1 | 10.50 | 10.15 | 10.82 | 8.82 |
| 357.3-02.0 | PPA1747-3215 | M1 | 9.77 | 9.65 | 8.62 | 7.20 |
| 357.5-02.4 | PPA1749-3216 | M1 | 10.19 | 9.87 | 10.61 | 8.45 |
| 359.1-01.7 | M 1-29 | ACK | 8.19 | 6.99 | 6.93 | 4.57 |
| 359.4-03.4 | H 2-33 | ACK | 11.30 | 10.17 | 10.60 | 7.57 |

**Figure Captions**

**Figure 1**

Images of six planetary nebulae observed during the Spitzer GLIMPSE 3D survey, where 3.6 μm emission is represented as blue, 4.5 μm emission is green, and 8.0 μm emission is red. Three of the sources appear to have internal ring-like structures (viz. IC 4673, H 1-6 and M 1-29), whilst the MASH I source G009.8-01.1 (PHR1811-2100) has a bipolar morphology.

**Figure 2**

Contour maps of Pe 2-12, where contour levels [A, B, C] are given by [9, 3.65, 0.1780] at 3.6 μm, [10, 2.58, 0.1111] at 4.5 μm, [10, 4.44, 0.1780] at 5.8 μm, and [17.5, 10.33, 0.1780] at 8.0 μm. it will be noted that the dimensions of the source appear to increase with increasing MIR wavelength, and that the internal contours are highly elliptical, perhaps indicative of a barely resolved bipolar structure.

**Figure 3**

Profiles along the major axis of Pe 2-12, where we show the variation in surface brightness with nebular position (upper panel), and of the flux ratios 3.6μm/4.5μm, 5.8μm/4.5μm, and 8.0μm/4.5μm (lower panel). The inserted images show the directions and widths of the traverses used for these trends. It will be noted that the longer wave profiles have a somewhat double-humped appearance, as might be expected where we are observing the two lobes of a bipolar structure – and there is indeed some further indication of this structure in the inserts, and in the contour maps in Fig. 2. Note the evidence for an increase in flux ratios with distance from the nucleus.

**Figure 4**

Mapping of IRAC flux ratios over the projected envelope of Pe 2-12, where the contour parameters [D,E] are given by [0.5, 0.0618] for 3.6μm/4.5μm, [1.6, 0.0972] for 5.8μm/4.5μm, and [2.0, 0.1307] for 8.0μm/4.5μm. Notice how ratios are lowest (lighter grey colouration) at the centre of the source, and increase outwards towards the perimeter. The region of lower ratios to the southwest (lower right-hand side) results from contamination by an unrelated field star.

**Figure 5**

Contour mapping of IC 4673 in the MIR, where the contour parameters [A, B, C] are given by [3.9, 2.48, 0.0478] at 3.6 μm, [3.2, 1.50, 0.0972] at 4.5 μm, [5.7, 3.57, 0.0450] at 5.8 μm, and [17, 13.04, 0.0521] at 8.0 μm. The shell is seen to have an elliptical morphology, with higher levels of surface brightness along the minor axis of the source. The fragmentary appearance of the envelope at 3.6 and 5.8 μm is due to lower levels of S/N.

**Figure 6**

As for Fig. 3, but for the case of IC 4673. The profiles are taken along the major axis of the source, and show a double-humped profile. There is also evidence for lower 3.6μm/4.5μm and 8.0μm/4.5μm ratios at the positions of peak surface brightness.

**Figure 7**

As for Fig. 6, but for an orthogonal direction through the nucleus of the source.

**Figure 8**

IRAC band ratio mapping for IC 4673, where the parameters [D, E] are given by [0.2, 0.0865] for 3.6μm/4.5μm, [0.2, 0.0777] for 5.8μm/4.5μm, and [1.7, 0.0943] for 8.0μm/4.5μm. The structure of the 5.8μm/4.5μm map is highly complex, and shows little evidence for systematic changes with position in the shell. The 3.6μm/4.5μm and 8.0μm/4.5μm maps, on the other hand, show evidence for ratio minima along the minor axis of the source, at the positions of maximum surface brightness.

**Figure 9**

IRAC contour mapping of NGC 6578, where parameters [A, B, C] are given by [11, 2.56, 0.2680] at 3.6 μm, [6, 1.56, 0.3051] at 4.5 μm, [12.5, 6.50, 0.2513] at 5.8 μm, and [25, 19.67, 0.1947] at 8.0 μm. The map shows evidence for a bright interior elliptical shell, and a more circular halo at longer wavelengths. The point source close to $\alpha$ = 18:16:16.65, $\delta$ = -20:27:04.7 is likely to correspond to an unrelated field star.

**Figure 10**

As for Fig. 3, but for the case of NGC 6578. The profiles avoid the region of contamination by the MIR point source. Note the increase in 5.8μm/4.5μm and 8.0μm/4.5μm close to the limits of the bright interior shell.

**Figure 11**

Mapping of Sp 1 in the four IRAC bands, where the contour parameters [A, B, C] are given by [1.18, 0.74, 0.1119] at 3.6 μm, [1, -0.05, 0.0939] at 4.5 μm, [6, 3.89, 0.0442] at 5.8 μm, and [19, 17.33, 0.0048] at 8.0 μm. This source is close to the limits of detection for the GLIMPSE 3D survey, and levels of S/N are extremely low. Most of the fragmentary appearance can be attributed to noise. The apparent extension of the 8.0 μm shell to the east results from contamination by background emission.

**Figure 12**

Mapping of H 1-6 in the four IRAC bands, where the contour parameters [A, B, C] are given by [3.3, 1.64, 0.1126] at 3.6 μm, [2.4, 0.63, 0.1250] at 4.5 μm, [9.5, 6.40, 0.0427] at 5.8 μm, and [30, 25.62, 0.0227] at 8.0 μm.

**Figure 13**

As for Fig. 3, but for the case of H 1-6. The peaks in the 3.6-5.8 μm channels close to RP ~ -5 arcsec are mostly attributable to emission by an unrelated field star. This also leads to minima in the 8.0μm/4.5μm and 5.8μm/4.5μm ratios (lower panel), and a corresponding maximum in 3.6μm/4.5μm. There is, apart from this component, not too much evidence for variations in the IRAC ratios.

**Figure 14**

IRAC mapping of H 1-7, where the contour parameters [A, B, C] are given by [5.8, 1.75, 0.2313] at 3.6 μm, [4.3, 0.57, 0.2150] at 4.5 μm, [12, 5.71, 0.1592] at 5.8 μm, and [30.6, 17.23, 0.1720] at 8.0 μm. Although the innermost contours are elliptical, the outer structure appears to be more circular. Note the strong variation in dimensions between shorter and longer MIR wavelengths.

**Figure 15**

As for Fig. 3, but for the case of H 1-7; where in this case, the upper profiles are represented in terms of logarithmic surface brightness. The profiles have a mostly Gaussian emission fall-off, whilst there is evidence for an increase in 8.0μm/4.5μm and 5.8μm/4.5μm ratios towards the limits of the interior shell.

**Figure 16**

Mapping of M 1-29 in the four IRAC channels, where the contour parameters [A, B, C] are given by [12, 5.12, 0.1502] at 3.6 μm, [12, 3.10, 0.1192] at 4.5 μm, [12, 6.55, 0.0950] at 5.8 μm, and [25, 17.32, 0.1242] at 8.0 μm. Note again how the dimensions of the source increase as the wavelength of observation becomes larger.

**Figure 17**

As for Fig. 3, but for the case of M 1-29, where it is seen that the longer wave profiles have a double-humped appearance. The 8.0μm/4.5μm ratio in the inner portions of the shell increases with increasing distance from the nucleus.

**Figure 18**

IRAC band ratio mapping over the projected envelope of M 1-29, where the contour parameters [D, E] are given by [0.2, 0.0930] for 3.6µm/4.5µm, [0.5, 0.0669] for 5.8µm/4.5µm, and [2.4, 0.0884] for 8.0µm/4.5µm. There is evidence for an increase in ratios with radial distance from the nucleus, in conformity with the profiles illustrated in Fig. 17.

**Figure 19**

IRAC mapping of MeWe 1-6, where the contour parameters [A, B, C] are given by [2.28, 1.33, 0.1094] at 3.6 µm, [2.0, 1.00, 0.1179] at 4.5 µm, [7.8, 5.83, 0.0528] at 5.8 µm, and [23.0, 19.62, 0.0128] at 8.0 µm. It is clear that the source has a ring-like morphology at 4.5 and 8.0 µm which is closely comparable to that observed in the visible. However, the high field star densities, and low surface brightness of the envelope make it much more difficult to discern at 3.6 and 5.8 µm.

**Figure 20**

As for Fig. 3, but for profiles through MeWe 1-6. The double-humped structure arises from the ring-like morphology of the source.

**Figure 21**

IRAC band mapping of the MASH I source PN G009.8-01.1 (PHR1811-2100), where the contour parameters [A, B, C] are given by [2.8, 2.35, 0.0995] at 3.6 µm, [3.8, 1.57, 0.0909] at 4.5 µm, [12.9, 11.00, 0.0374] at 5.8 µm, and [39, 34.77, 0.0398] at 8.0 µm. The nebula has a highly elongated appearance at 8.0 µm, and an orthogonal bright interior band. Both of these features are consistent with a bipolar structure.

**Figure 22**

As for Fig. 3, but for the major axis of PN G009.8-01.9 (PHR1811-2100). The centre of the source has a Gaussian emission fall-off, whilst ratios 8.0µm/4.5µm and 5.8µm/4.5µm have maxima at ± 4 arcsec from the nucleus.

**Figure 23**

As for Fig 20, but for a minor axis traverse through the nucleus. The longer-wave profiles have a slightly double-humped appearance, perhaps indicating a toroidal internal structure.

**Figure 24**

IRAC band ratio mapping over the envelope of PN G009.8-01.1 (PHR1811-2100), where the contour parameters [D, E] are given by [0.4, 0.0569] for 3.6μm/4.5μm, [0.5, 0.0865] for 5.8μm/4.5μm, and [2.0, 0.0752] for 8.0μm/4.5μm. It would seem that ratios vary over quite small spatial scales, with little systematic variation from one part of the shell to the other. The central ratio values are in all cases low.

**Figure 25**

IRAC mapping of the MASH II PN G344.4+01.8 (MPA1654-4041), where the contour parameters [A, B, C] are given by [1.7, 0.65, 0.0777] at 3.6 μm, [1.5, 0.59, 0.0854] at 4.5 μm, [4.27, 2.92, 0.0409] at 5.8 μm, and [15.8, 13.29, 0.0231] at 8.0 μm. This is an extremely low surface brightness structure in most of the bands, although the elliptical morphology is clearly visible at 8.0 μm.

**Figure 26**

Major and minor axis profiles though PN G344.4+01.8 (MPA1654-4041). Note the presence of a bright and unresolved core source, and the steep and monotonic fall-off in emission along the minor axis (lower panel). This latter variation is untypical of elliptical PNe.

**Figure 27**

Profiles through the nuclei of four compact PNe. These sources have an elliptical/circular appearance, and are typical of a large fraction of the PNe detected in the GLIMPSE 3D survey.

**Figure 28**

The distribution of GLIMPSE 3D sources in the [3.6]-[4.5]/[5.,8]-[8.0] colour plane, where the Acker et al. (1992) sources are represented as disks, MASH I sources are squares, and the sole MASH II PN is represented by a triangle. The solid lines indicate the approximate limits for Galactic PNe deduced from previous photometry of these sources. There is a tendency for MASH I sources to have somewhat lower indices [5.8]-[8.0], and for the Acker et al. sources to have larger values for these parameters; a situation which, if confirmed, might indicate a secular evolution in MIR colours.

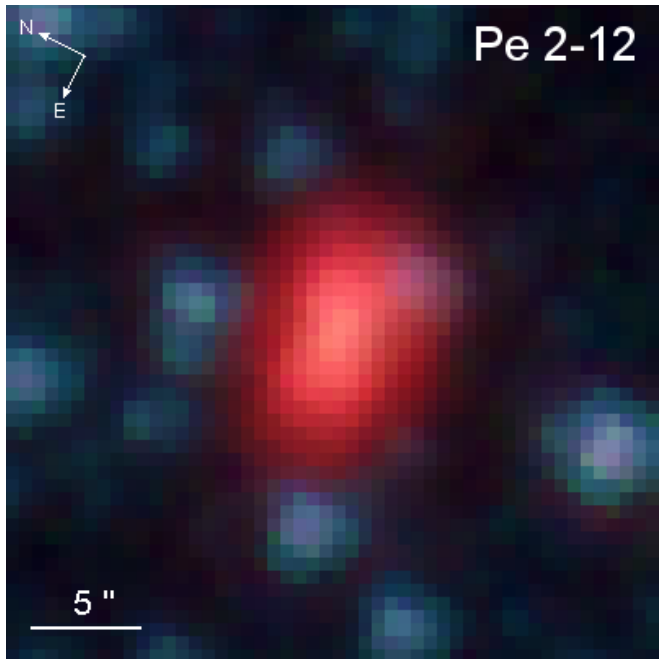 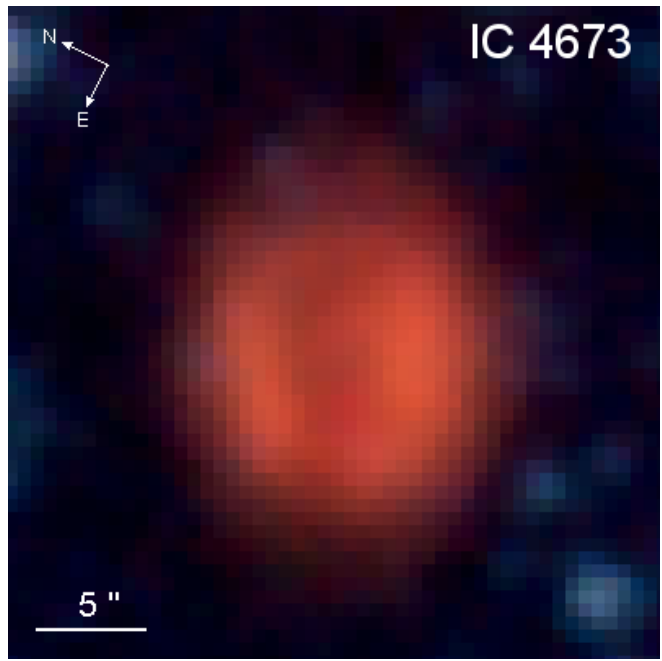 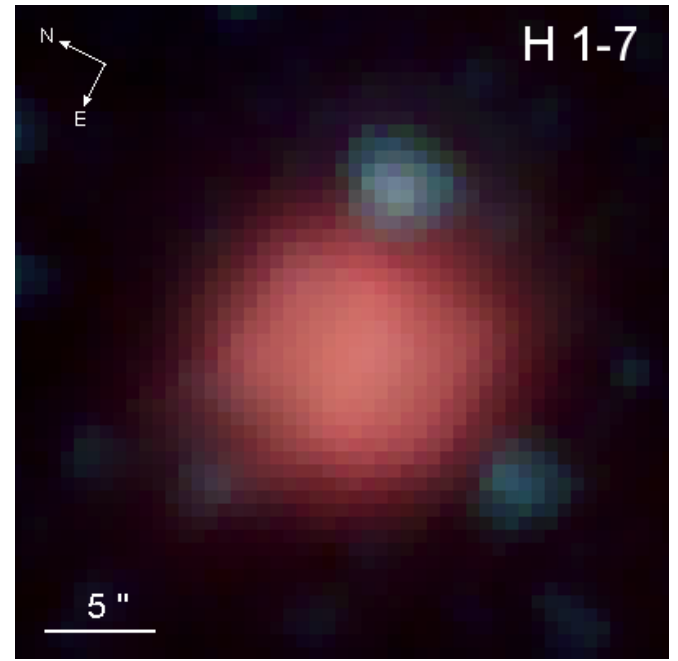
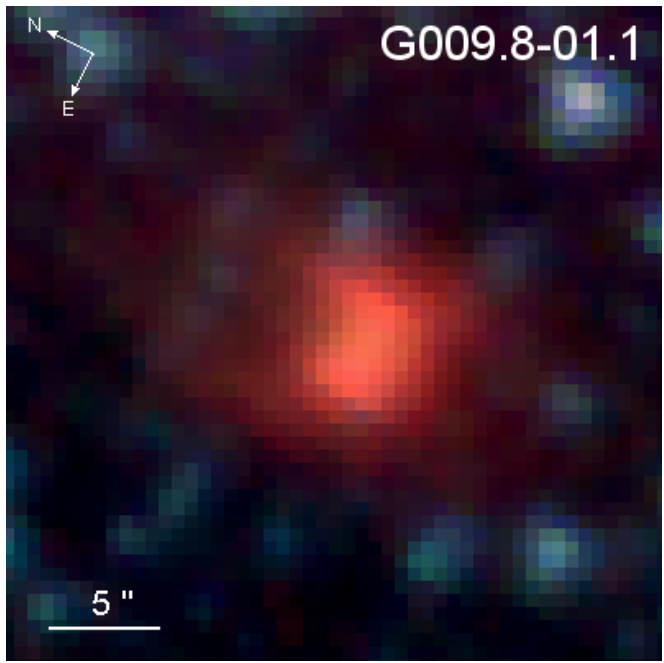 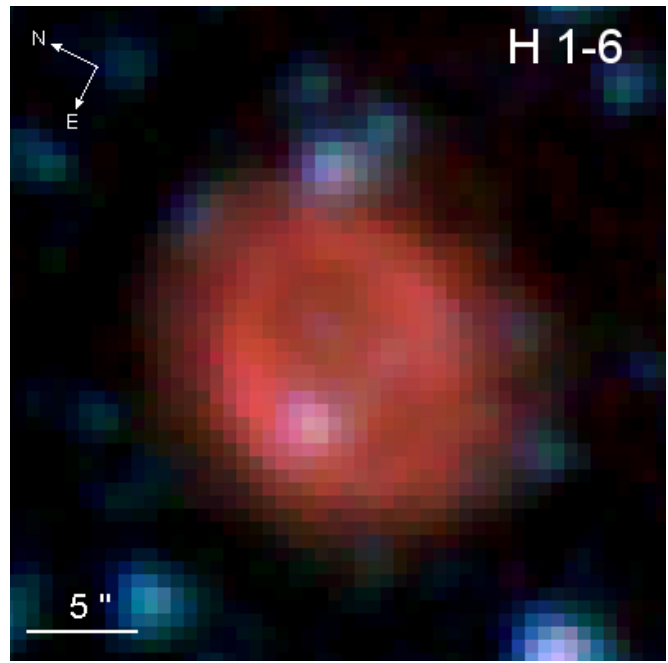 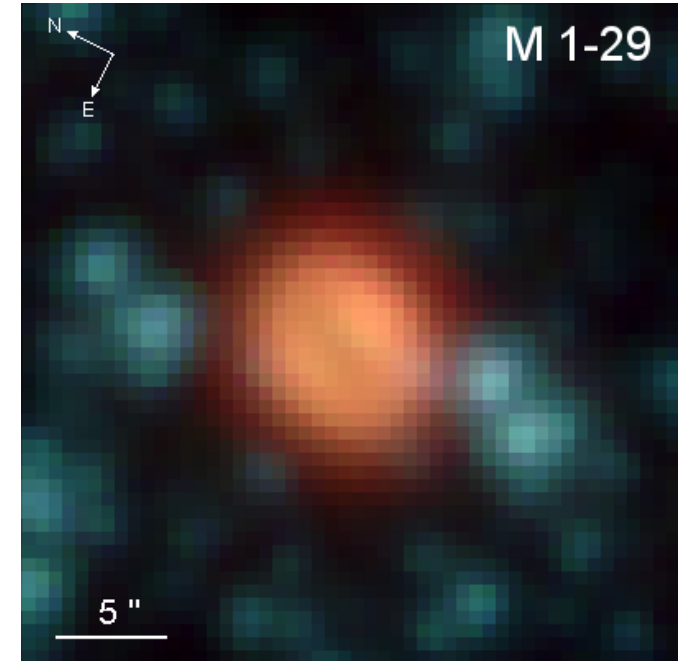

FIGURE 1

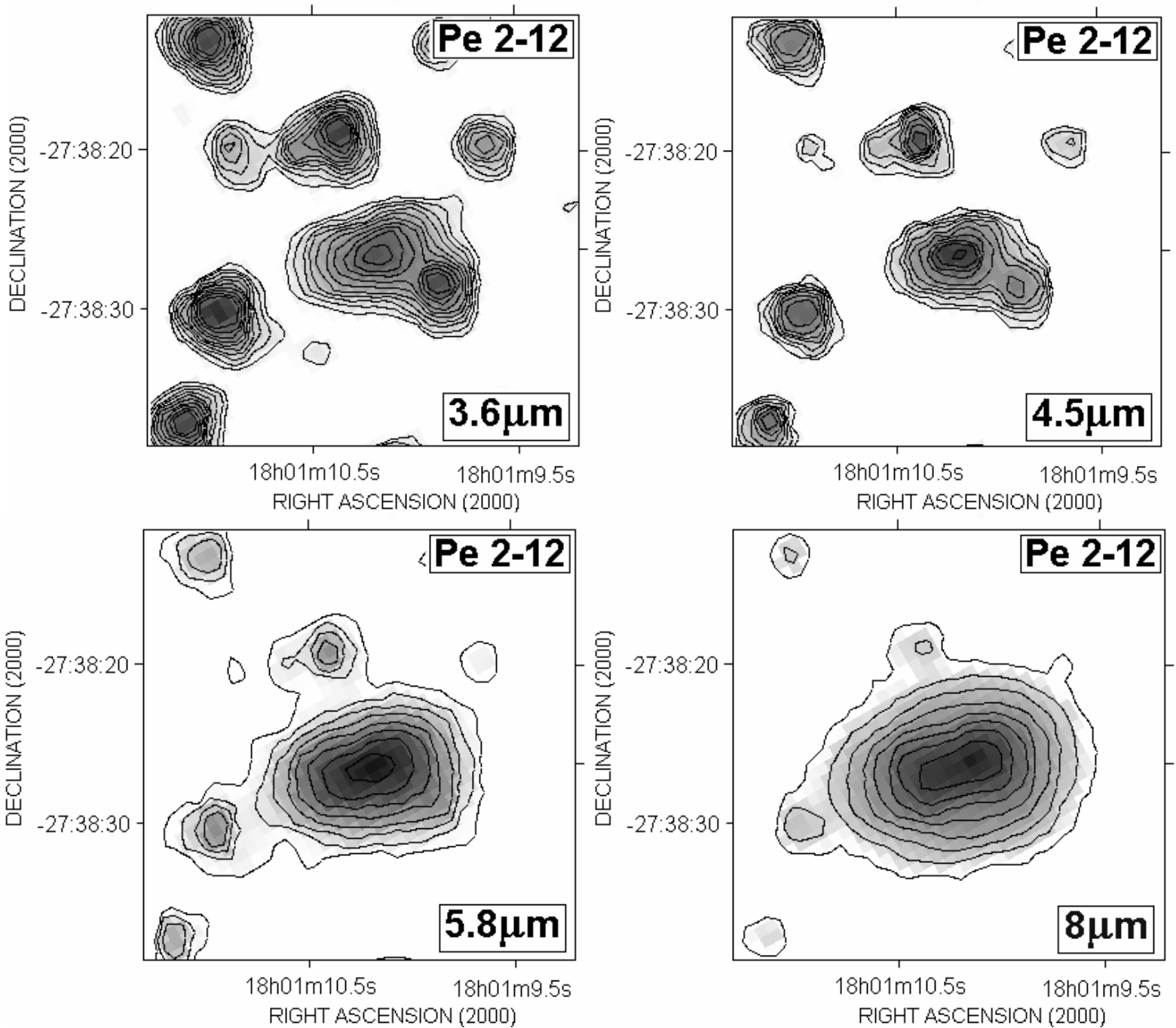

FIGURE 2

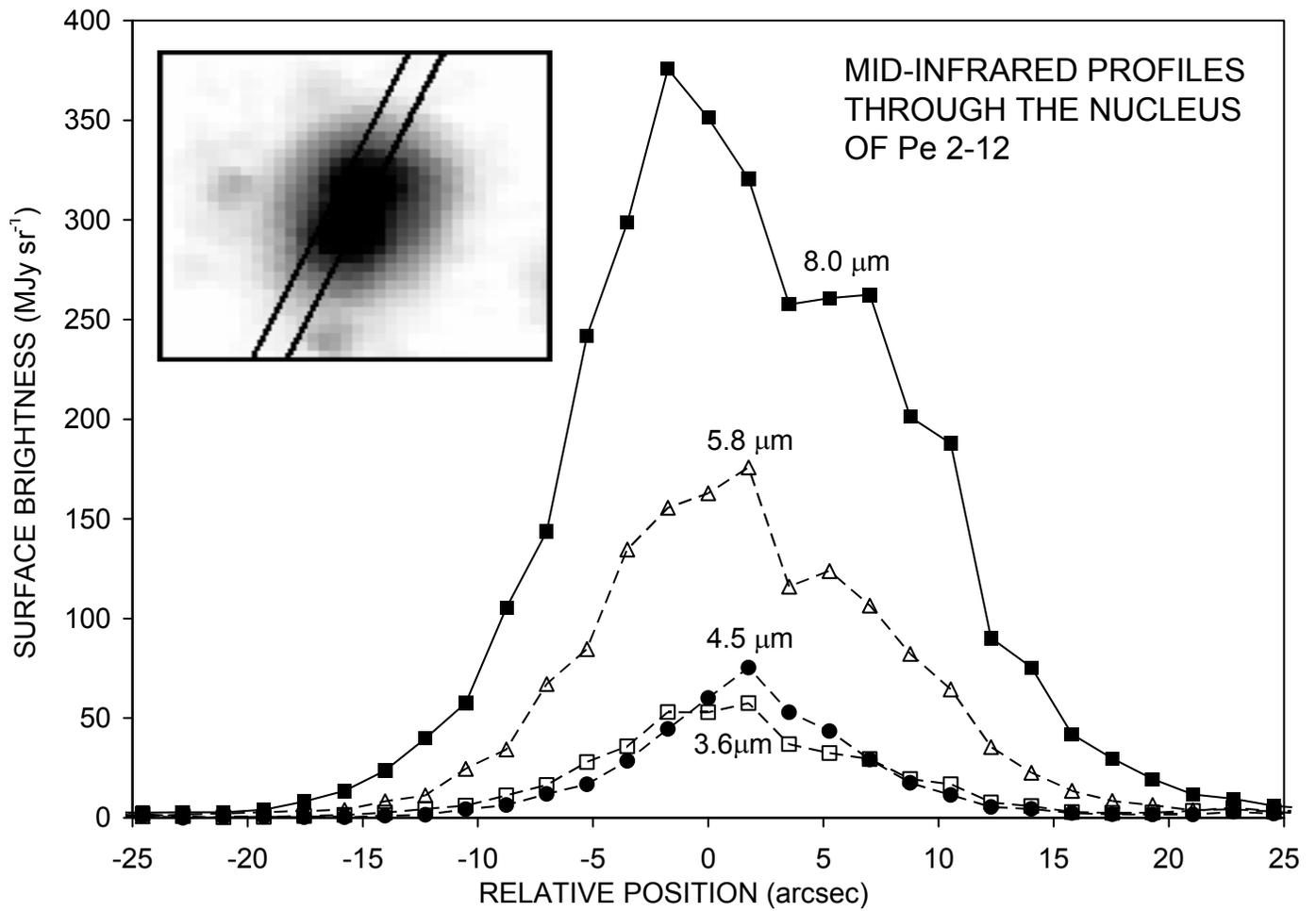
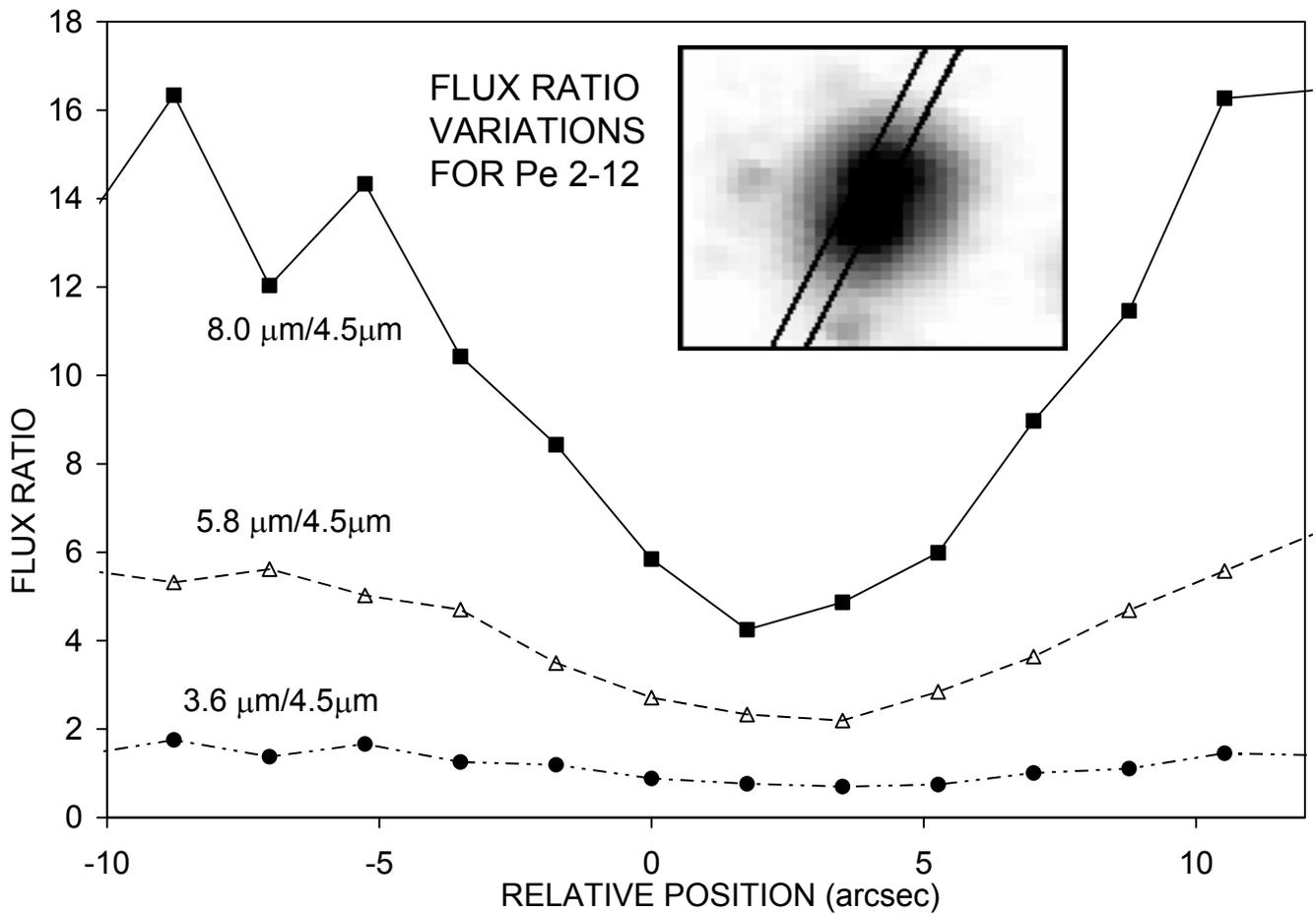

FIGURE 3

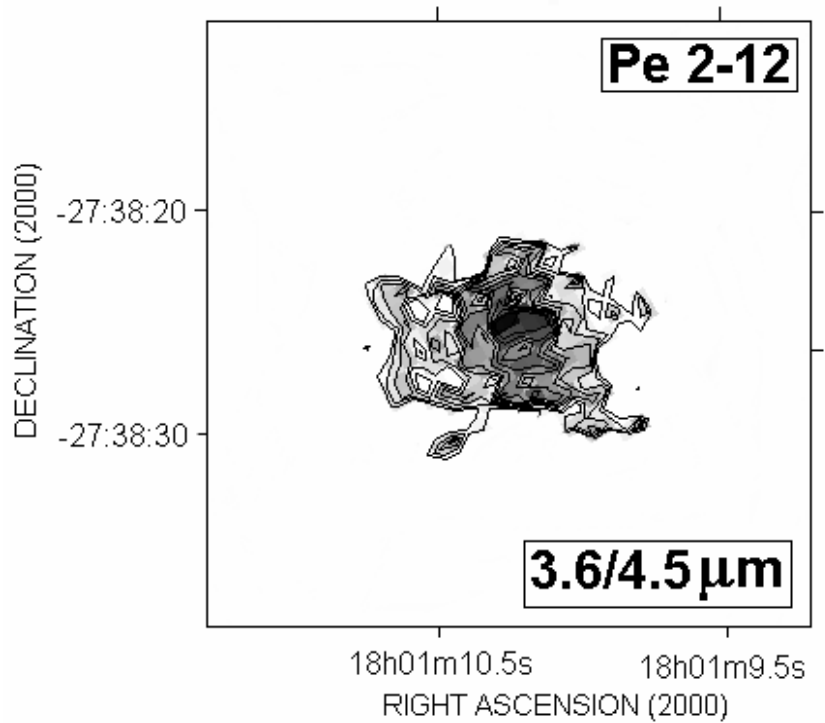
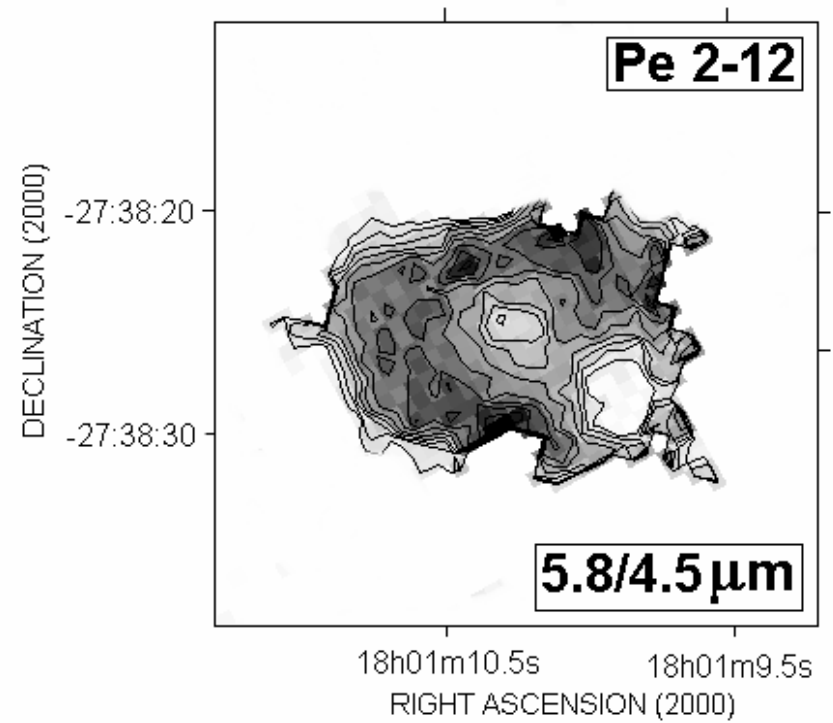
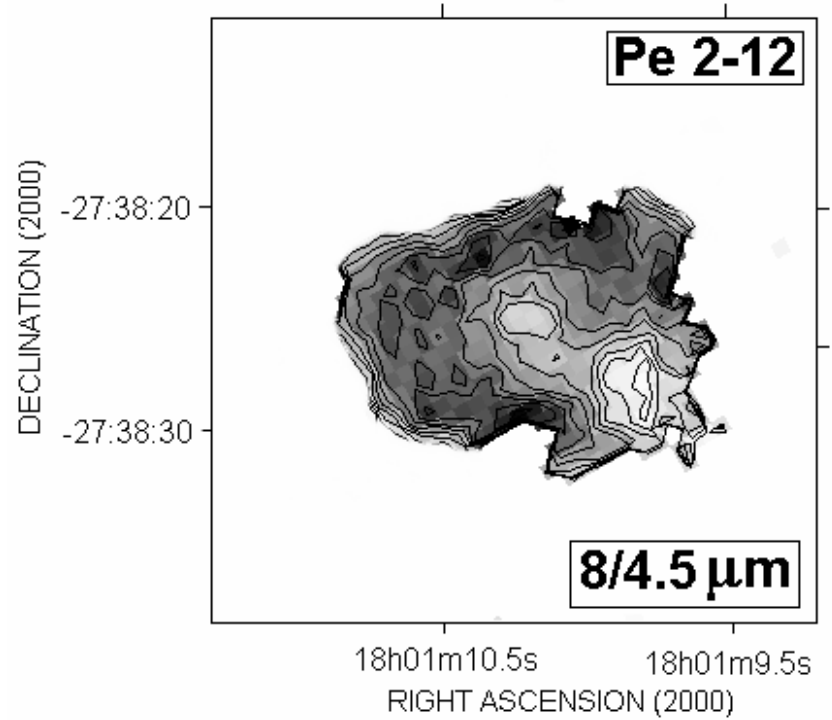

FIGURE 4

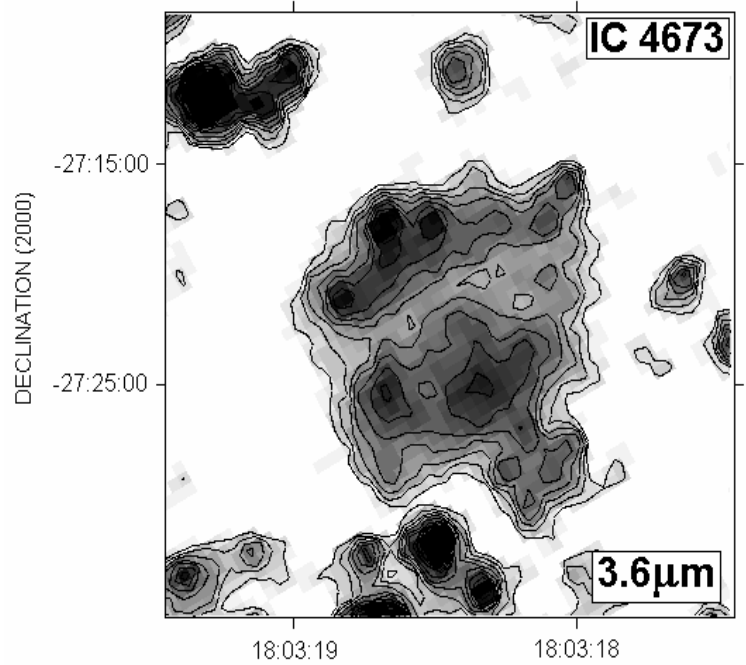
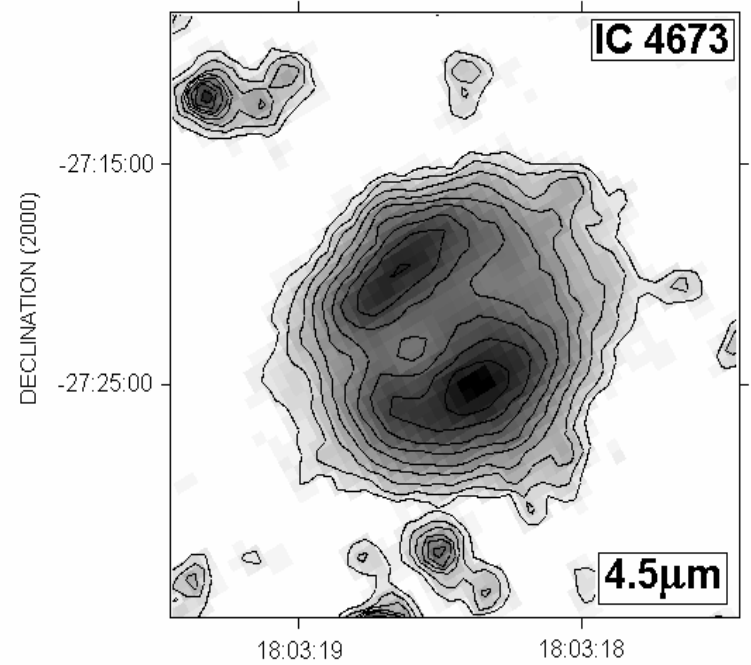
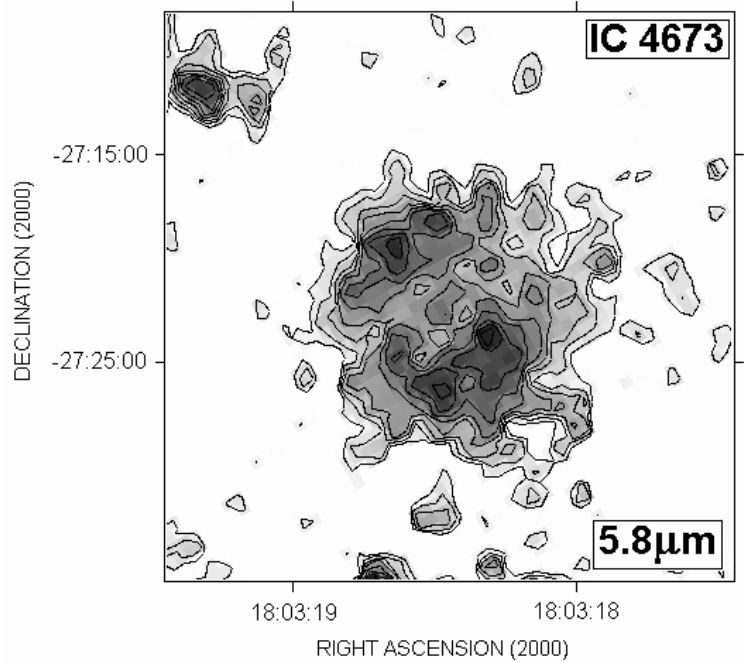
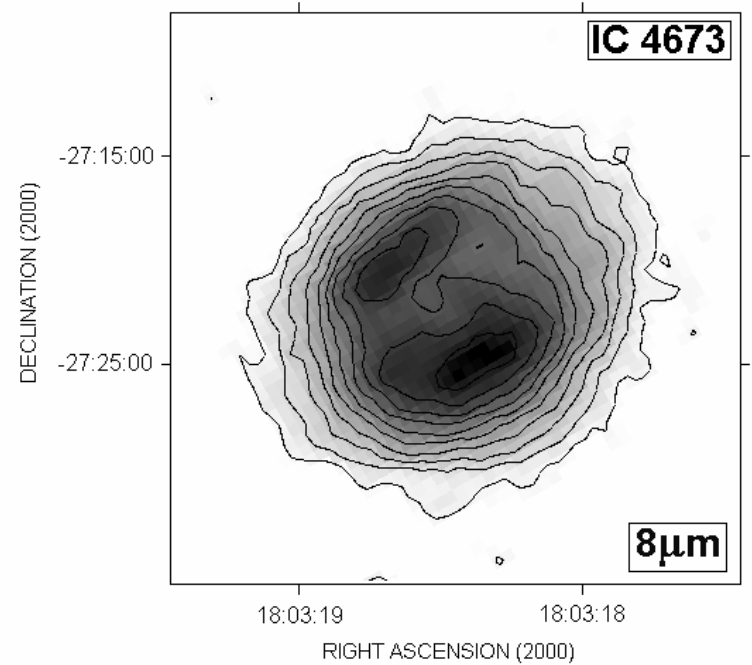

FIGURE 5

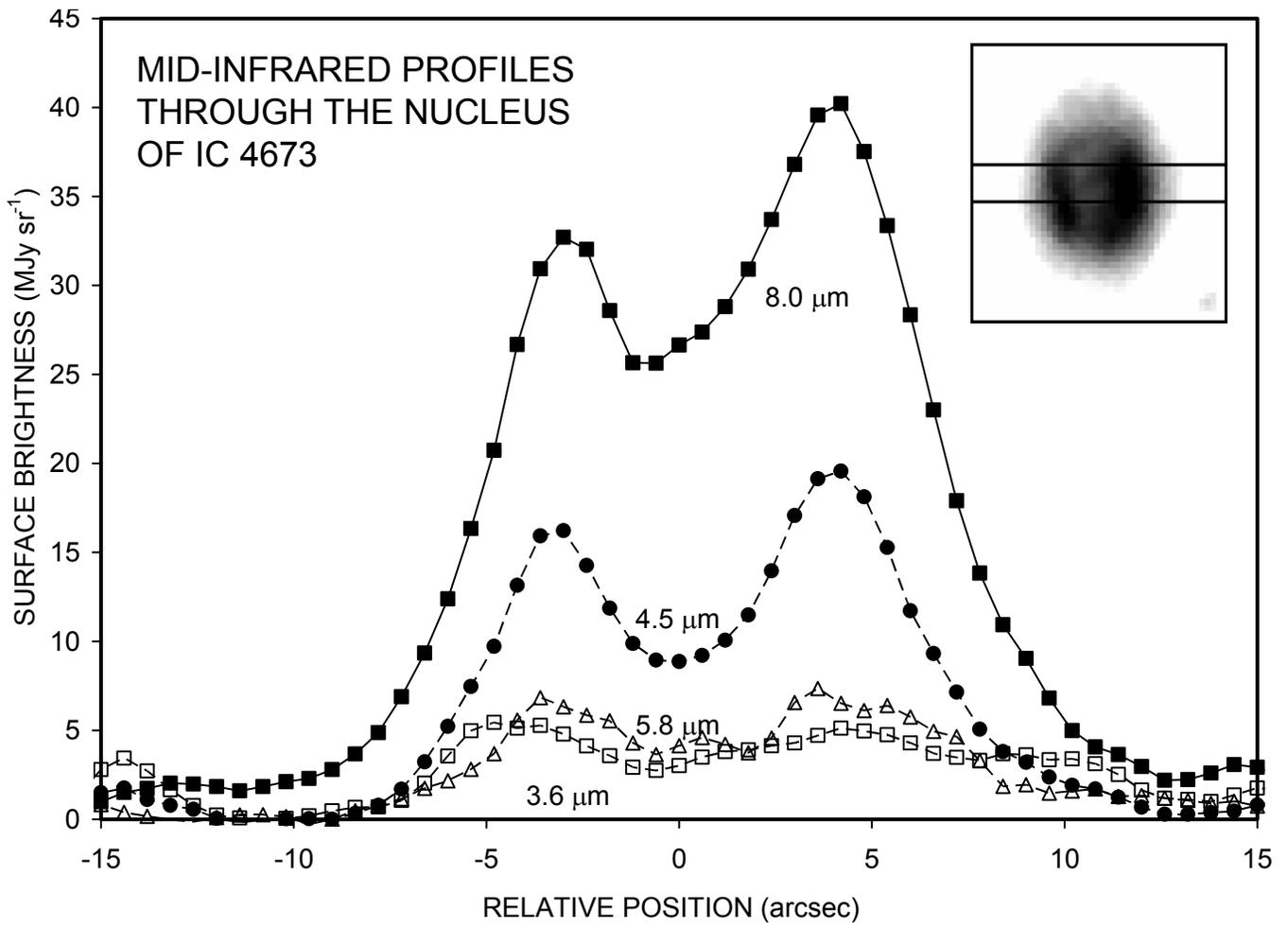
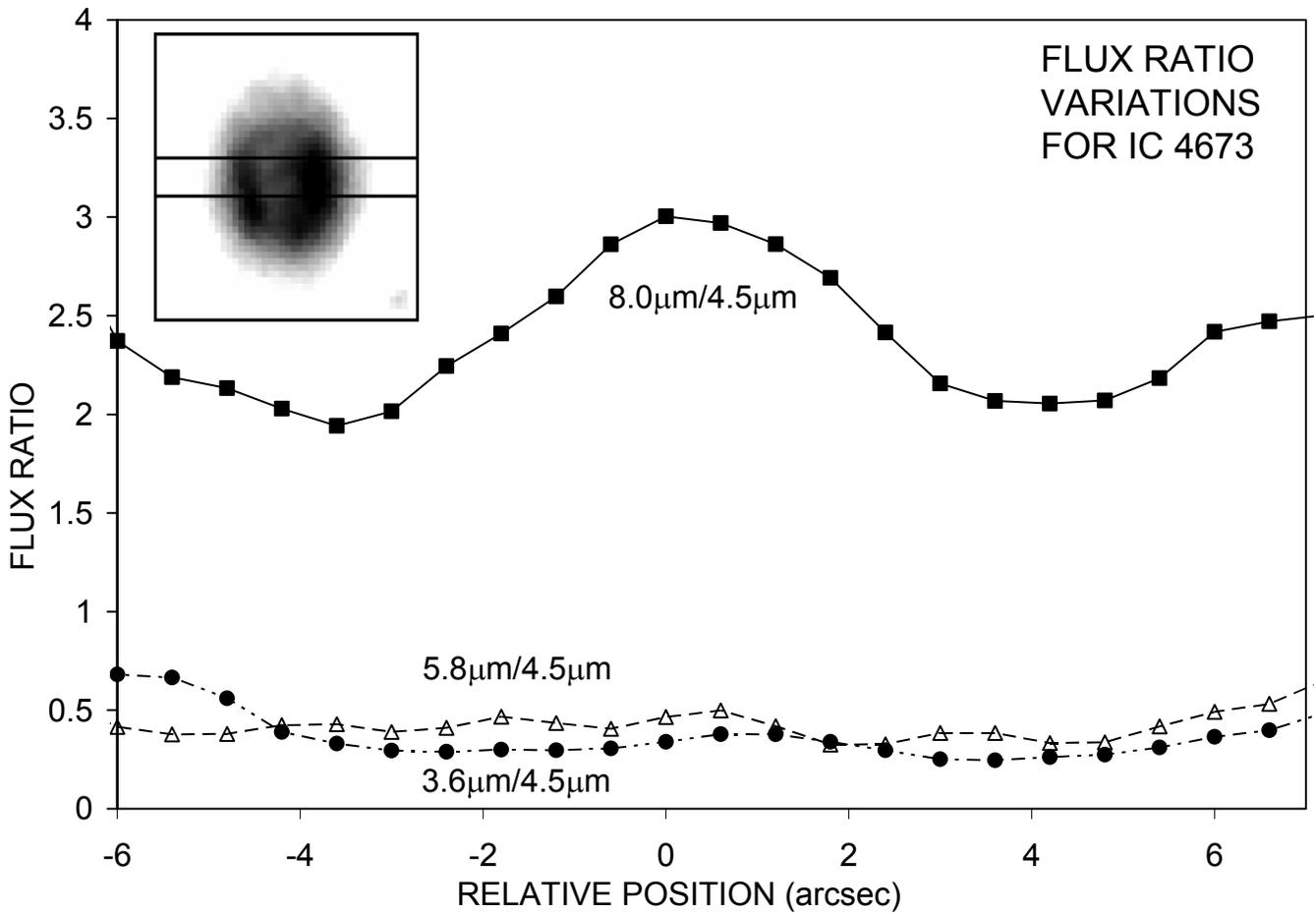

FIGURE 6

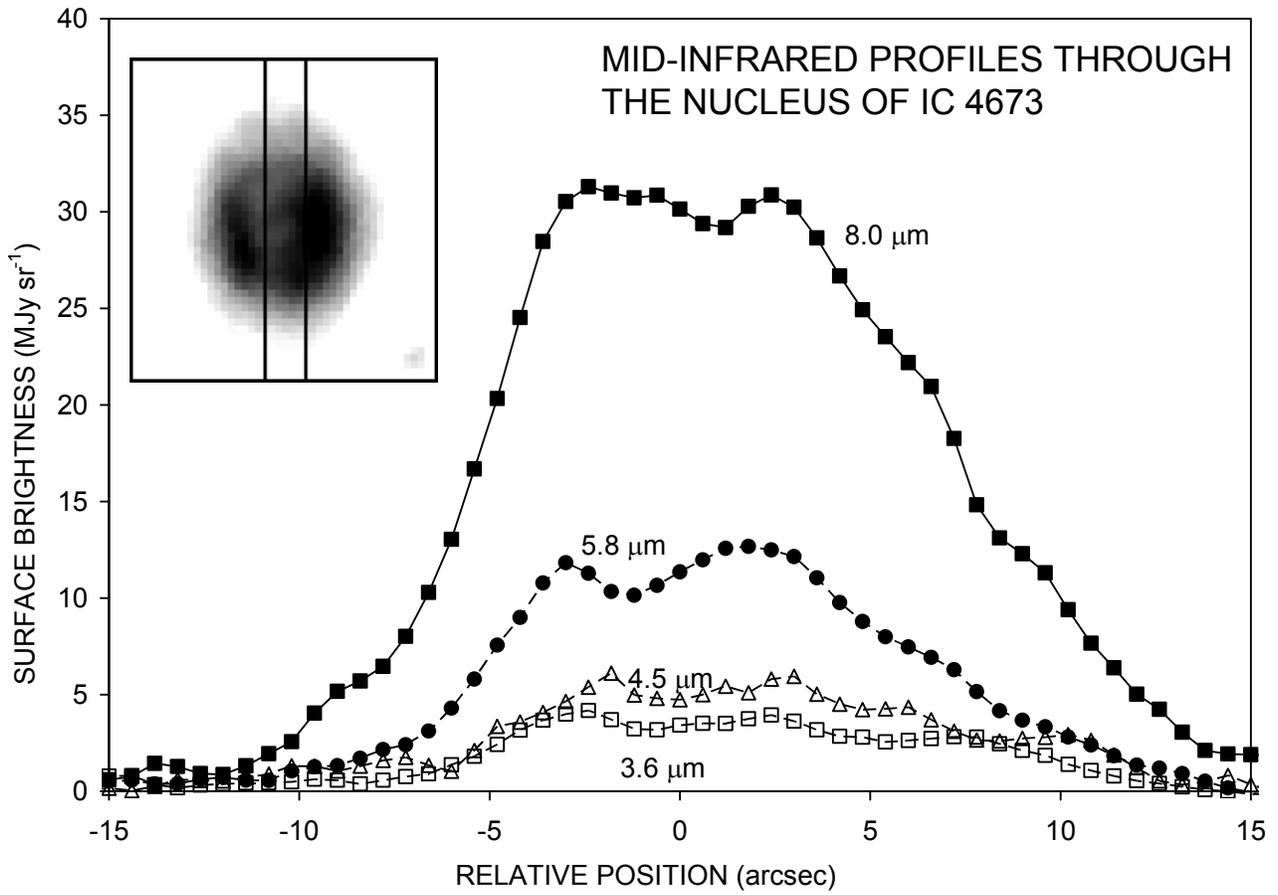
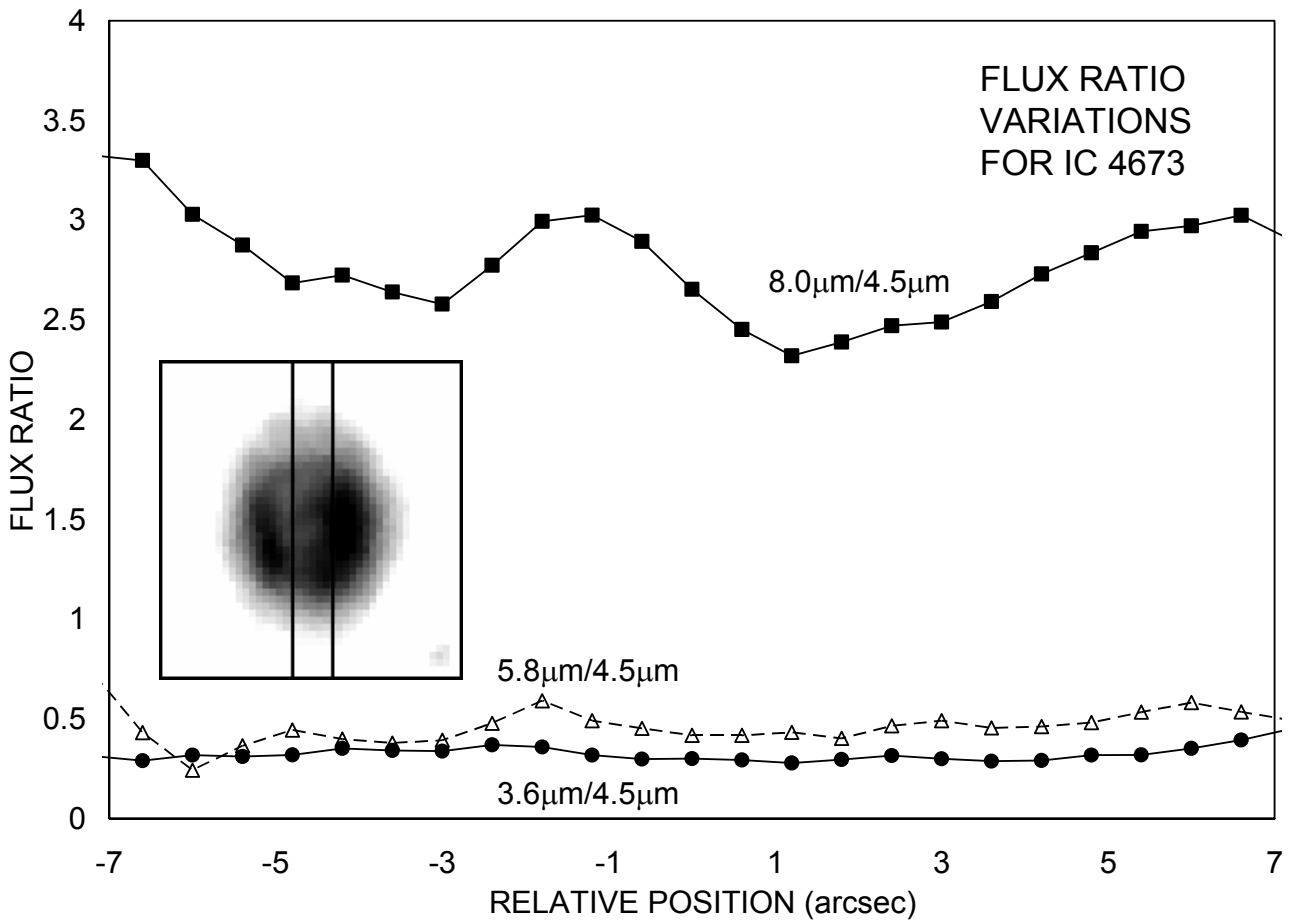

FIGURE 7

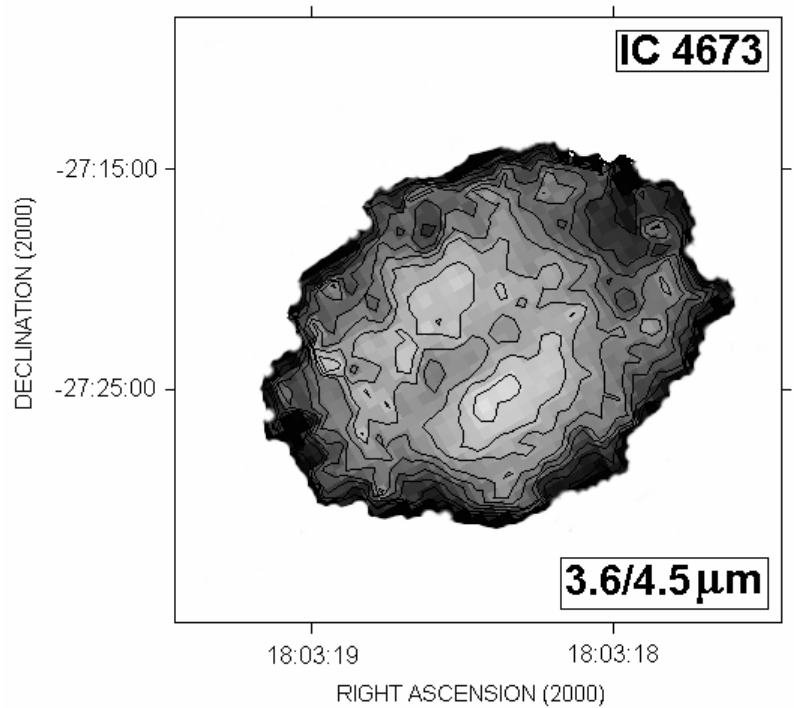
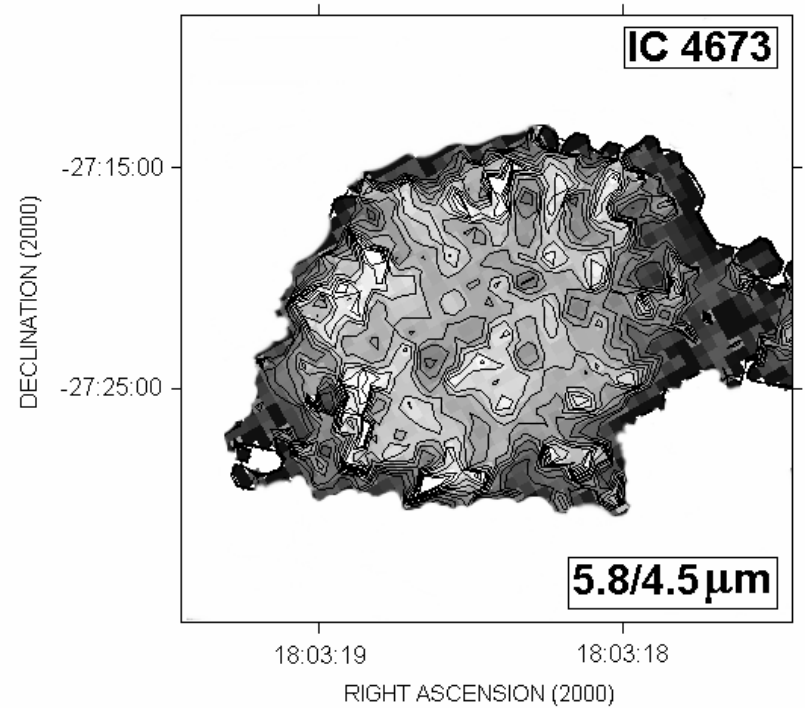
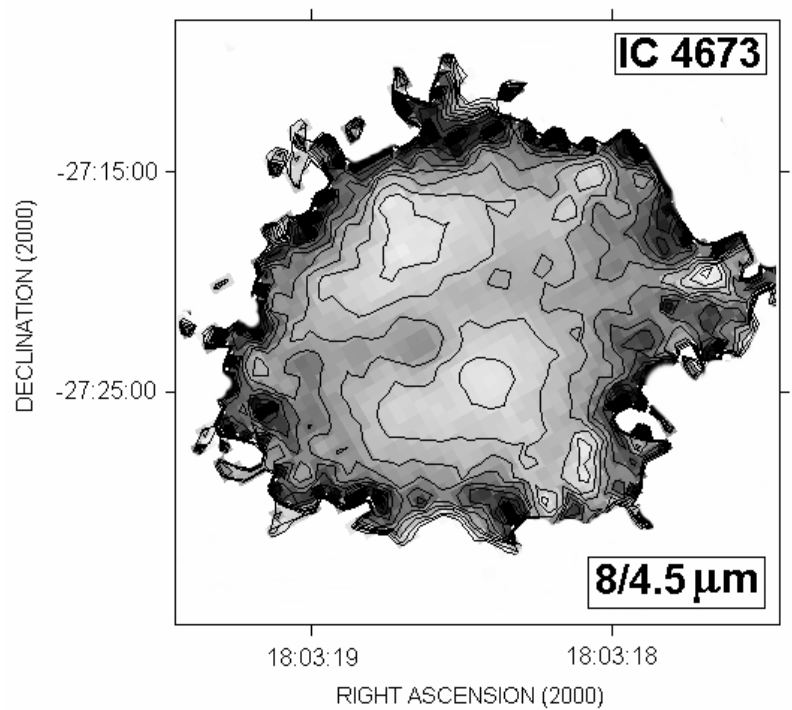

FIGURE 8

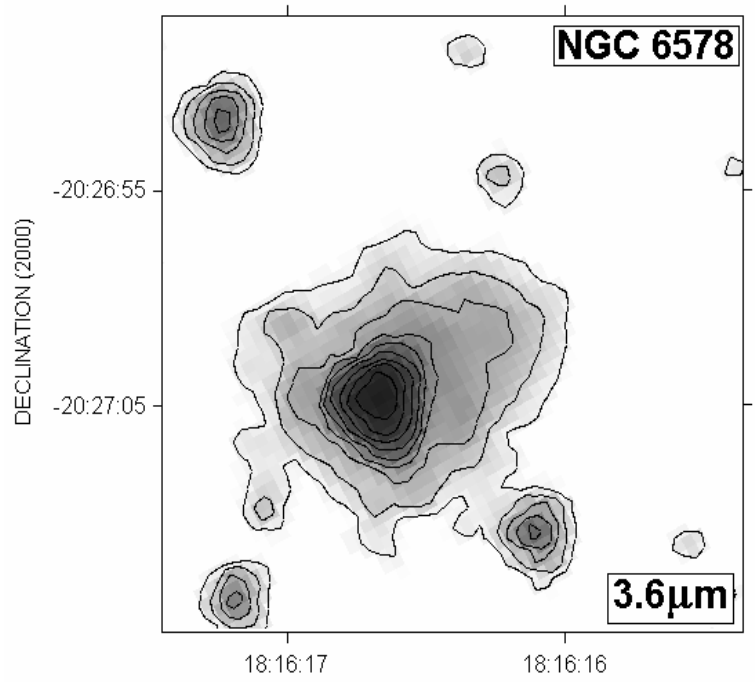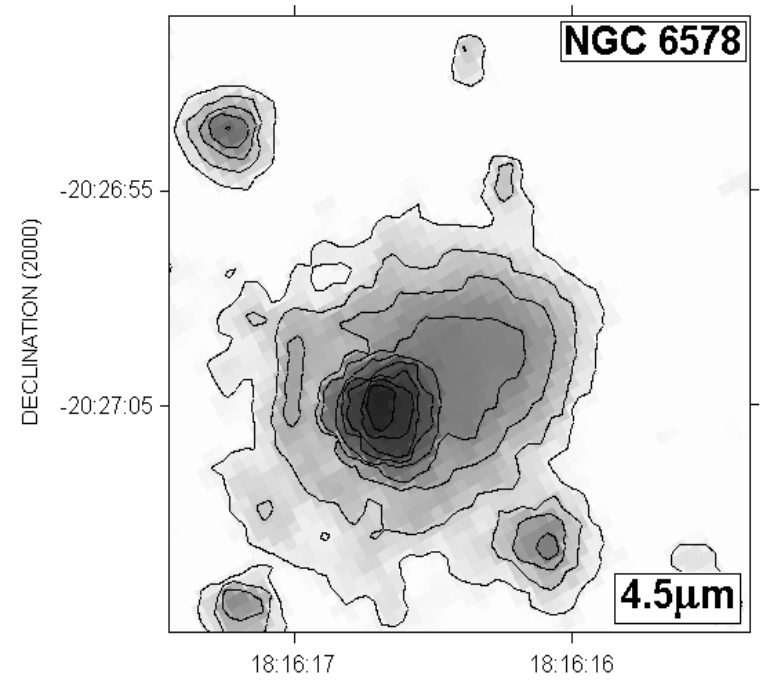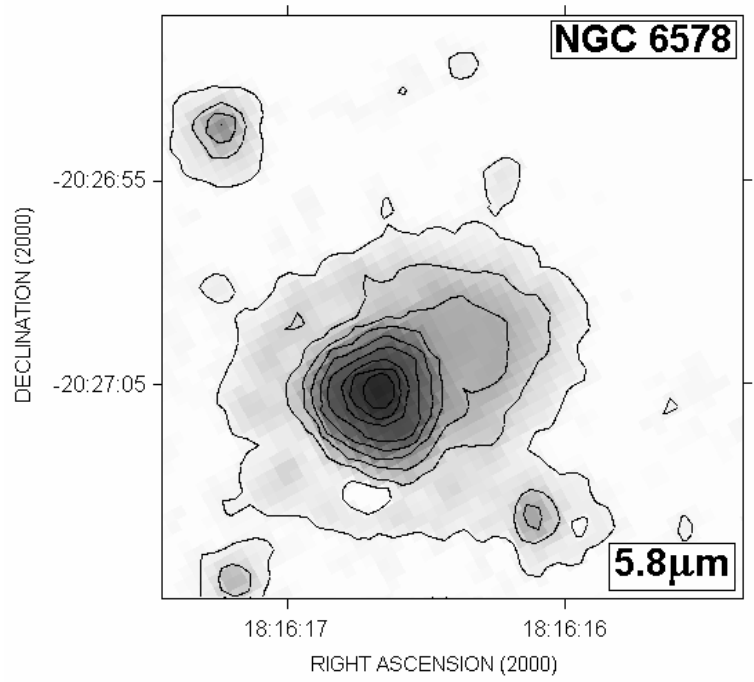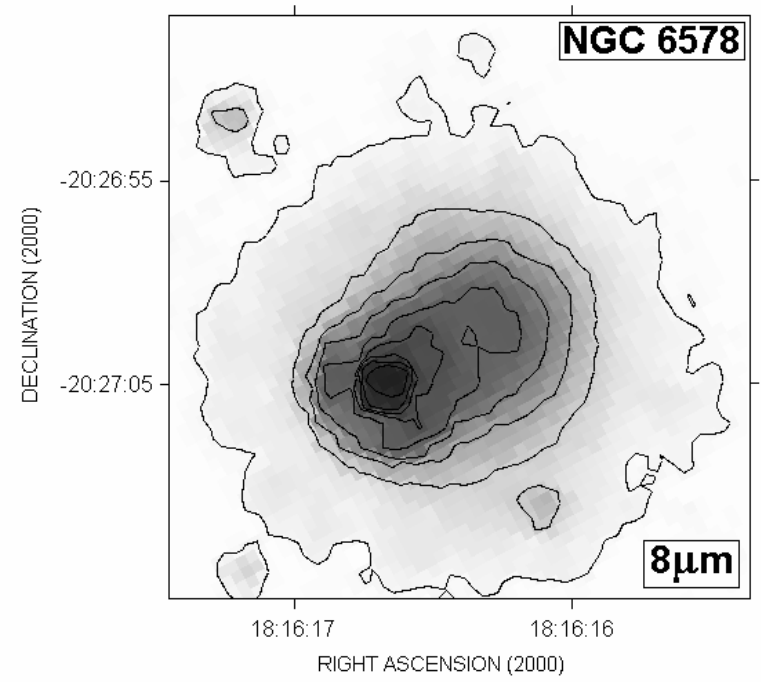

FIGURE 9

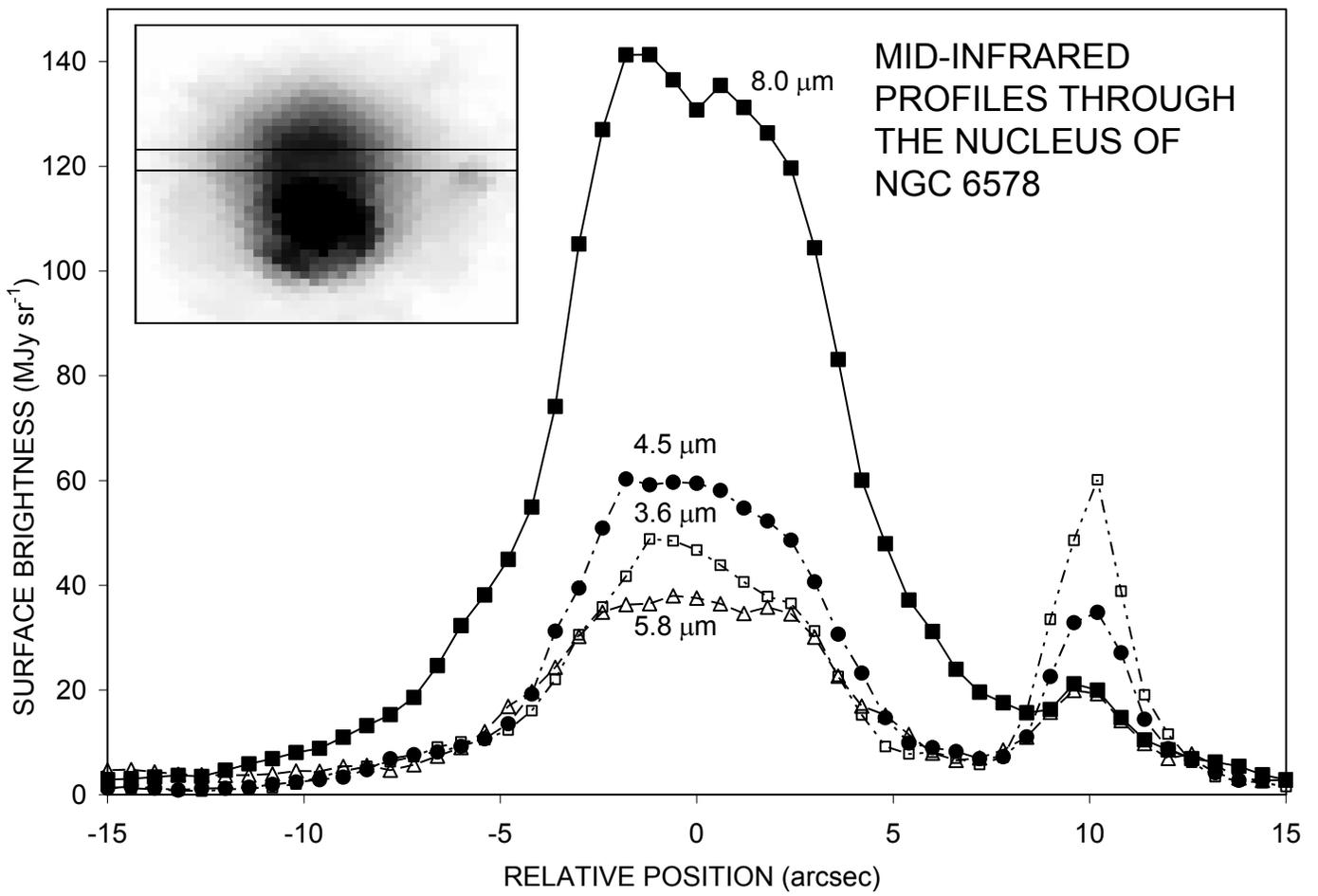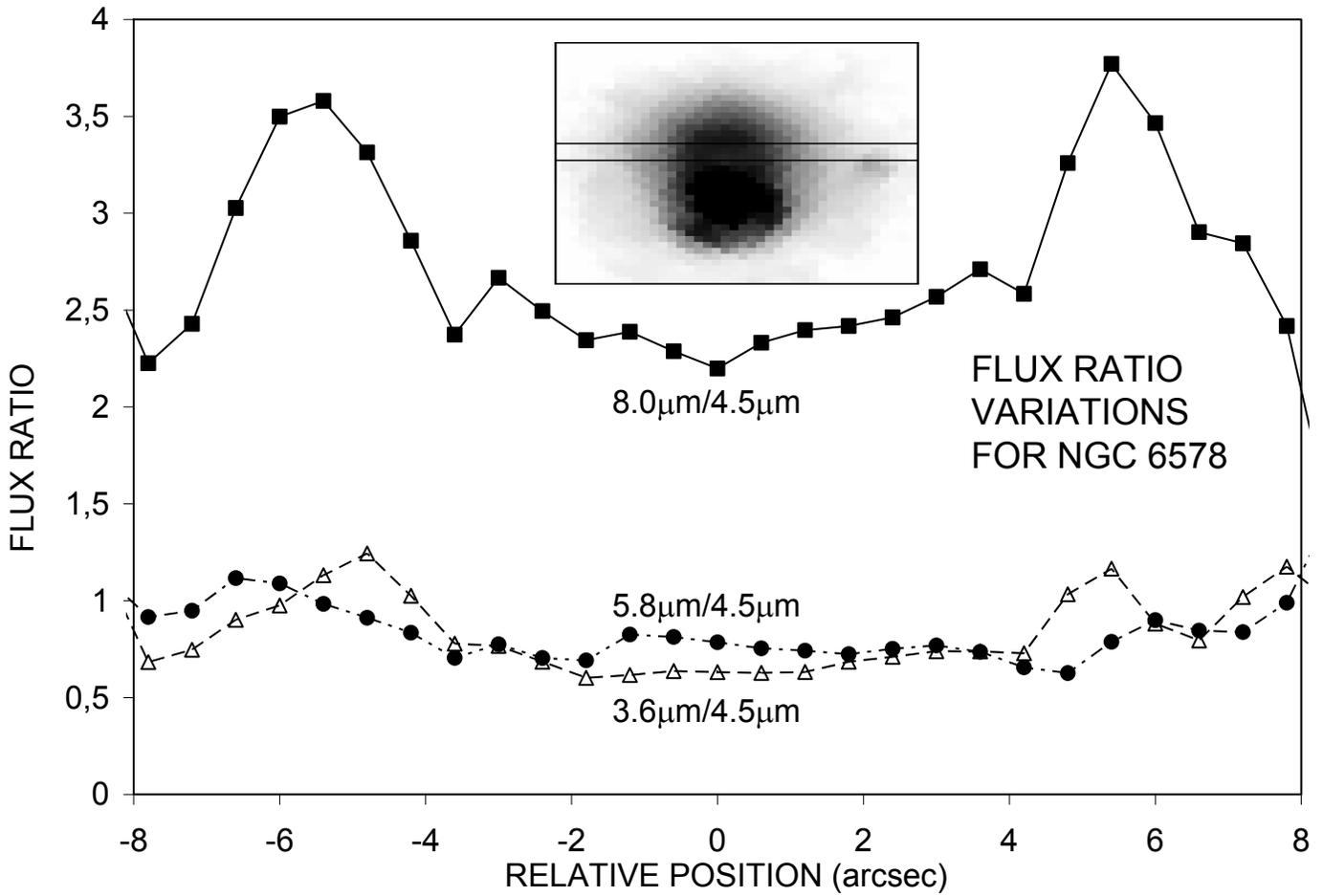

FIGURE 10

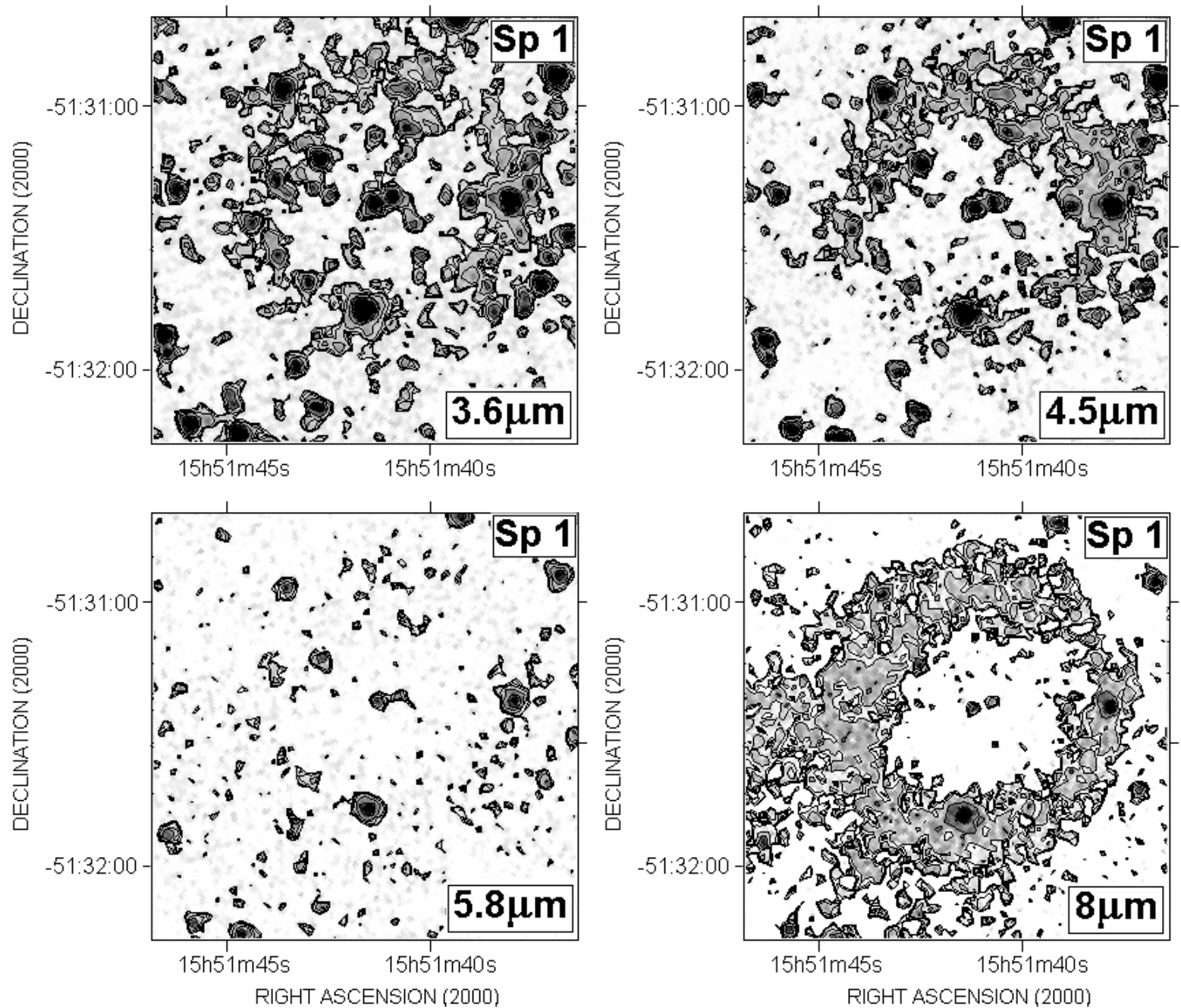

FIGURE 11

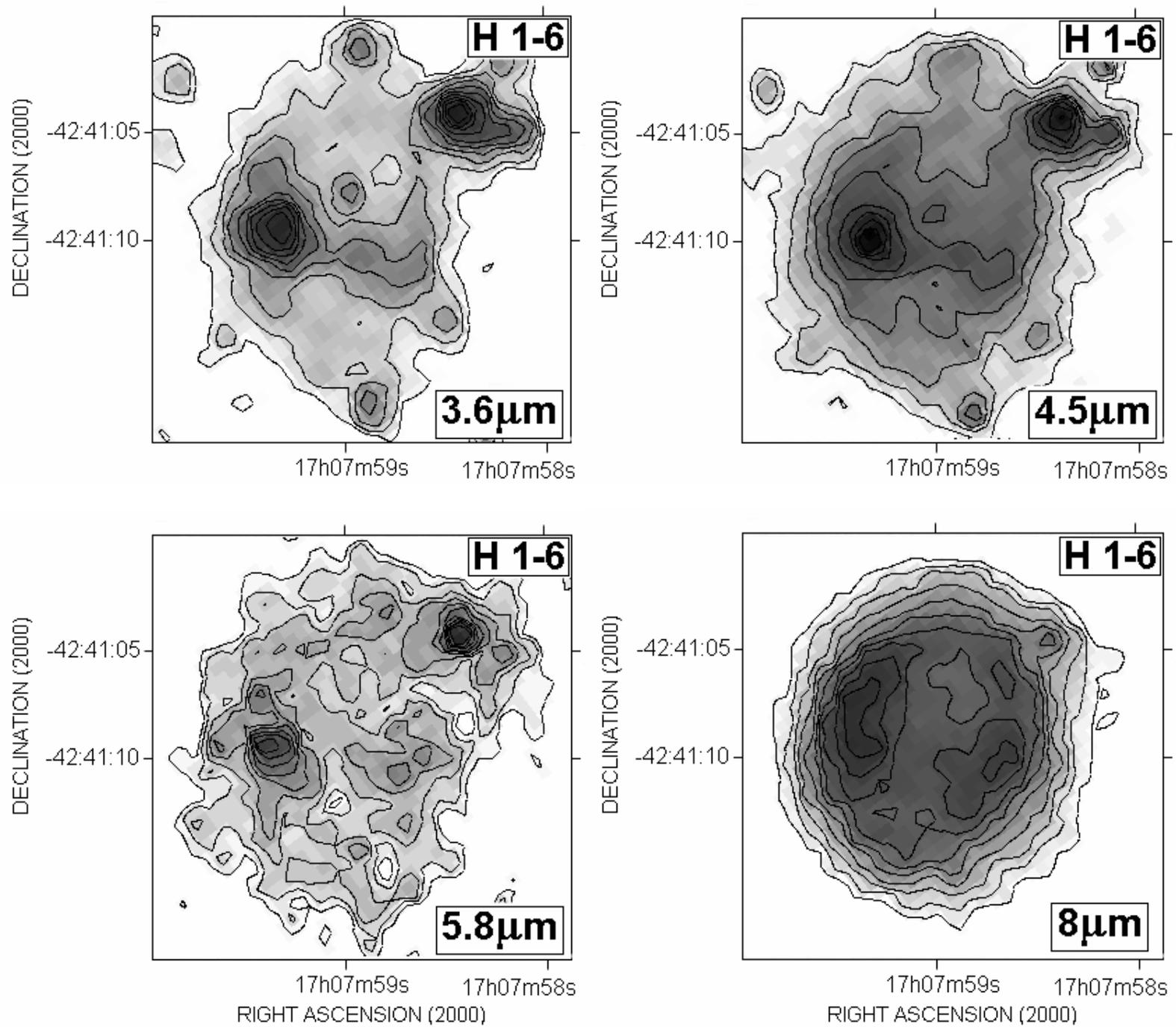

FIGURE 12

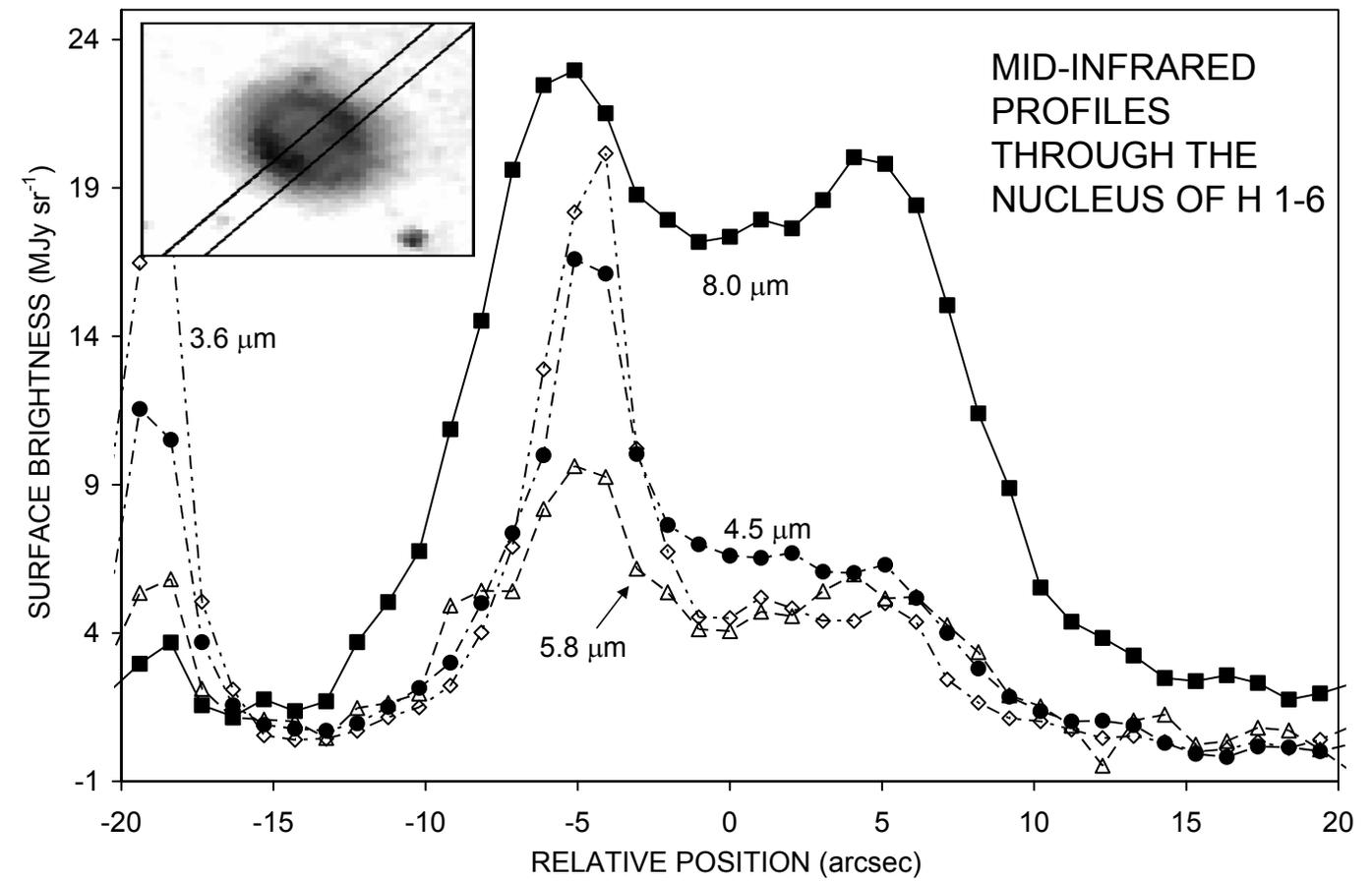
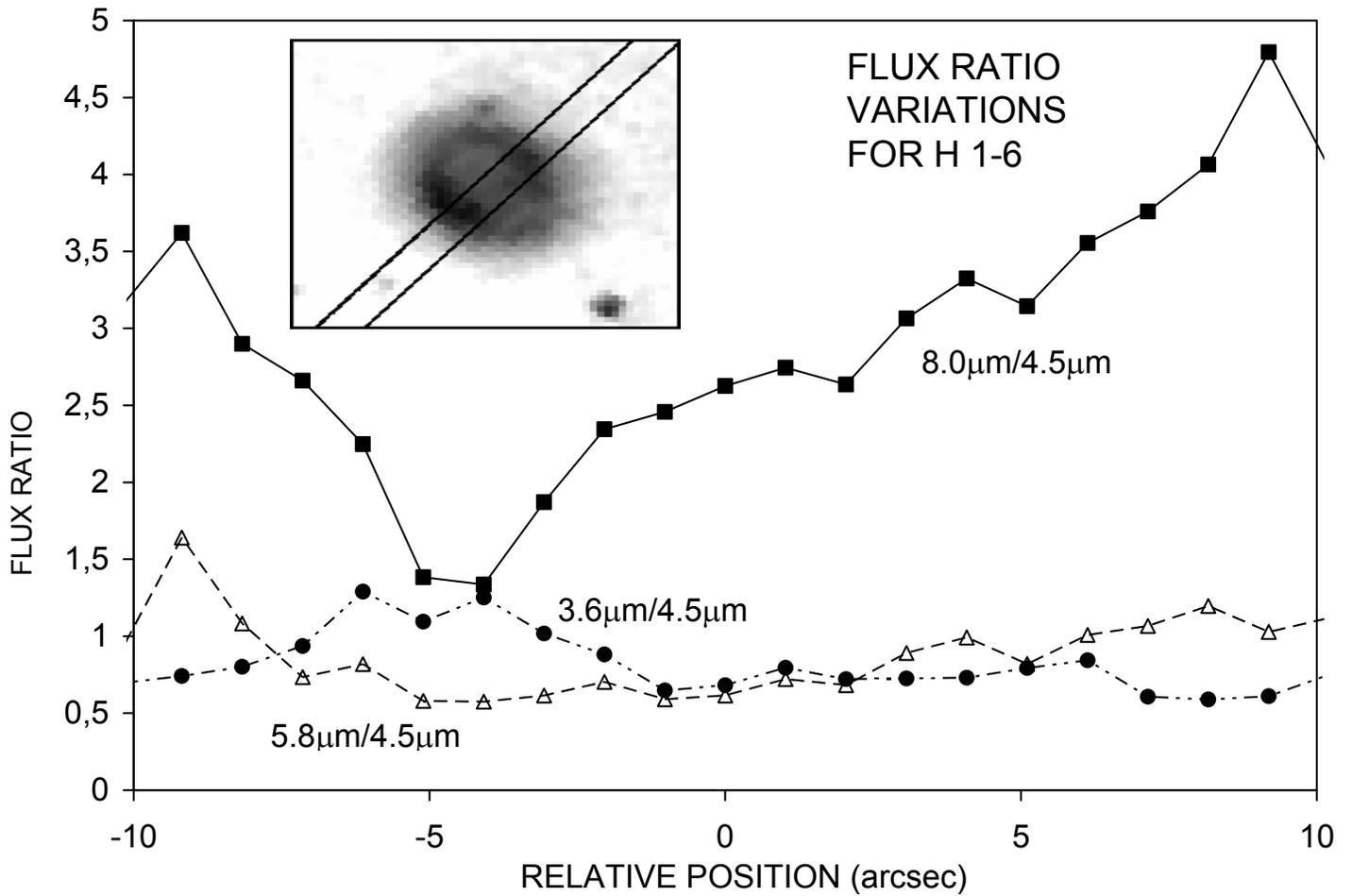

FIGURE 13

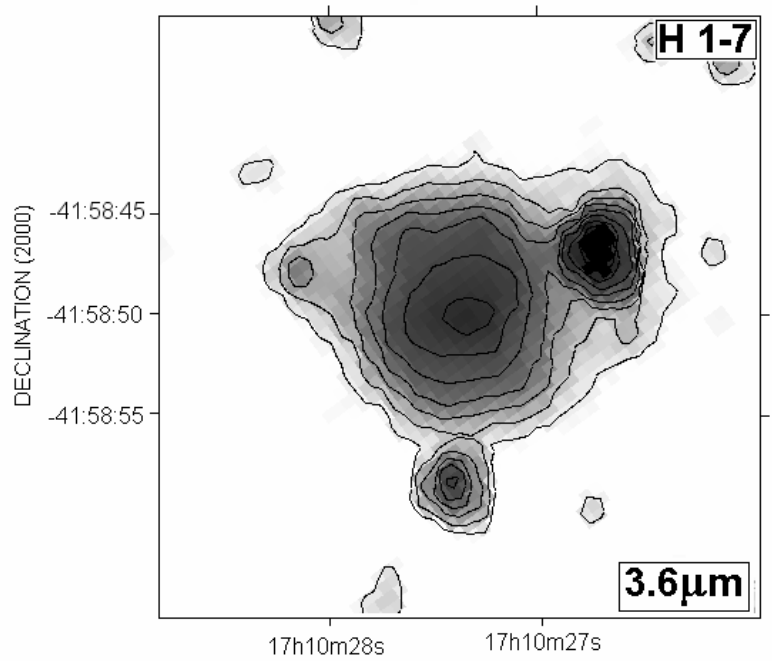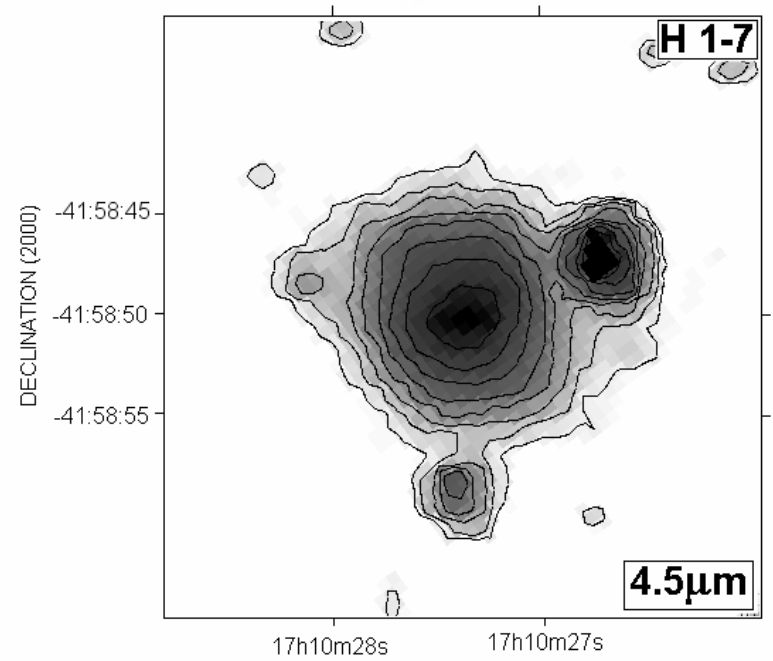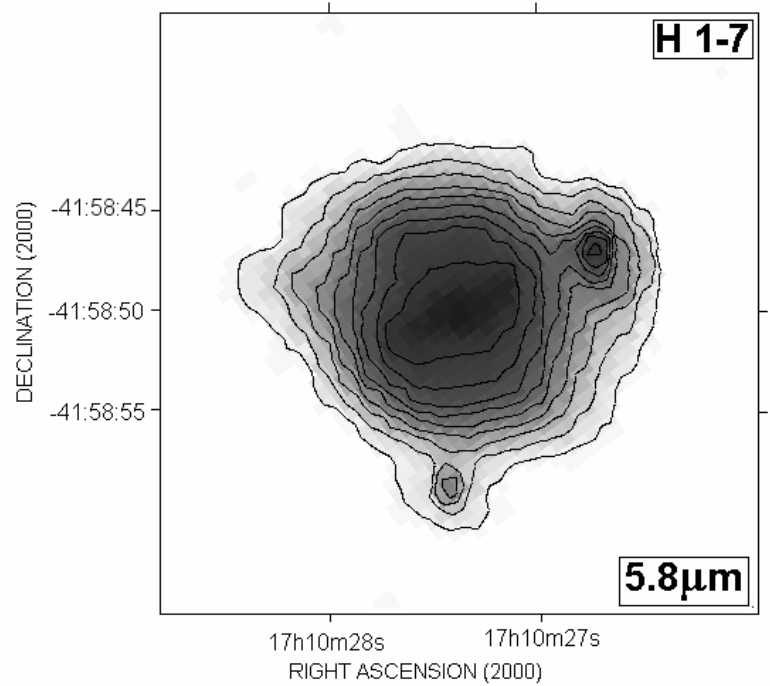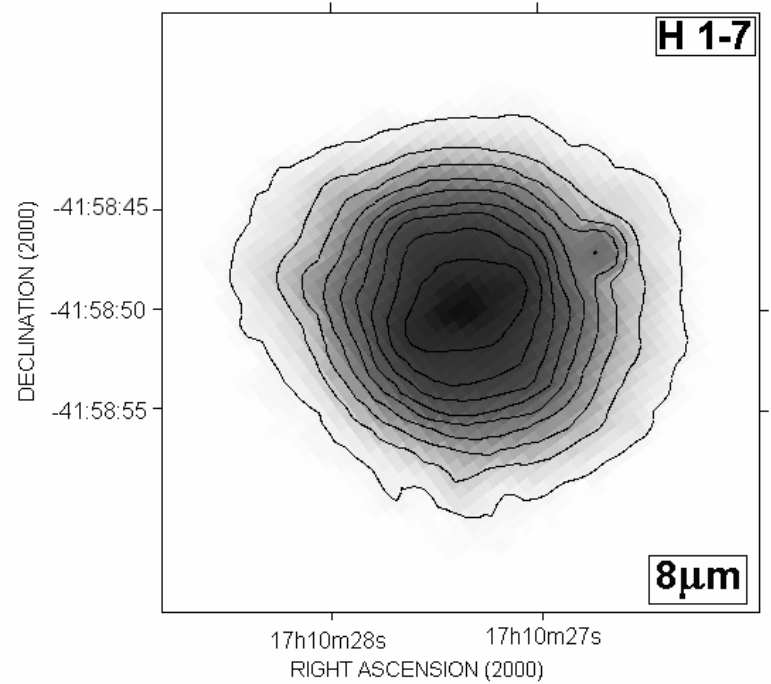

FIGURE 14

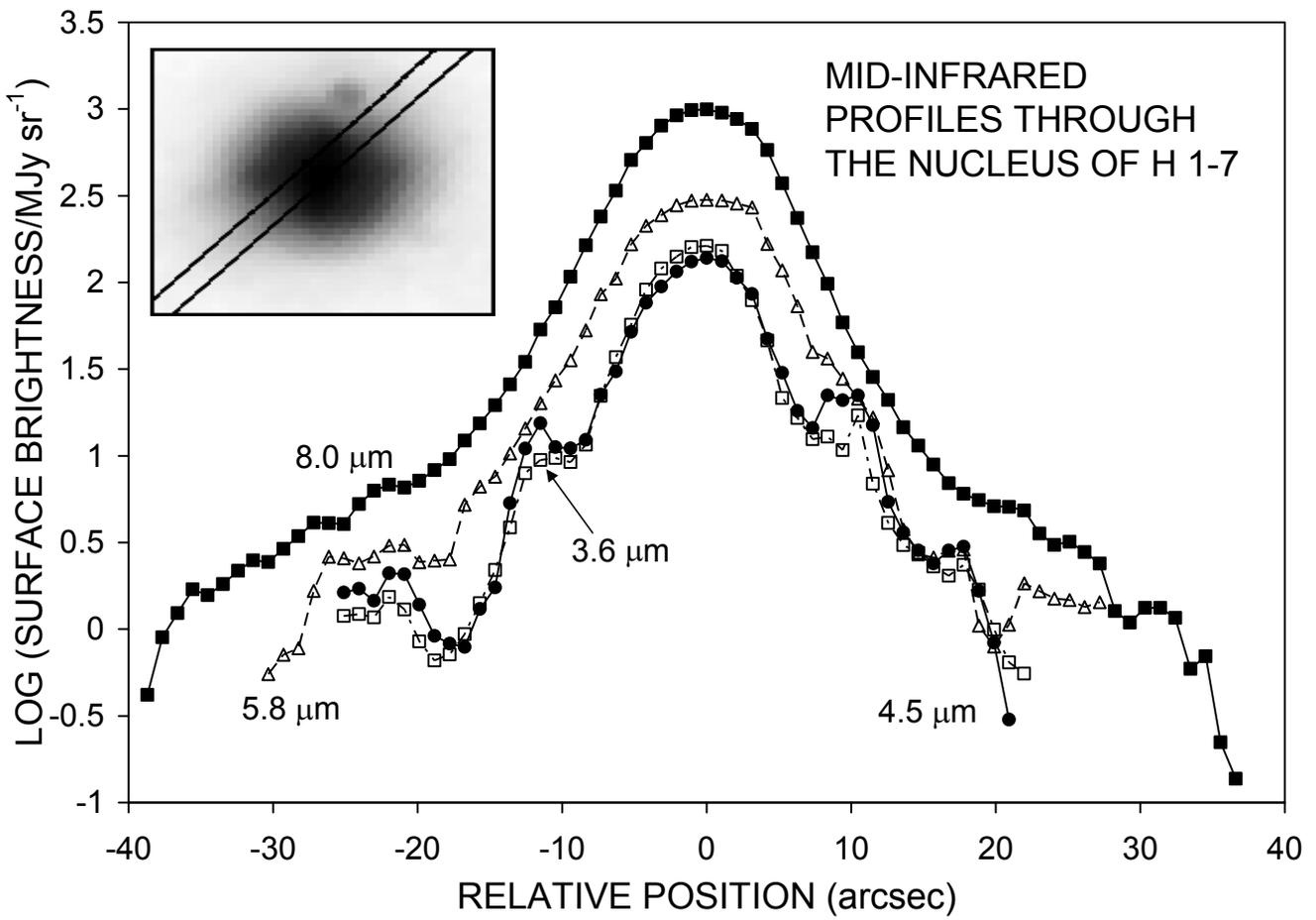
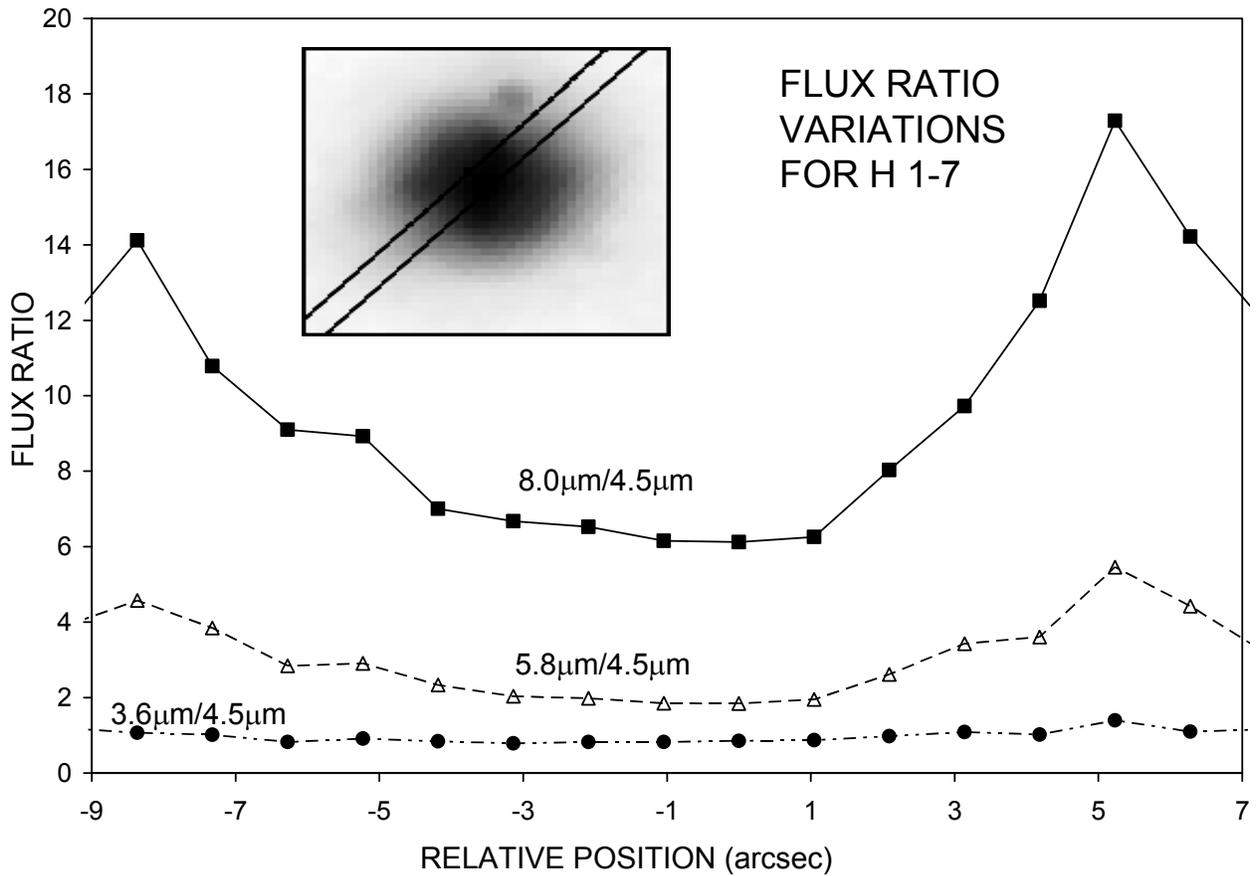

FIGURE 15

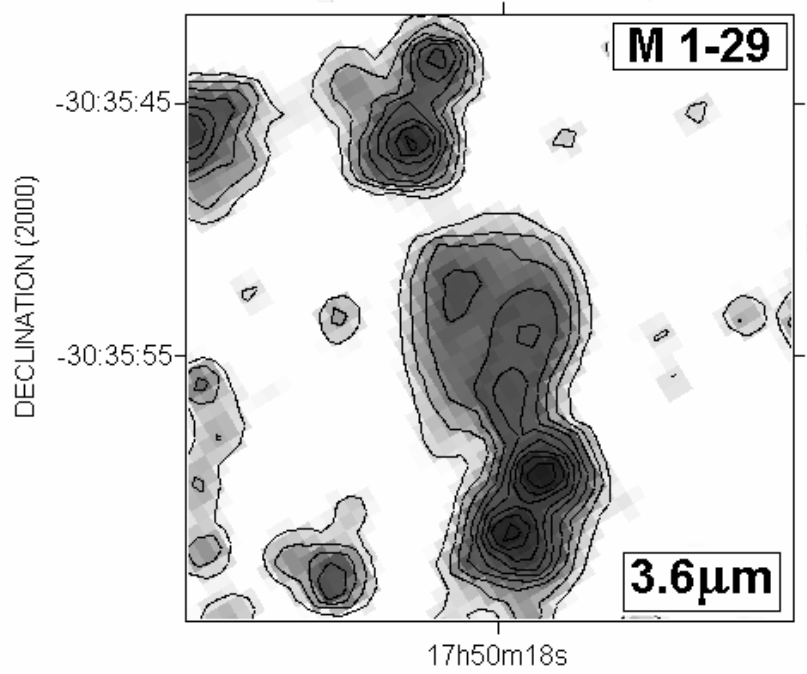
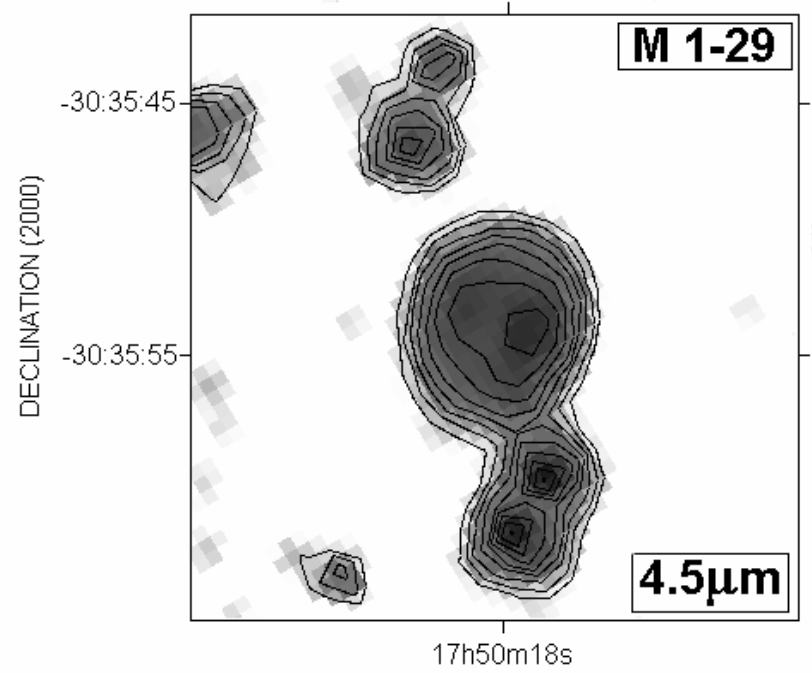
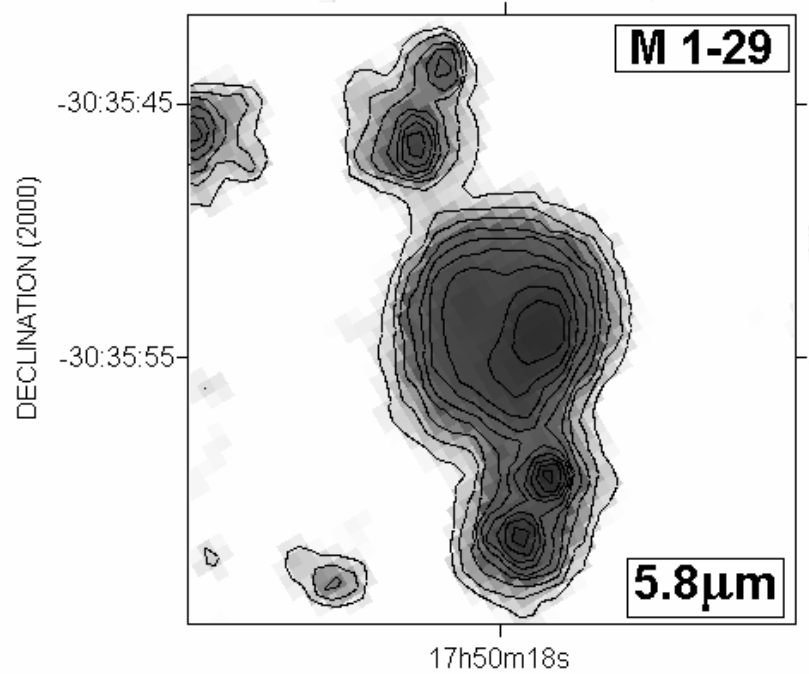
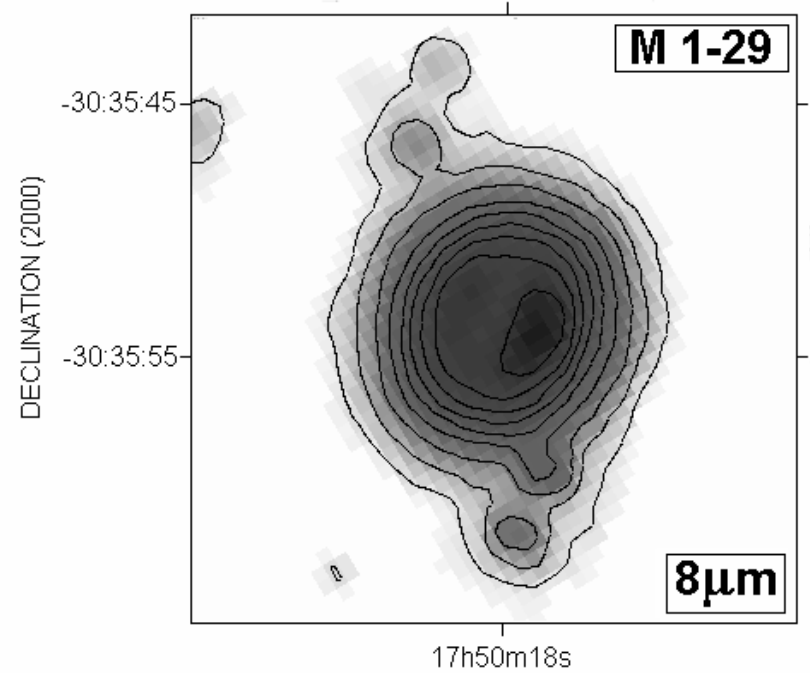

FIGURE 16

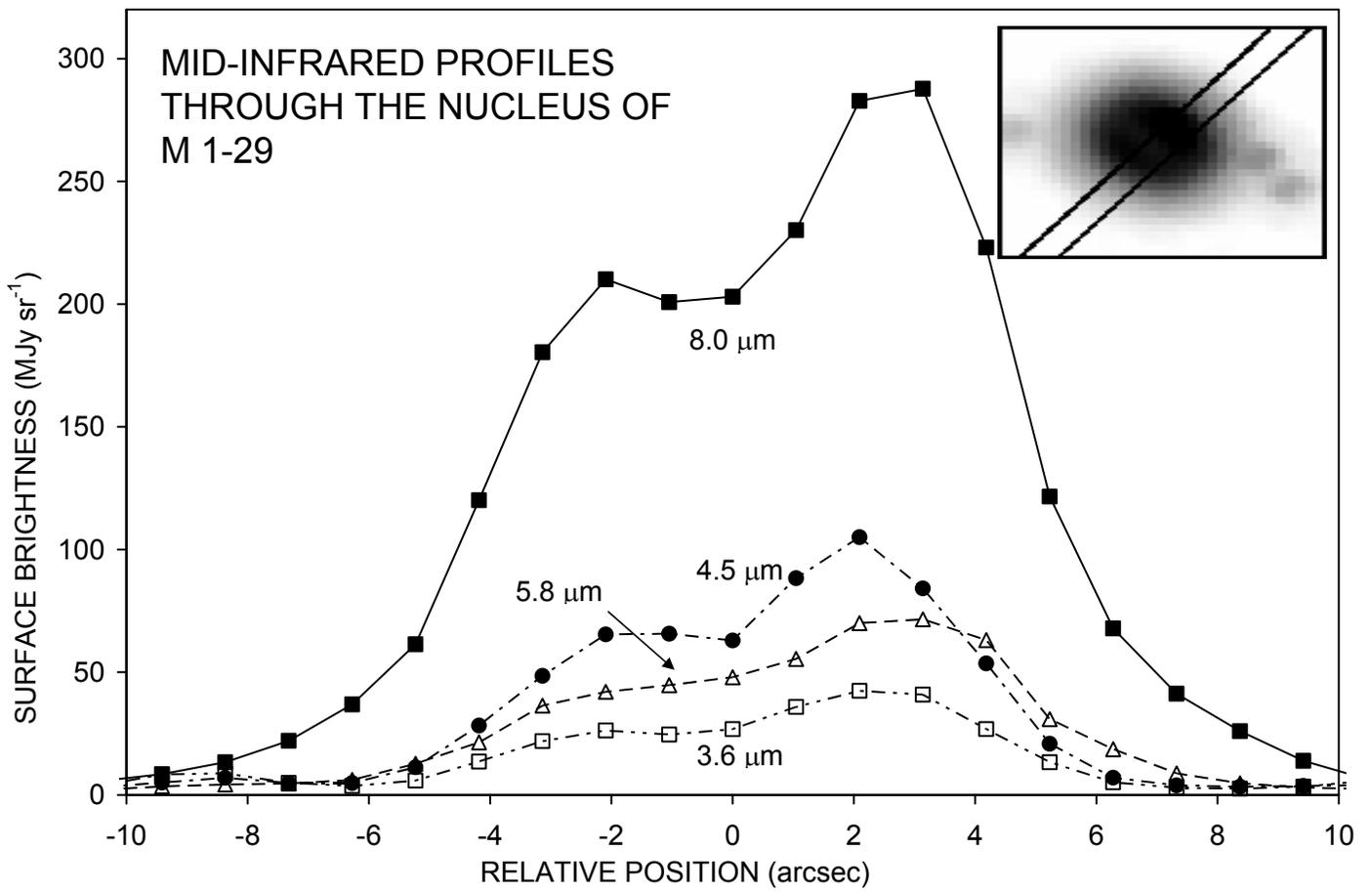
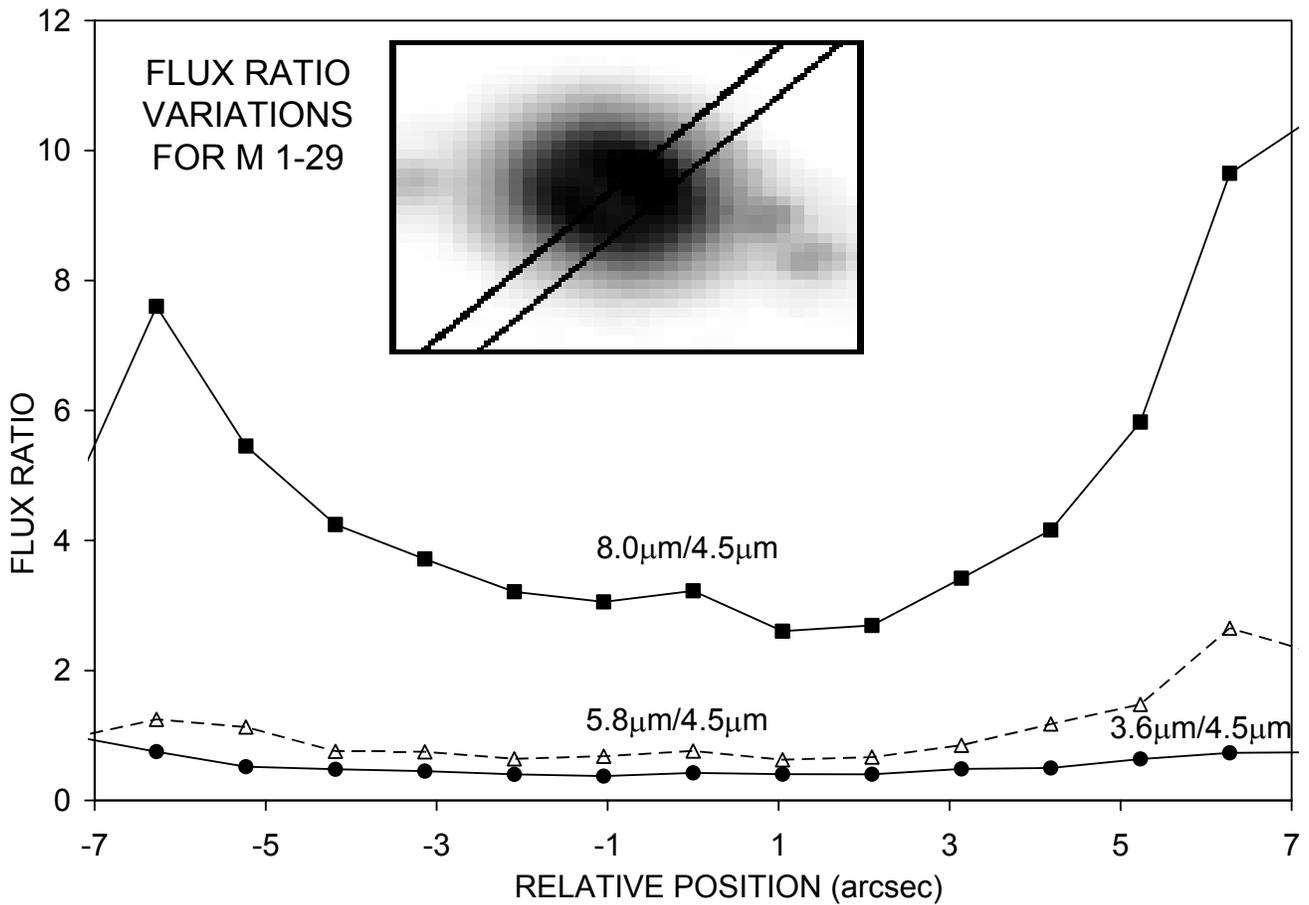

FIGURE 17

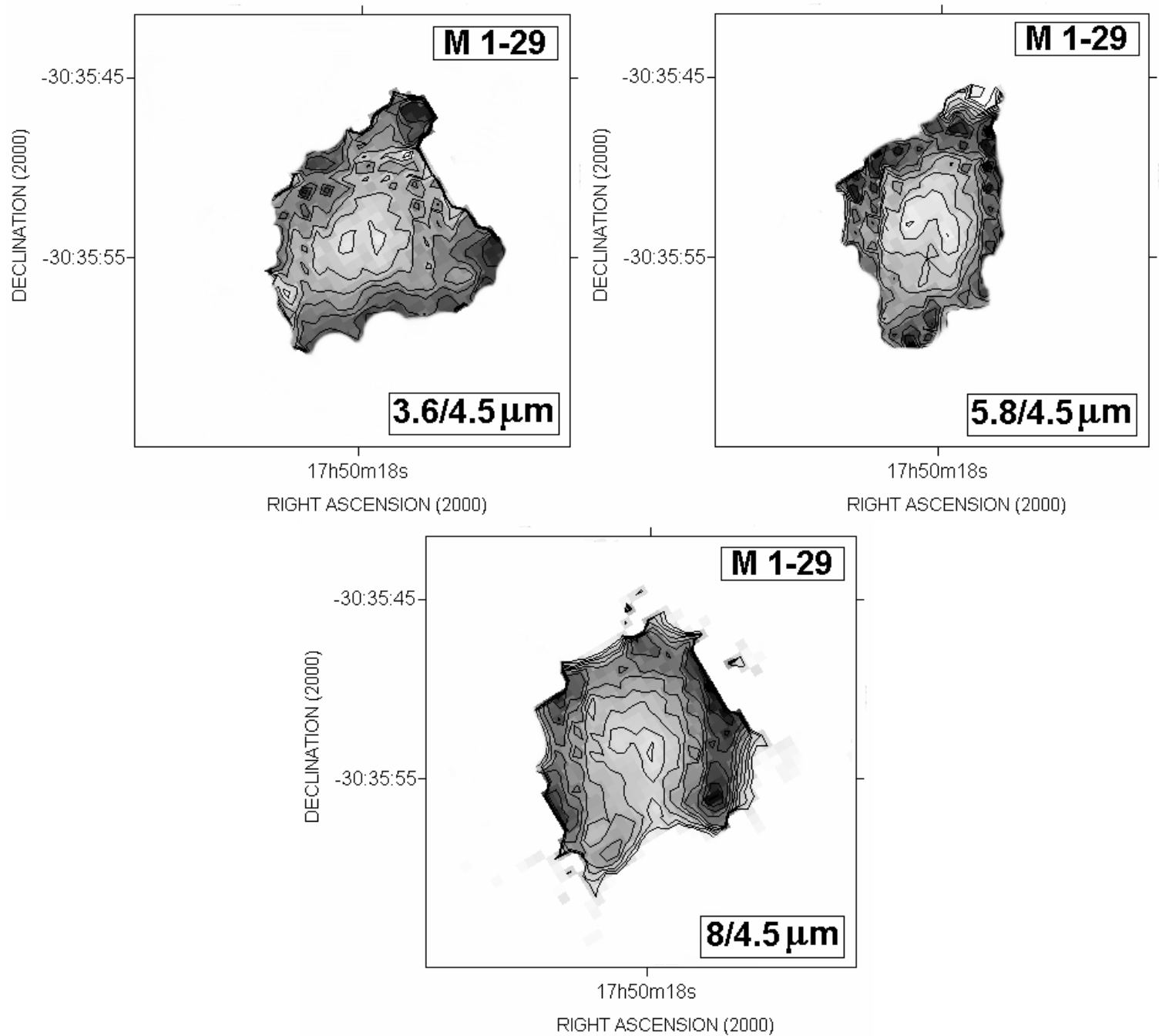

FIGURE 18

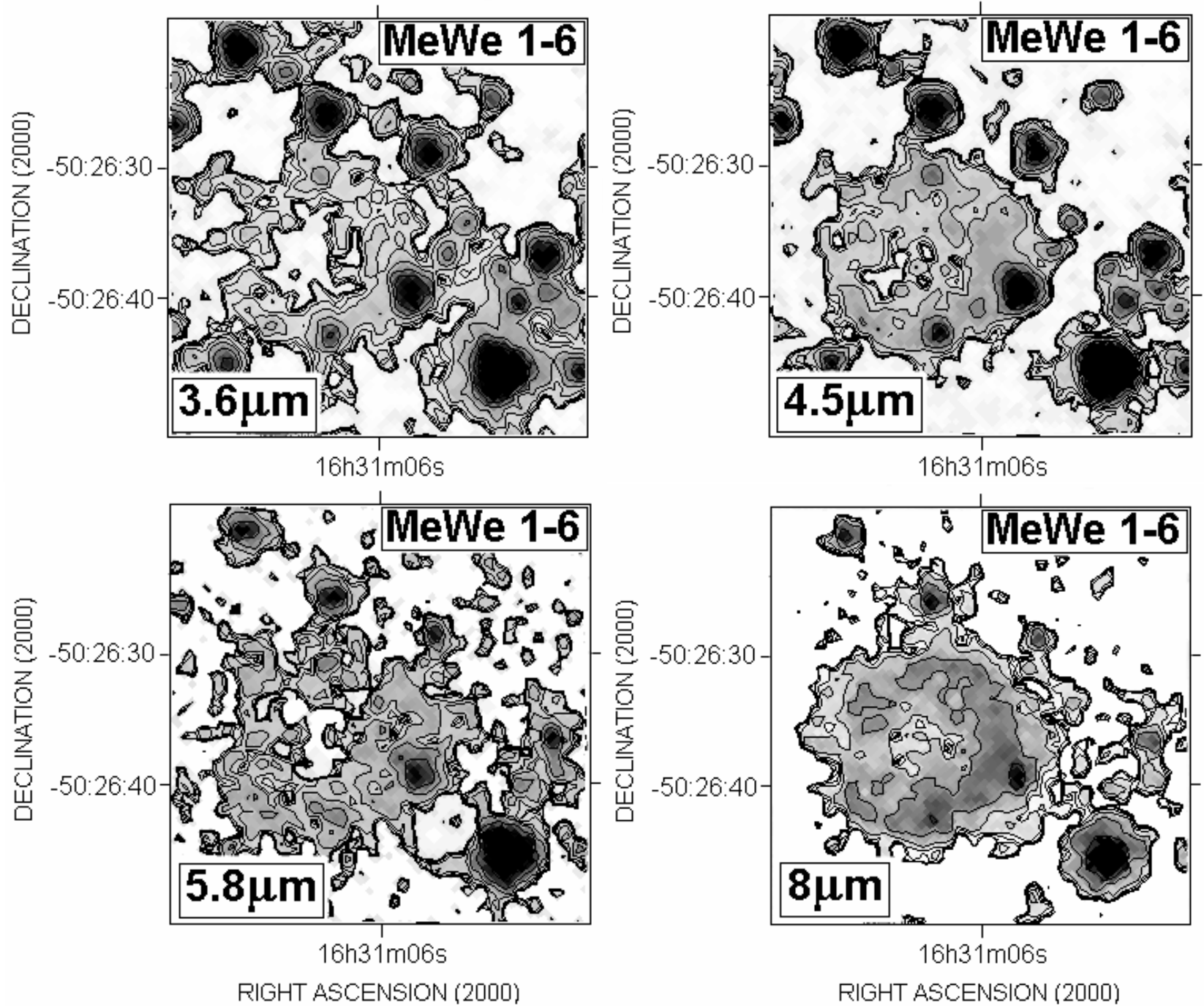

FIGURE 19

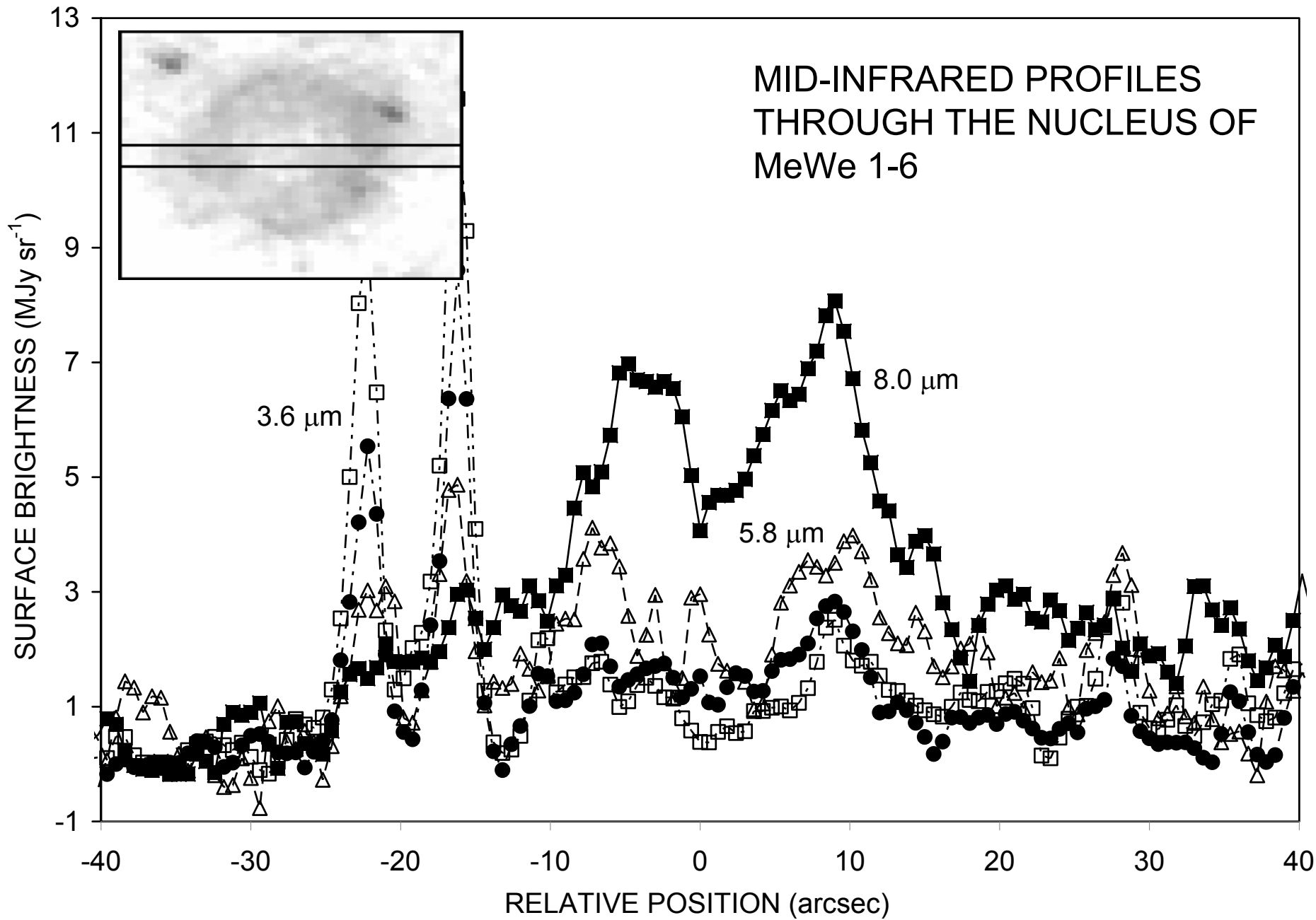

FIGURE 20

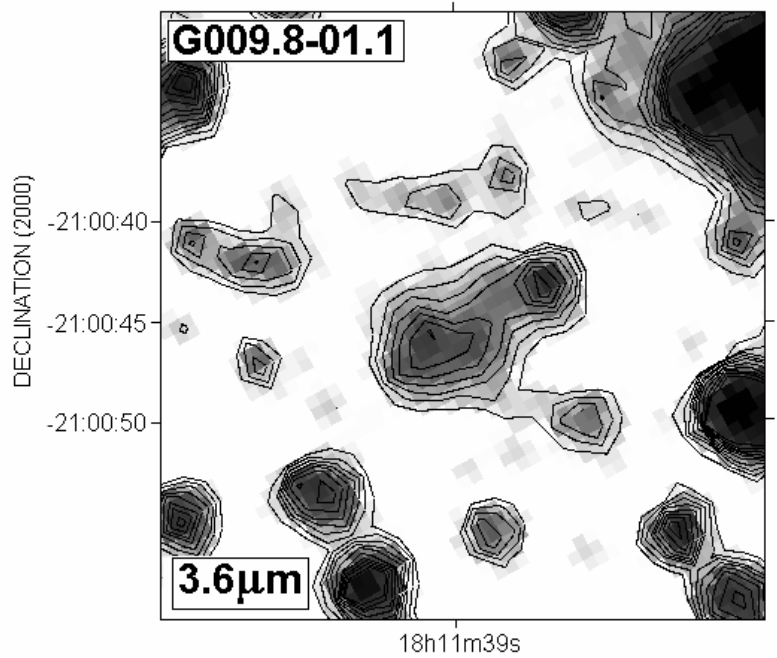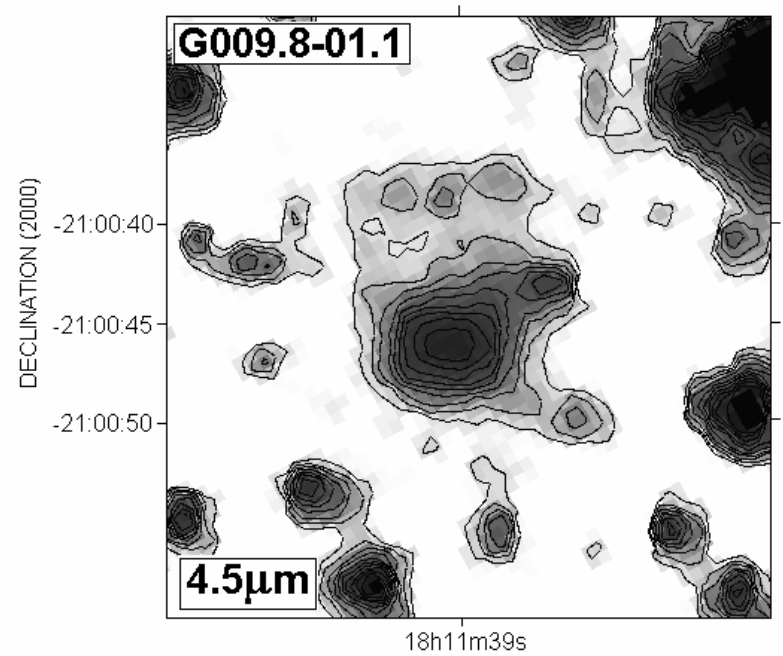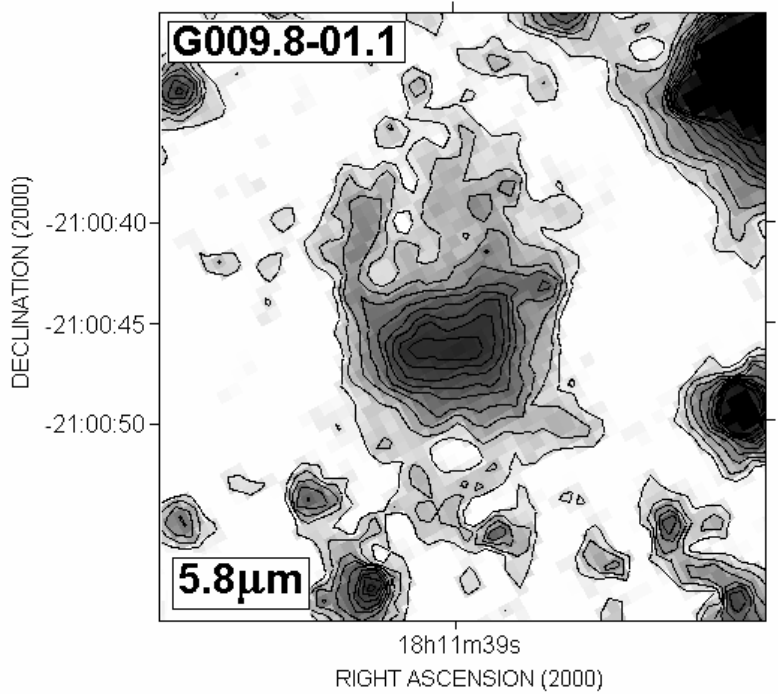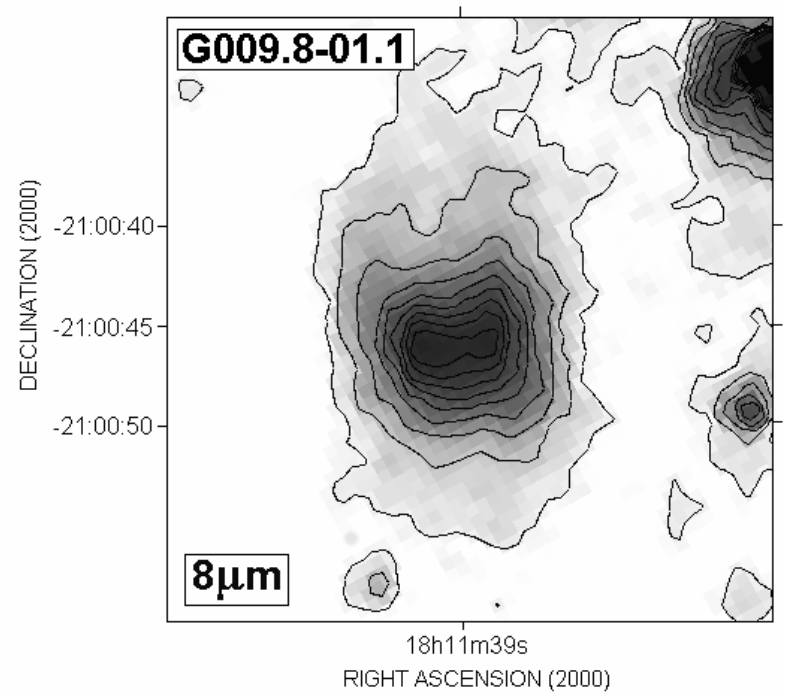

FIGURE 21

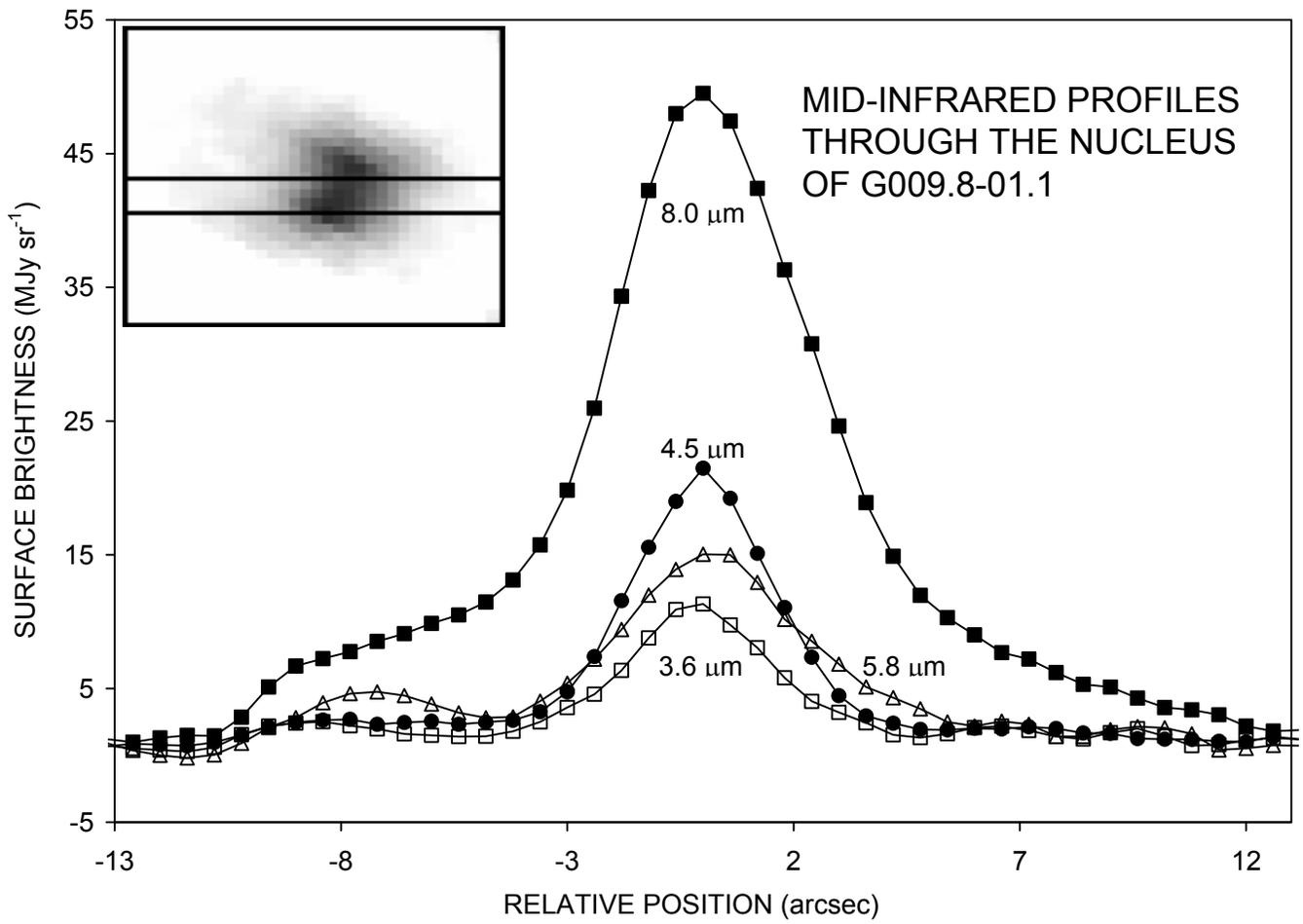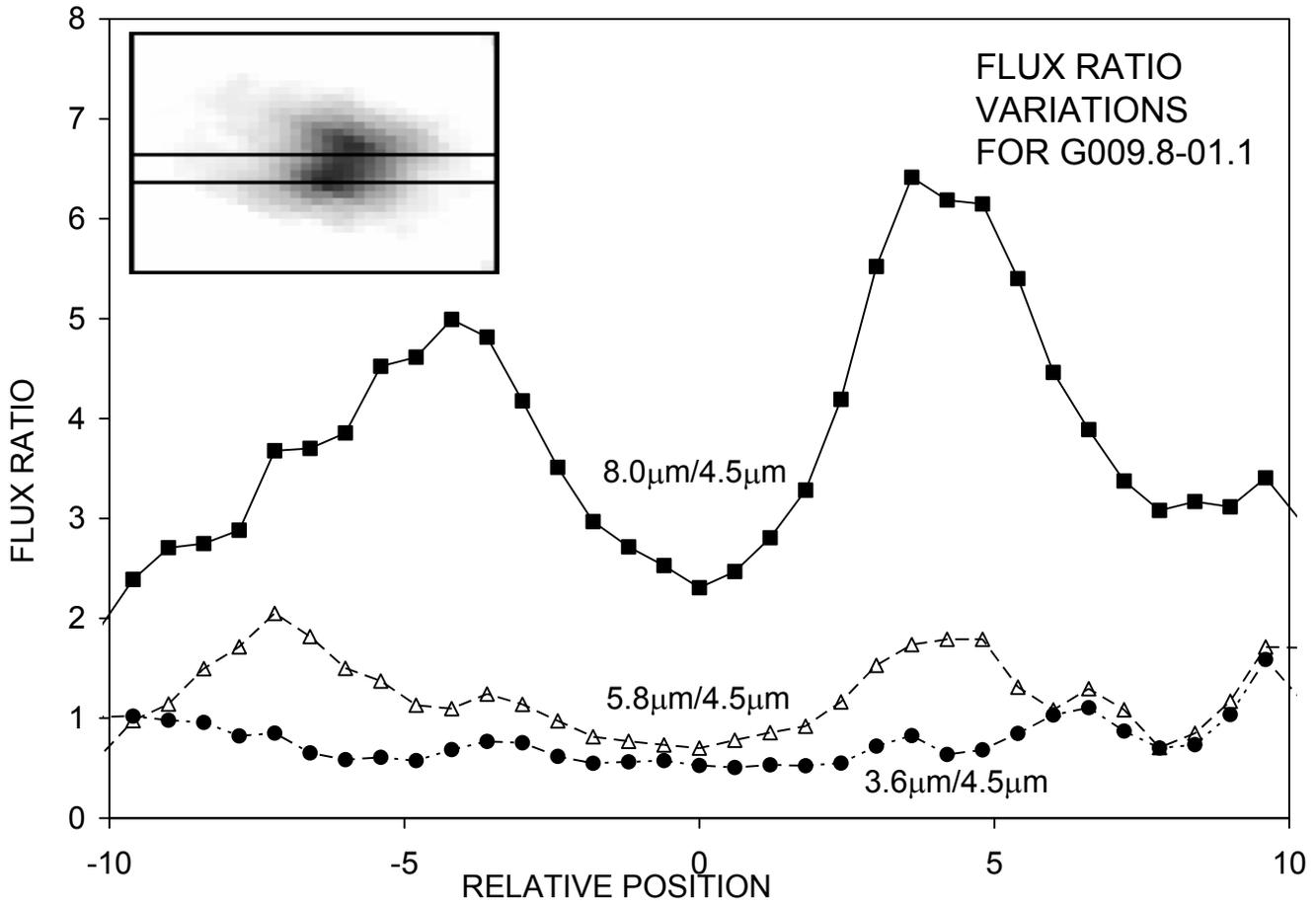

FIGURE 22

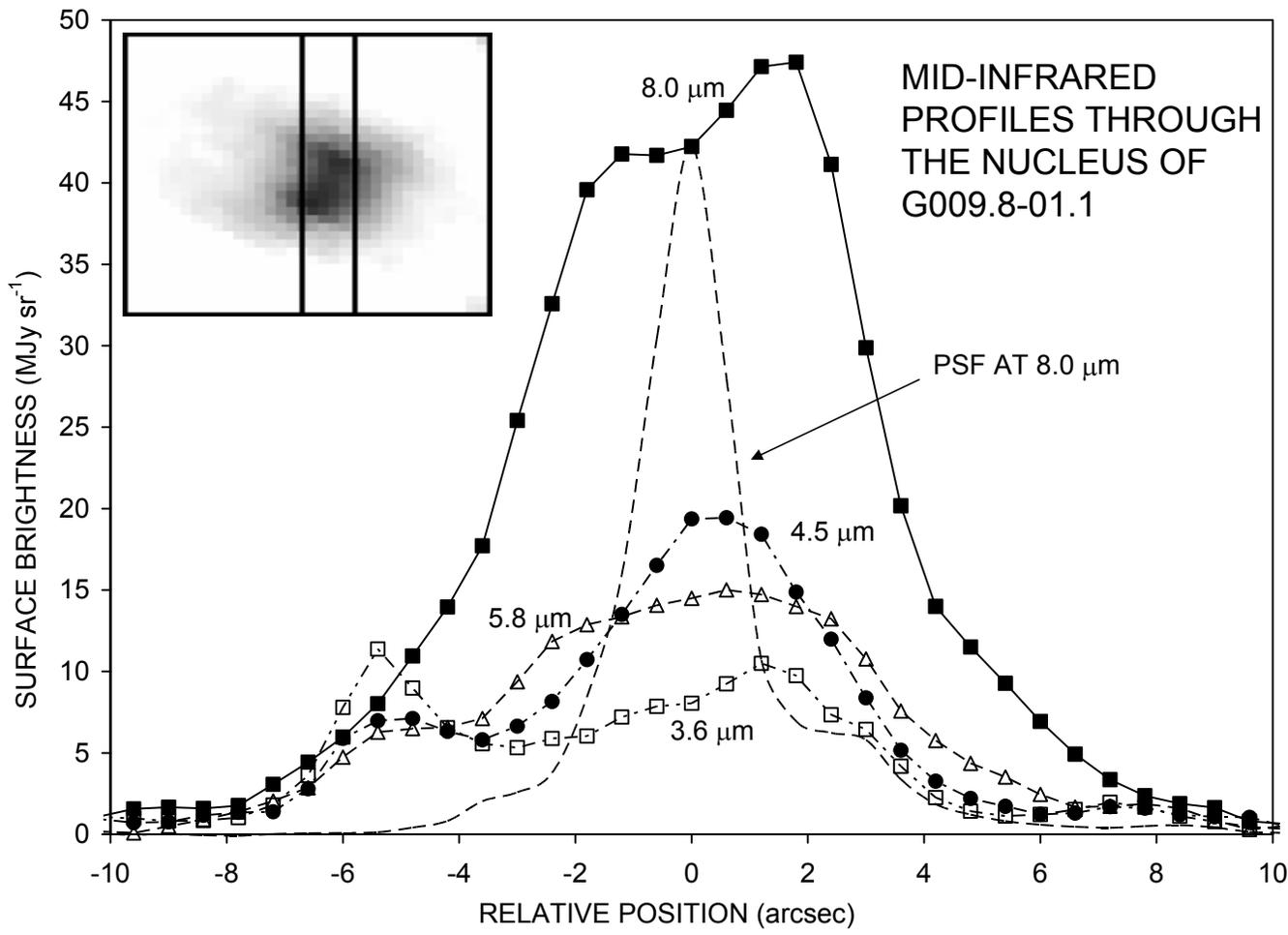
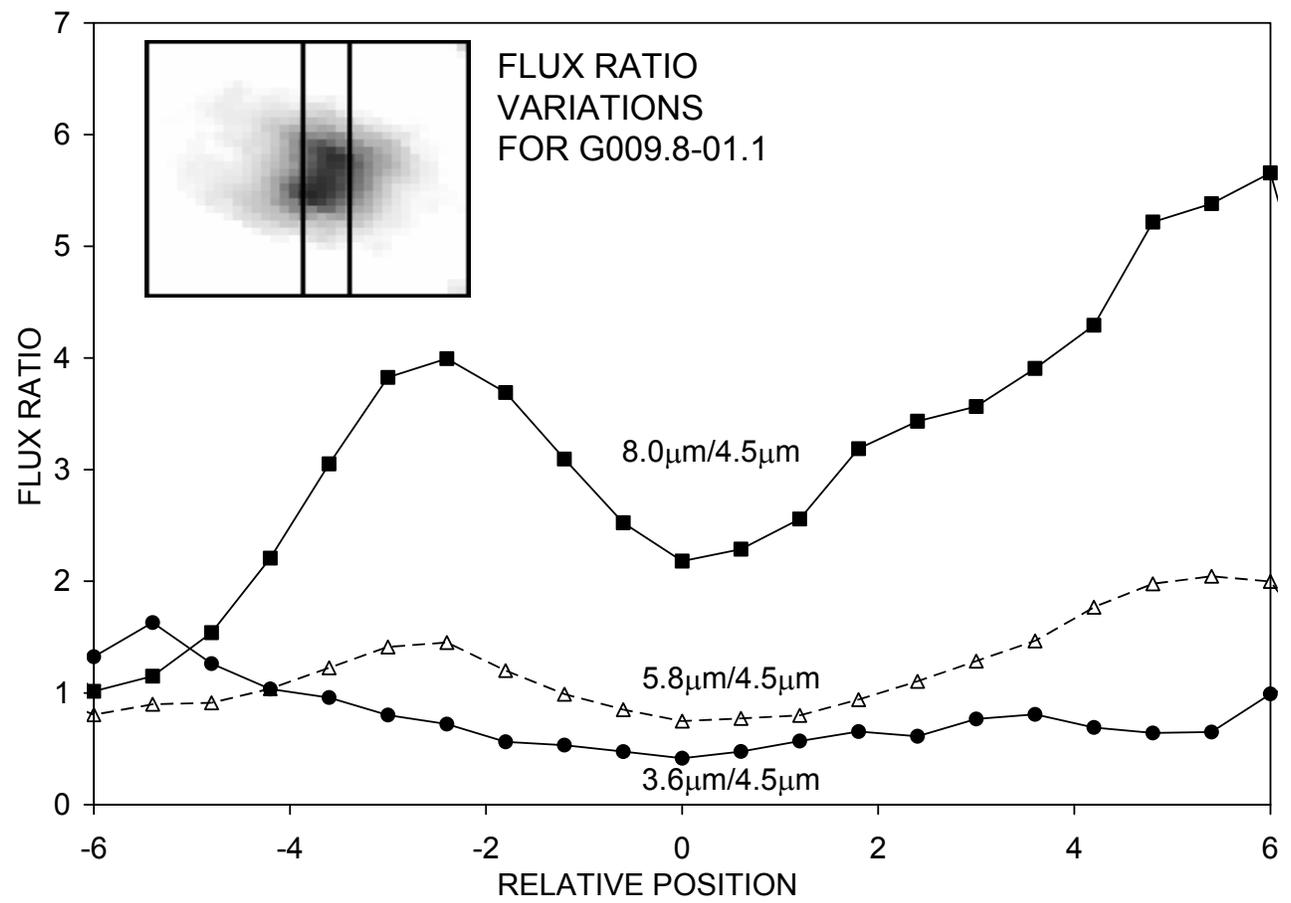

FIGURE 23

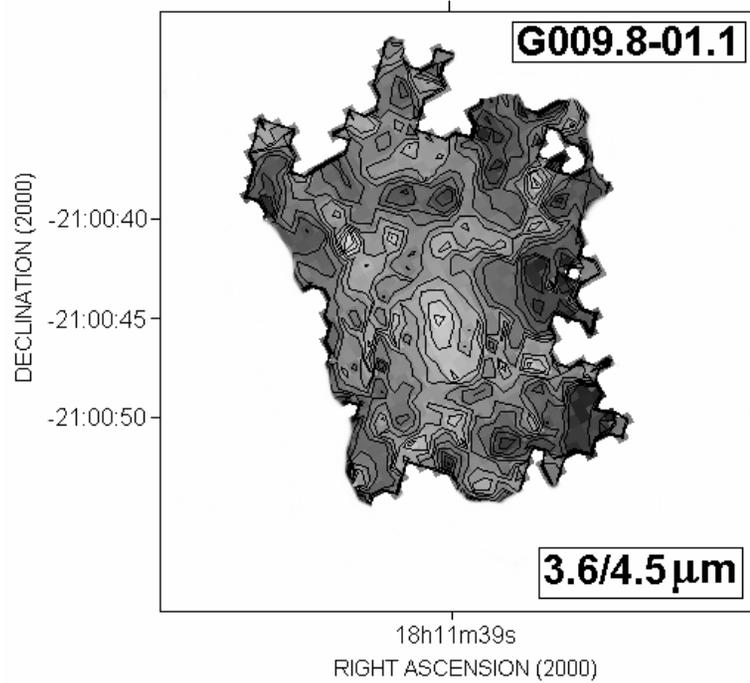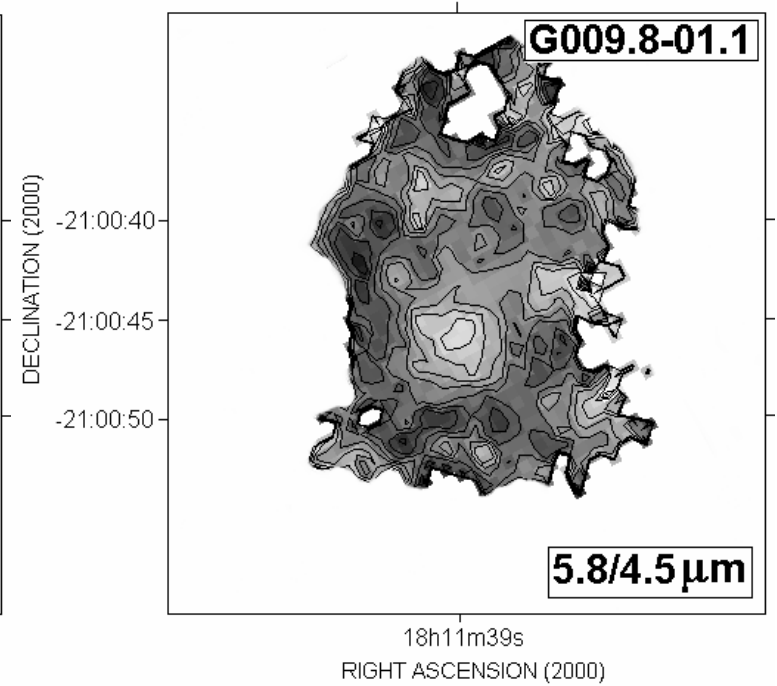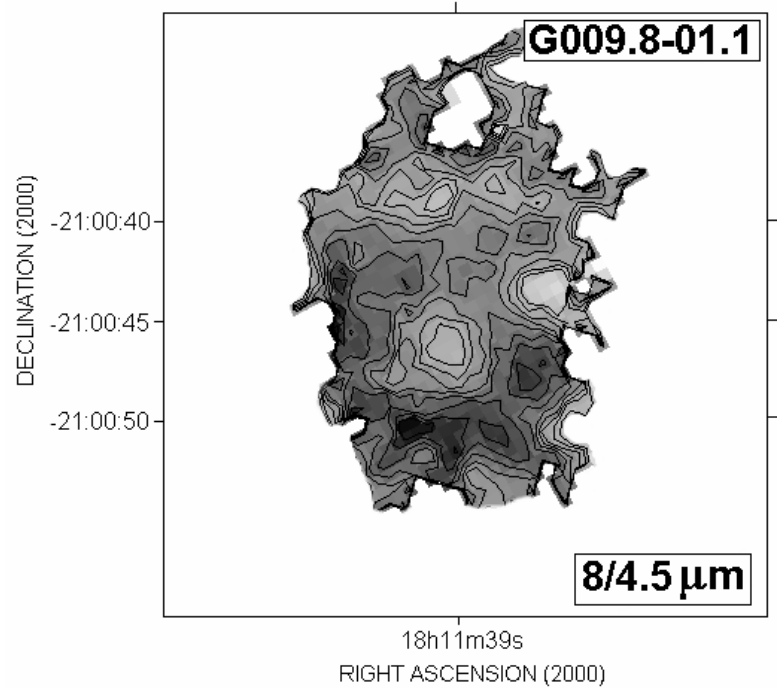

FIGURE 24

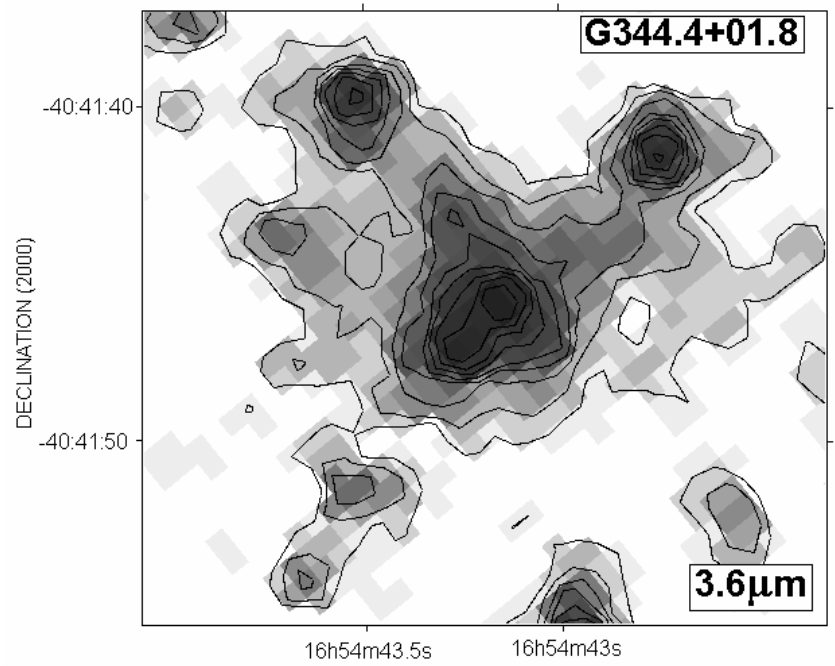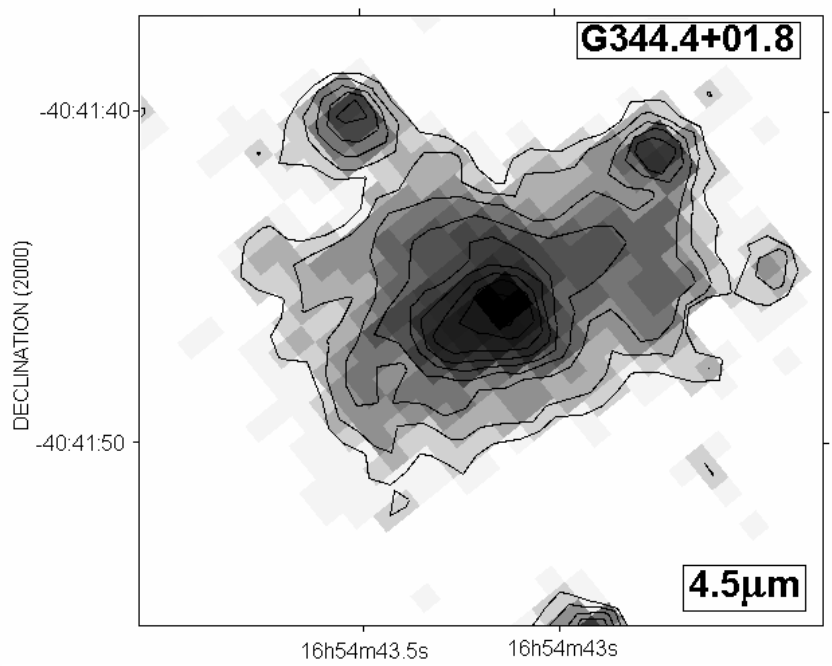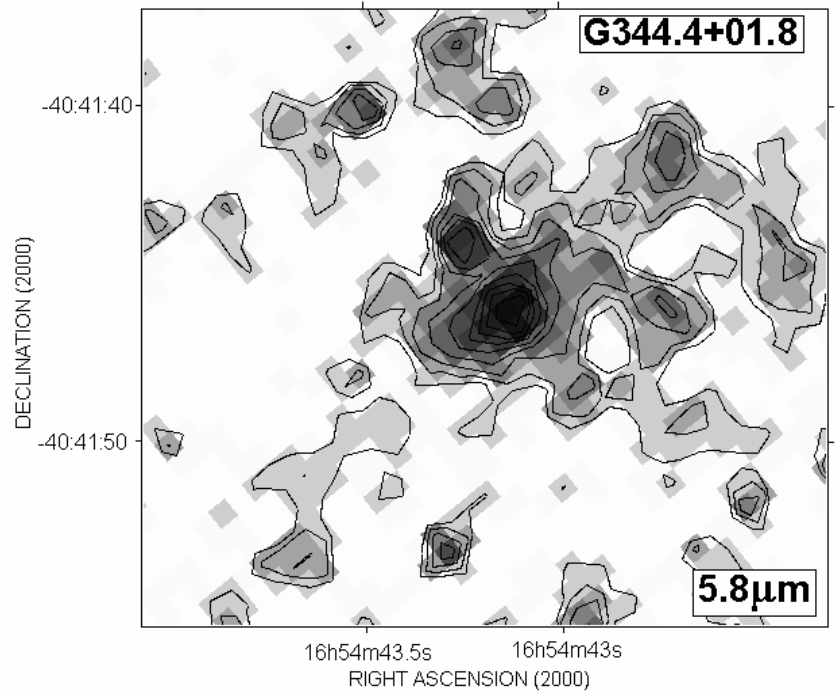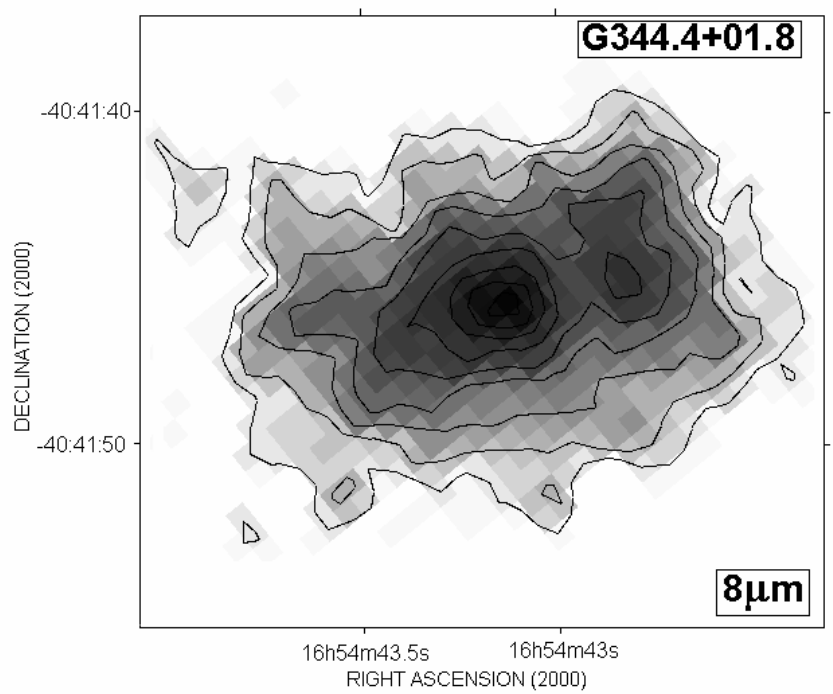

FIGURE 25

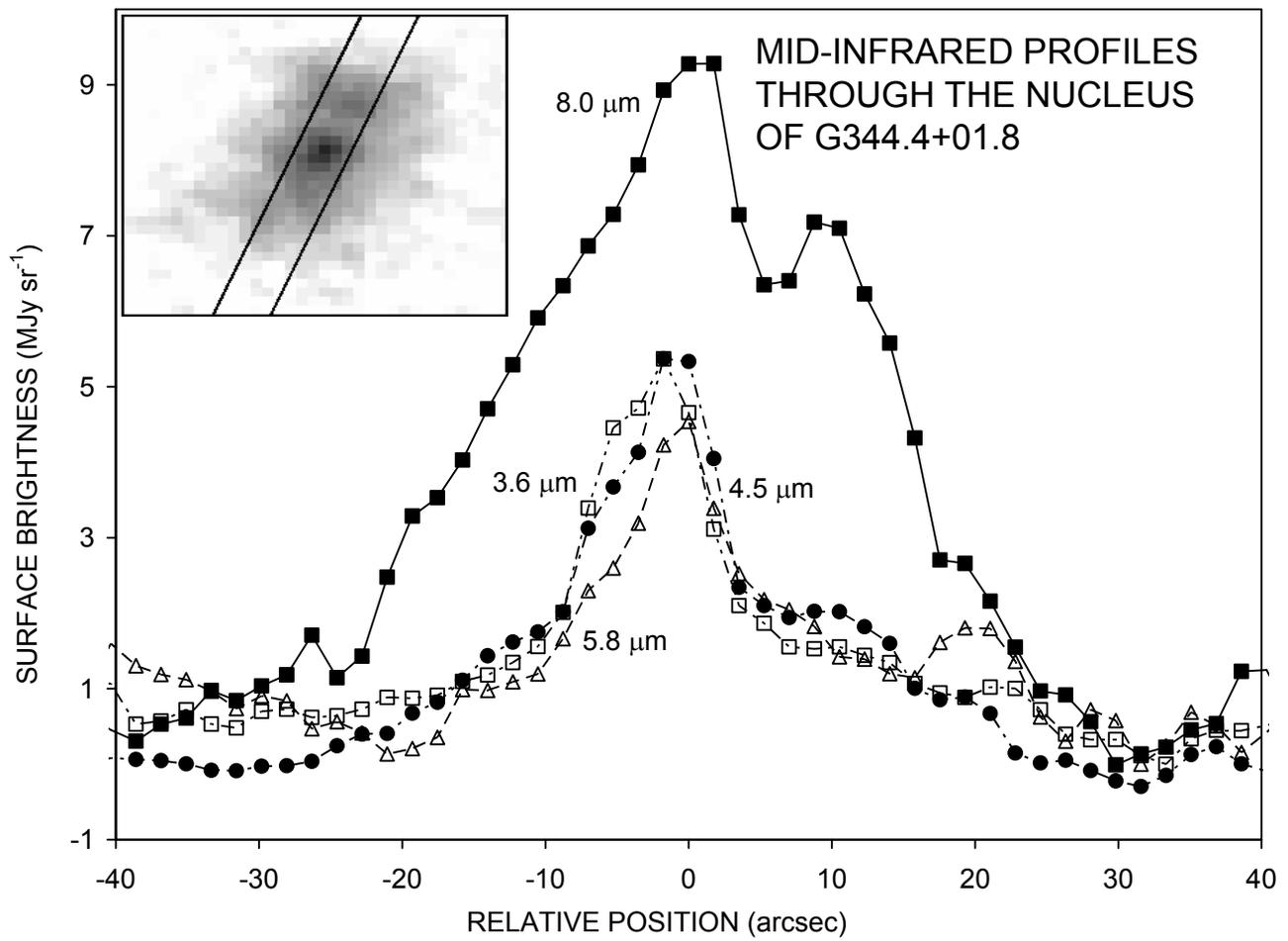

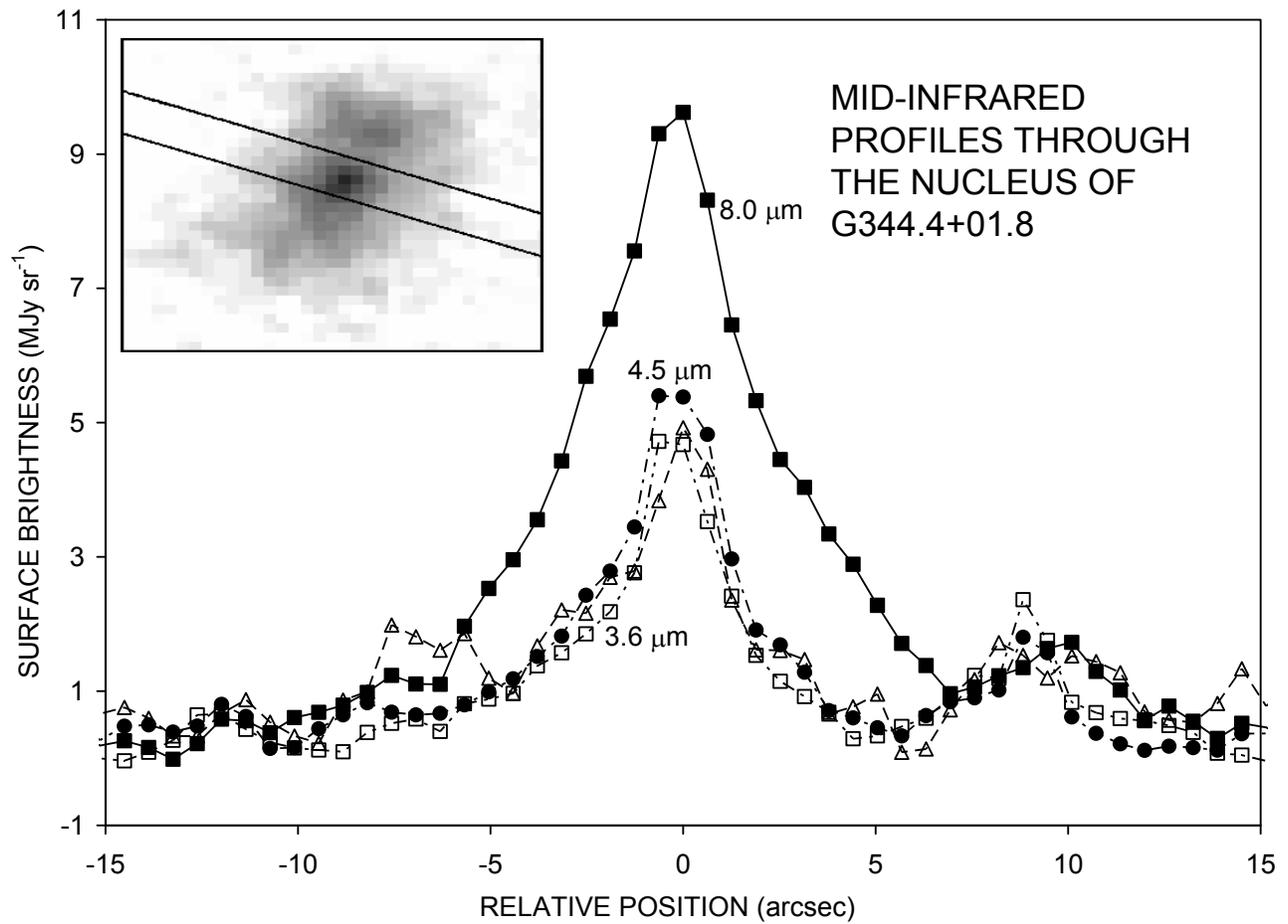

FIGURE 26

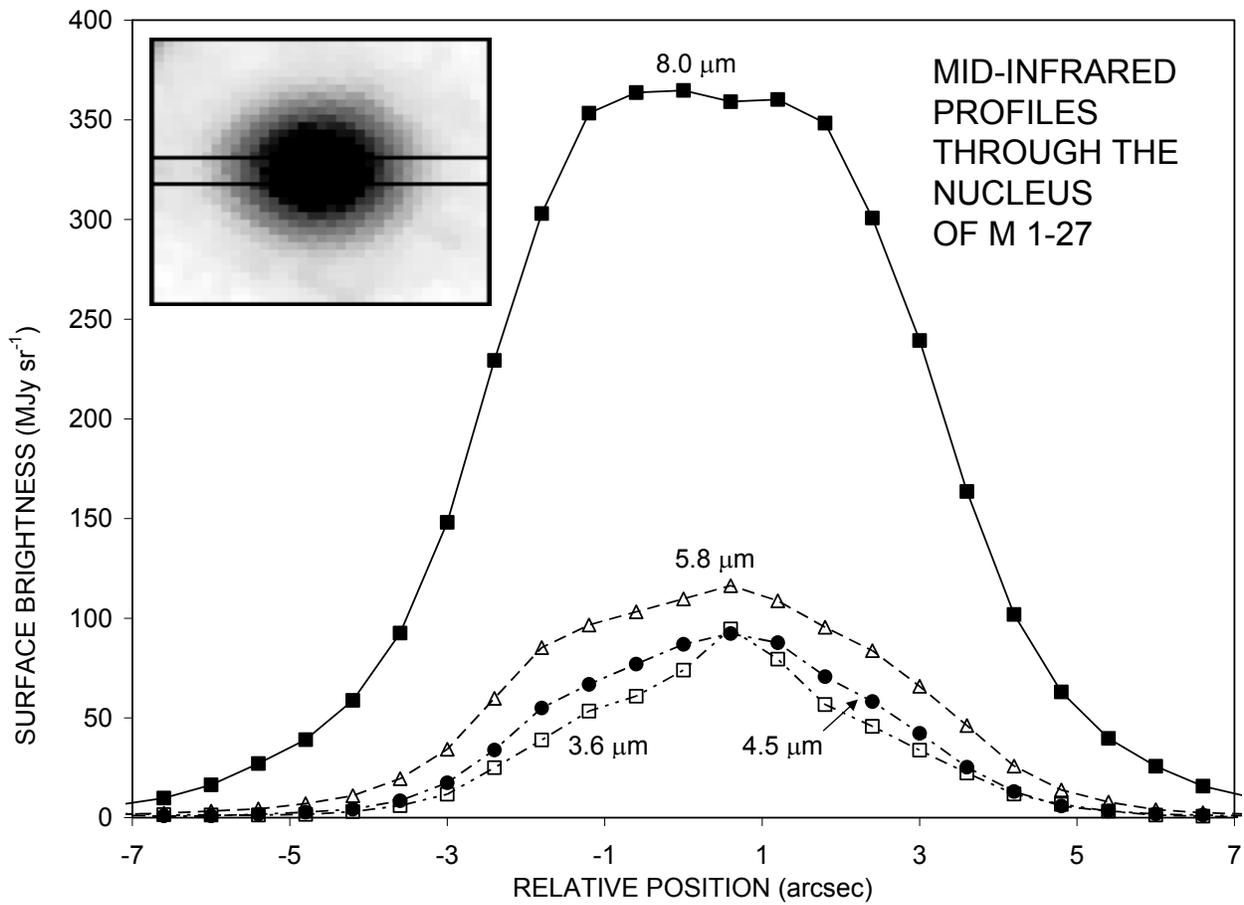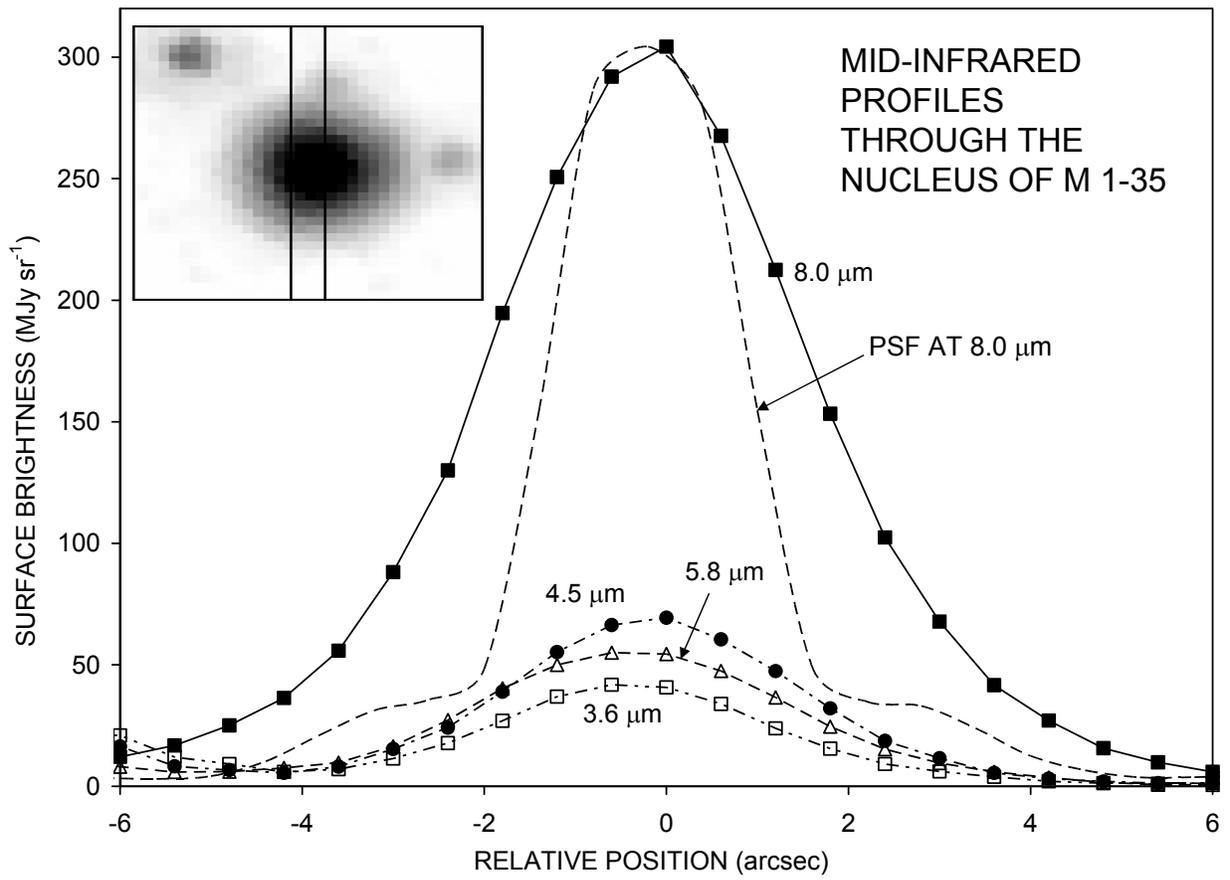

FIGURE 27

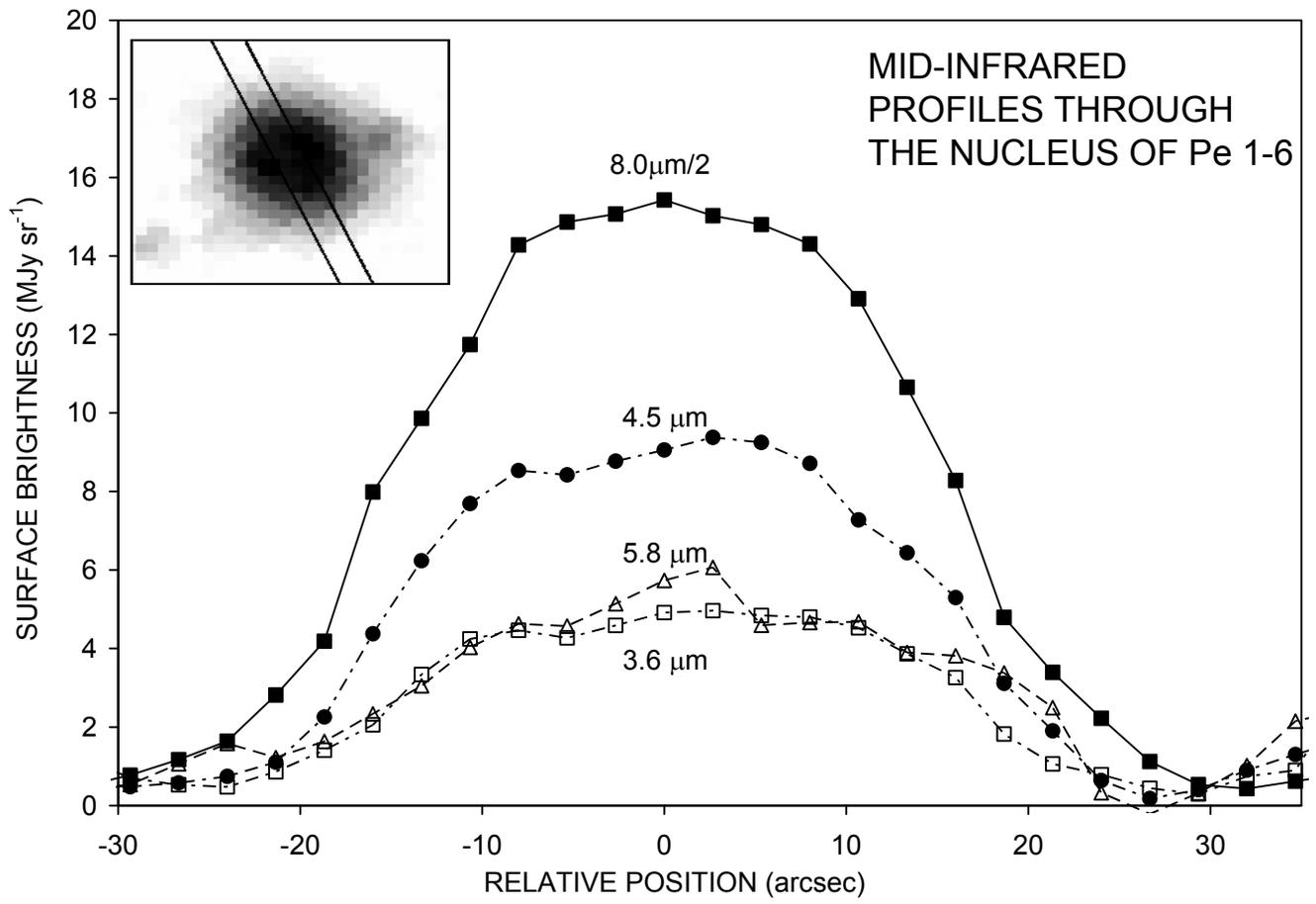

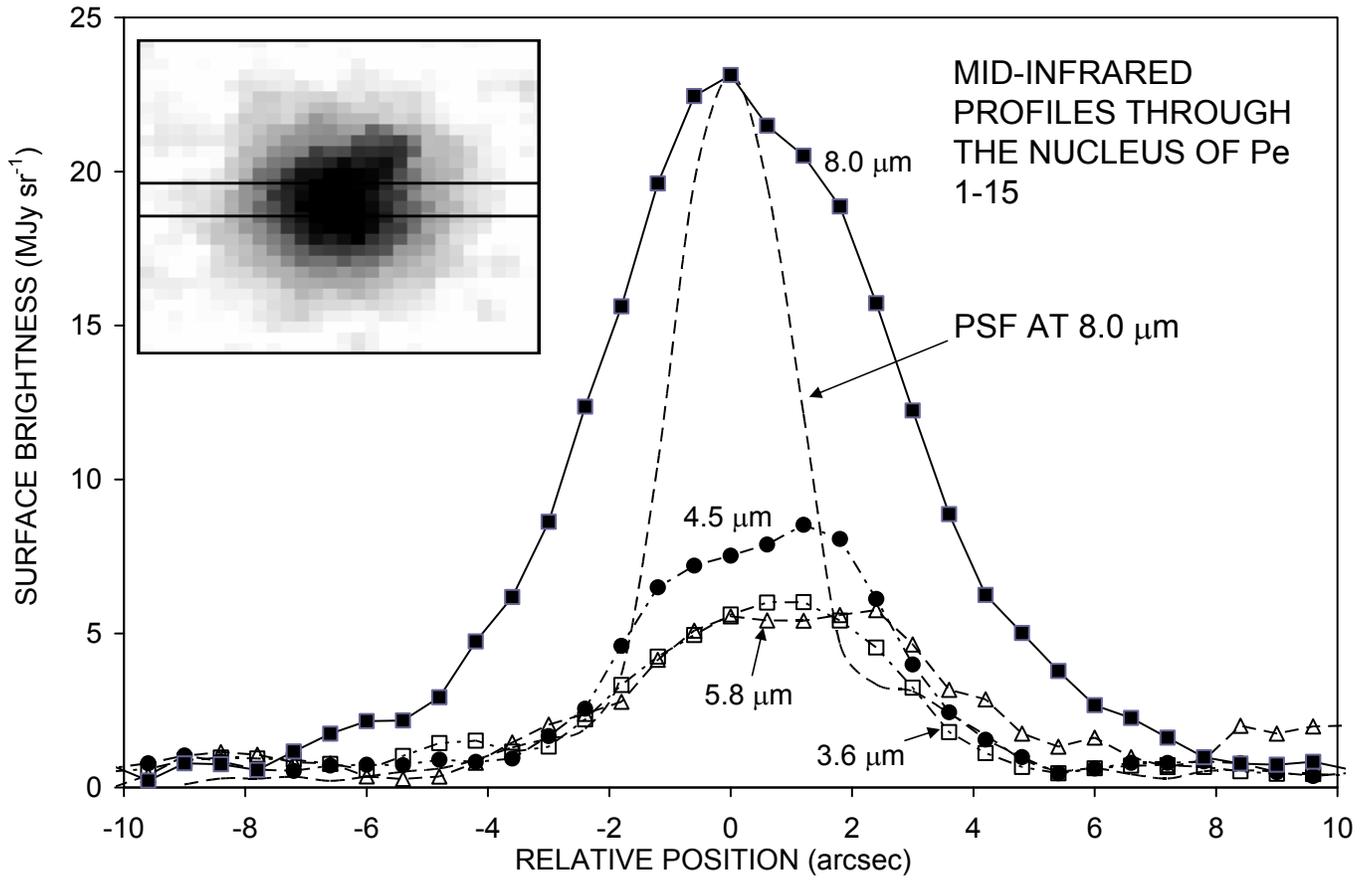

FIGURE 27 (CONT.)

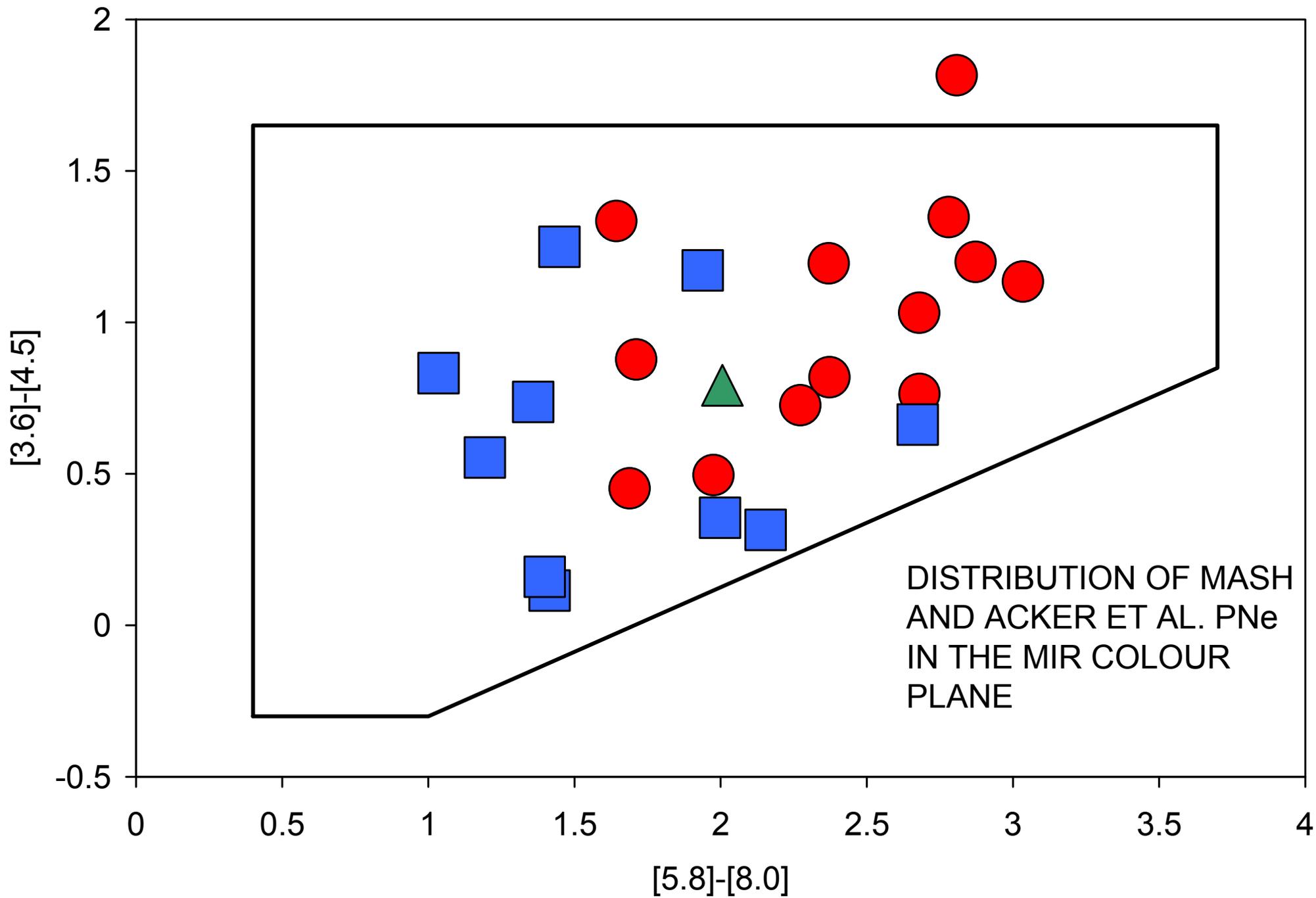

FIGURE 28
61